%% file: paper.tex
\documentclass[a4paper]{article}
\usepackage{etex}
\usepackage{a4wide}
\usepackage{graphicx}
\usepackage{amsbsy,amssymb,amsgen,amsfonts}
\usepackage{amsmath,amsfonts,color,latexsym}
\usepackage[pdftex,bookmarks=true,colorlinks=true,linkcolor=blue,citecolor=blue]{hyperref}
\usepackage[toc,title]{appendix}
\usepackage[authoryear,round,longnamesfirst]{natbib}
\shortcites{amoura:10,caporaloni:75,eames:11,gerashenko:08,glowinski:99,lsvezzo:10,mordant:02,qureshi:08,qureshi:07,tencate:04,calzavarini:09}
\usepackage{sectsty}
\allsectionsfont{\sffamily}
\let\LaTeXmaketitle\maketitle
\renewcommand{\maketitle}{{\sf\LaTeXmaketitle}}

\newcommand{\ben} {\begin{equation}}
\newcommand{\een} {\end{equation}}
\newcommand{\be} [1] {\begin{equation} \label{#1}}
\newcommand{\ee} {\end{equation}}
\newcommand{\bse} [1] {\begin{subequations} \label{#1}}
\newcommand{\ese} {\end{subequations}}
\newcommand{\ban} {\begin{eqnarray*} }
\newcommand{\ean} {\end{eqnarray*} }
\newcommand{\bea} {\begin{eqnarray}}
\newcommand{\eea} {\end{eqnarray}}

\newcommand{\solid}{---$\!$---$\!$--}

\newcommand{\dashed}{\hbox{{--}\,{--}\,{--}\,{--}}}

\newsavebox{\mybox}
\sbox{\mybox}{\dashed}

\newcommand{\comment} [1] {}
\newcommand{\MU} [1] {}
\newcommand{\MGV} [1] {}
\newcommand{\revision}[2]{#2}

\begin{document}
%

\title{
  DNS of vertical plane channel flow with finite-size particles:
  Voronoi analysis, acceleration statistics and particle-conditioned
  averaging 
}


\author{Manuel Garc\'ia-Villalba$^a$, 
  Aman G.\ Kidanemariam$^b$ and
  Markus Uhlmann$^b$
  \footnote{Corresponding author: {\tt markus.uhlmann@kit.edu}}
  \\[1ex]
  {\it\small
    $^a$ Bioingenier\'{\i}a e Ingenier\'{\i}a Aeroespacial, Universidad
    Carlos III de Madrid, Legan\'es 28911, Spain}\\[0ex]
  {\it\small $^b$ Institute for Hydromechanics, Karlsruhe Institute of
    Technology (KIT), 76131 Karlsruhe, Germany}
}
%
\date{
}
\maketitle
\input{abstract}
%
%
\input{intro}
\input{numa}
%
\input{results_a}
%
\input{results_b}
%
\input{results_c}
%
\input{results_d}
%
\input{conclusion}
\input{ack}
%
%
\addcontentsline{toc}{section}{Bibliography}
%

\input{paper.bbl}
\clearpage
\begin{appendices}
  \input{app_averaging}
  \input{app_pure_stats}
\end{appendices}
%
%
%
%
%
%
%
%
%
\end{document}

%% file: abstract.tex
\begin{abstract}
  %
  We have performed a direct numerical simulation of dilute turbulent
  particulate flow in a vertical plane channel, fully resolving the phase
  interfaces. The flow conditions are the same as those in the main 
  case of ``Uhlmann, M., {\it Phys.\ Fluids}, vol.\ 20, 2008,
  053305'', with the exception of the computational 
  domain length which has been doubled in the present study. The
  statistics of flow and particle motion are not significantly altered by
  the elongation of the domain. The large-scale columnar-like structures
  which 
  had previously been identified do persist and they  
  are still only marginally decorrelated in the prolonged domain. 
  Voronoi analysis of the spatial
  particle distribution shows that the state of the dispersed phase
  can be characterized as slightly more ordered than random tending
  towards a homogeneous spatial distribution. 
  It is also found that
  the p.d.f.'s of Lagrangian particle accelerations for wall-normal and
  spanwise directions follow a lognormal distribution as observed in
  previous experiments
  of homogeneous flows. The streamwise component deviates from this law
  presenting significant skewness.  
  Finally, a statistical analysis of the flow in the near field around
  the particles reveals that particle wakes present two regions, a
  near wake where the velocity deficit decays as $x^{-1}$ and a far
  wake with a decay of approximately $x^{-2}$. 
  \\[1ex]
\end{abstract}

%% file: intro.tex
\section{Introduction}\label{sec-intro}
Fluid flow with suspended solid particles 
is encountered in a multitude of natural and industrial systems. 
Examples include the motion of sediment particles in rivers, 
fluidized beds and blood flow. 
%
Despite the great technological importance of these systems 
our understanding of the dynamics of fluid-particle interaction is
still incomplete at the present date.  
%
Recently, however, significant progress has been made based on data
provided by new experimental methods as well as numerical simulations.  
While most past investigations of numerical type have been performed
in the context of the point-particle approach, it has now become
possible to simulate the motion of a considerable number of
finite-size particles including an accurate description of the
surrounding flow field on the particle scale
\citep{pan:97,kajishima:02,tencate:04,uhlmann:08a,lucci:10,lucci:11}. 
Although the complexity of these particle-resolved simulations (in
terms of Reynolds number, number of particles and computational domain
size) is still limited, 
new insight into the physics of fluid--particle systems is beginning to 
emerge from such studies. 

\cite{uhlmann:08a} has simulated turbulent flow in a
vertically-oriented plane channel seeded with heavy spherical
particles with a diameter corresponding to approximately $11$ wall
units at a solid volume fraction of $0.4$\%. 
The pressure-driven upward flow (at
constant flow rate) was found to be strongly modified due to the
particle presence, with increased wall-shear stress and strongly
enhanced turbulence intensity.
The average relative flow, corresponding to a Reynolds number (based
on particle diameter) of approximately $135$, lead to the
establishment of wakes behind individual particles.  
Additionally, the formation of very large-scale, streak-like flow structures
(essentially spanning the entire box-size), 
absent in corresponding single-phase flow, 
was observed. 
At the same time the dispersed phase did not exhibit any of the common
signs of preferential concentration.  
%

In the present study we are revisiting the same flow configuration of
vertical particulate channel flow, 
expanding upon the previous analysis of \cite{uhlmann:08a} 
by addressing several unanswered questions.
%
%
First, we wish to determine the influence of the streamwise length of the
computational domain upon the largest flow scales. For this purpose we
have performed new simulations analogous to the ones conducted by
\cite{uhlmann:08a}, but with twice the value of the original 
streamwise period, while keeping all remaining parameters unchanged. 

%
Second, 
we intend to provide a more complete
description of the turbulent fluid-particle interaction in vertical
channel flow.  
To this end we have analyzed three aspects of the flow dynamics which
had previously not been considered by \cite{uhlmann:08a}: 
Voronoi analysis of the spatial structure of the dispersed phase,  
analysis of particle acceleration statistics, and 
particle-conditioned averaging of the fluid flow field.  

Voronoi analysis is a relatively recent addition to the arsenal of
tools for the description of particles suspended in fluids
\citep{monchaux:10b}. In the present flow configuration it turns out
that this methodology provides a more sensitive measure of the
particle phase geometry than previously employed criteria. 

The statistical properties of particle acceleration have 
received increasing attention in recent years \citep{toschi:09}.
Since particle acceleration is (up to particle mass) equivalent to the
resulting forces acting upon the particles, its analysis can be
instrumental in understanding turbulence-particle interaction
mechanisms. 
One application where the influence of turbulence upon particle
acceleration statistics is believed to be of key importance is
the growth of rain drops by collisions in atmospheric clouds 
\citep{warhaft:09}. 
Modern experimental results on the acceleration of {\it finite-size} 
particles \citep{qureshi:07,xu:08,brown:09} have only started to
emerge around the date of publication of the precursor paper
\citep{uhlmann:08a}.   
Therefore, such an analysis was not carried out 
therein. 
Here we present a statistical analysis of particle
acceleration/hydrodynamic forces, relating the
findings to available experimental results.

Finally, the understanding of the interaction between solid particles
and fluid turbulence does not seem complete without a statistical
analysis of the flow in the near-field around the particles. In order
to investigate the characteristics of particle-induced wakes 
and with the aim to provide data which might be useful for the purpose
of two-phase flow modelling, we have undertaken a study of
particle-conditioned averaging of the flow field. 
Reference data for fixed particles swept by (essentially) 
homogeneous-isotropic flow \citep{bagchi:04,amoura:10}
as well as wall-bounded shear flow \citep{wu:94b,legendre:06,zeng:10} 
is available 
and has been used 
for the purpose of comparison.
%

%% file: numa.tex
\section{Computational setup}\label{sec-numa}
\subsection{Numerical method}\label{sec-numa-meth}
The numerical method employed in the current simulations is identical
to the one detailed in \cite{uhlmann:04} which was already used
for the previous simulations of vertical particulate channel flow by 
\cite{uhlmann:08a}. The incompressible Navier-Stokes equations are
solved by a fractional step approach with implicit treatment of the
viscous terms (Crank-Nicolson) and a three-step Runge-Kutta scheme for
the non-linear terms. The spatial discretization employs second-order
central finite-differences on a staggered mesh. 
The no-slip condition at the surface of moving solid particles is
imposed by means of a specially designed immersed boundary technique
\citep{uhlmann:04}. The motion of the particles is computed from the
Newton equations for linear and angular motion of rigid bodies, driven
by buoyancy, hydrodynamic forces/torque and contact forces (in case of
collisions). 
Since the suspension under consideration is dilute, collisions are
treated by a simple repulsive force mechanism \citep{glowinski:99}
formulated such as to keep colliding particles from overlapping
non-physically. The same treatment is applied to particle-wall
encounters. 
It should be noted that the employed computational grid is uniform and
isotropic. The chosen grid width $\Delta x=\Delta y=\Delta z$ yields a
particle 
resolution of $D/\Delta x=12.8$, a resolution of the channel
half-width of $h/\Delta x=256$ and $\Delta x^+=0.67$ in terms of wall
units. 
Further information on our extensive validation tests and grid
convergence can be found in \cite{uhlmann:04,uhlmann:08a}
and further references therein. 
\begin{figure}
  \begin{center}
    \centerline{$(a)$\hspace*{.35\linewidth}$(b)$}
    \begin{minipage}[b]{.3\linewidth}
    \includegraphics[height=\linewidth,
    viewport=290 20 840 900,clip=true]
    {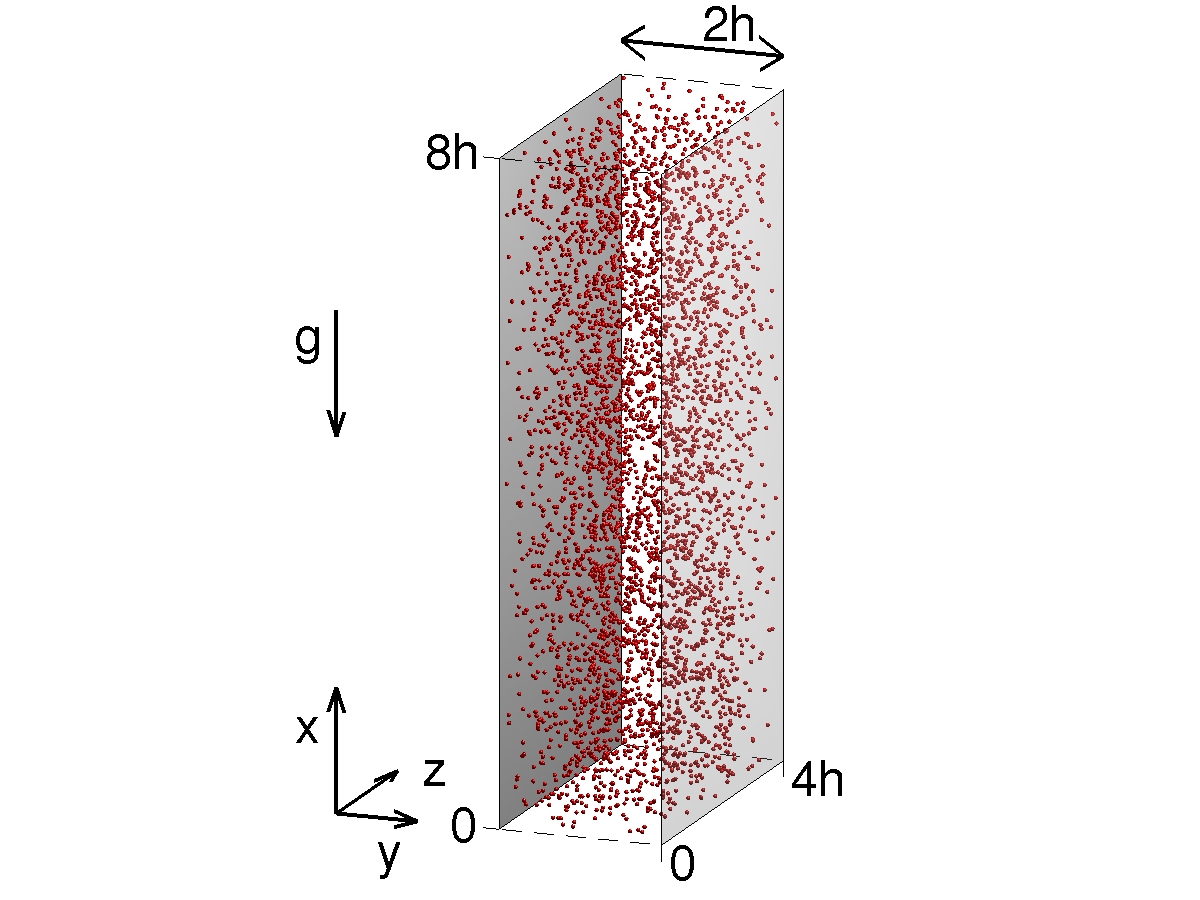}\\
    \end{minipage}
    \begin{minipage}[b]{.3\linewidth}
      \includegraphics[height=2\linewidth,clip=true,
      viewport=475 20 755 850]
      {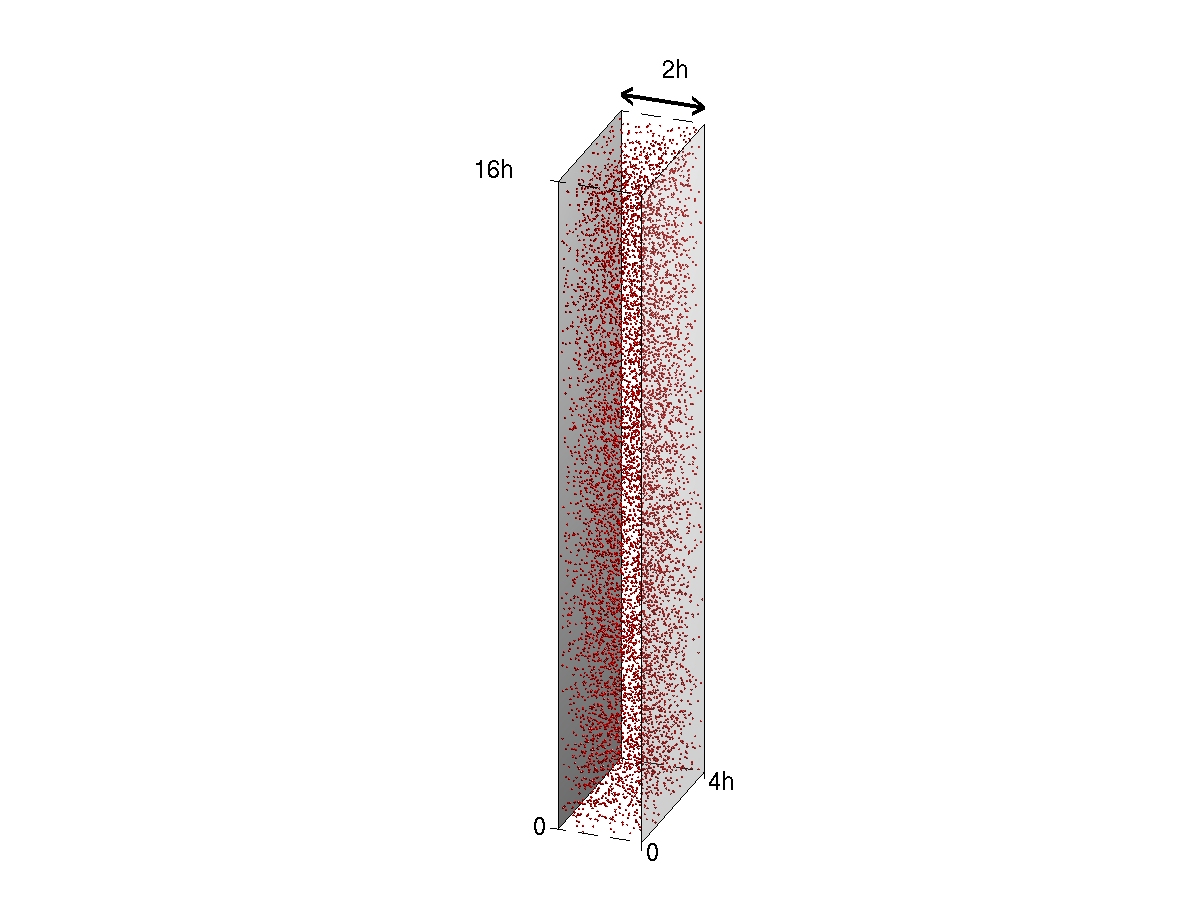}
    \end{minipage}
  \end{center}
  \caption{%
    Illustration of the computational domain, which is bi-periodic in
    the streamwise ($x$) and spanwise ($z$) directions. $(a)$ shows the
    domain used in \cite{uhlmann:08a}, $(b)$ the current domain which
    is a streamwise extension of the former (by a factor of two).
    The red spheres indicate actual instantaneous particle positions.  
  }
  \label{fig-domain}
\end{figure}
\begin{table}
  \centering
  \setlength{\tabcolsep}{5pt}
  \begin{tabular}{*{9}{c}}
    $Re_b$&
    $Re_\tau$&
    $\frac{\rho_p}{\rho_f}$&
    $|\mathbf{g}|h/u_b^2$&
    $St^+$&
    $St_{b}$&
    $h/D$&
    $D^+$&
    $\Phi_s$
    \\[1ex]
    $2700$&
    $220.9$&  
    $2.2077$&
    $12.108$&
    $15.5$&
    $0.83$&
    $20$&
    $11.25$&
    $0.0042$
    \\
  \end{tabular}
  \caption{Physical parameters for  particulate flow in a vertically
    oriented plane channel:
    bulk Reynolds number $Re_b$ (imposed quantity), 
    friction-velocity based Reynolds number $Re_\tau$ (derived
    quantity), 
    particle/fluid density ratio $\rho_p/\rho_f$, 
    gravitational parameter $|\mathbf{g}|h/u_b^2$,
    Stokes numbers based upon bulk units $St_b=\tau_p\,u_b/h$
    (imposed) and wall units $St^+=\tau_p\,u_\tau^2/\nu$ (derived), 
    length scale ratio $h/D$ (imposed) and $D^+$ (derived), 
    and global solid volume fraction $\Phi_s$. The particle relaxation
    time scale was defined as $\tau_p=\rho_pD^2/(\rho_f\,18\,\nu)$. 
  }    
  \label{tab-particles-channelp-params-phys}
\end{table}
\begin{table}
  \centering
  \setlength{\tabcolsep}{5pt}
  \begin{tabular}{*{5}{c}}
    case&
    $\Omega$&
    $N_x\times N_y\times N_z$&
    $N_p$&
    $t_{obs}u_b/h$
    \\[1ex] 
    \cite{uhlmann:08a}&
    $8h\times 2h\times 4h$&
    $2048\times513\times1024$&
    $4096$&
    $115$ 
    \\
    present&
    $16h\times 2h\times 4h$&
    $4096\times513\times1024$&
    $8192$&
    $90$  
    \\
  \end{tabular}
  \caption{Numerical parameters employed in the simulations: 
    computational domain size $\Omega$, 
    number of grid nodes $N_i$ in the $i$th
    coordinate direction, 
    number of particles $N_p$, 
    temporal observation interval $t_{obs}$ after
    discarding the initial transient.  
    The grid spacing in all cases is fixed at $\Delta x=h/256$,
    corresponding to $N_L=515$ Lagrangian force points distributed
    over the surface of each particle.
  }    
  \label{tab-particles-channelp-params-elong}
\end{table}
\begin{figure}
  \centering
  \begin{minipage}{3ex}
    $R_{uu}$
  \end{minipage}
  \begin{minipage}{.45\linewidth}
    \centerline{$(a)$}
    \includegraphics[width=\linewidth]
    {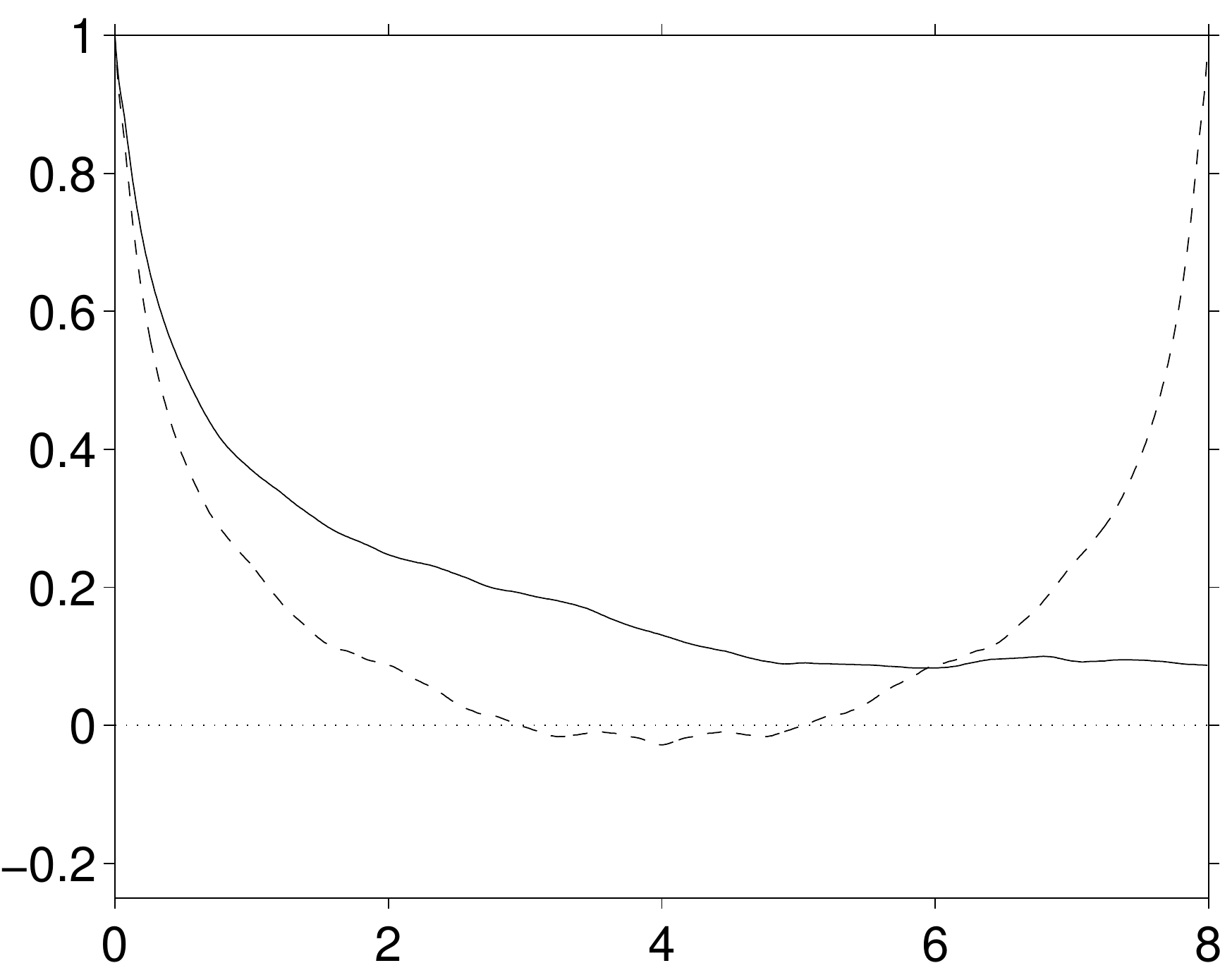}
    \\
    \centerline{$r_x/h$}
  \end{minipage}
  \begin{minipage}{4ex}
    $R_{ww}$
  \end{minipage}
  \begin{minipage}{.45\linewidth}
    \centerline{$(b)$}
    \includegraphics[width=\linewidth]
    {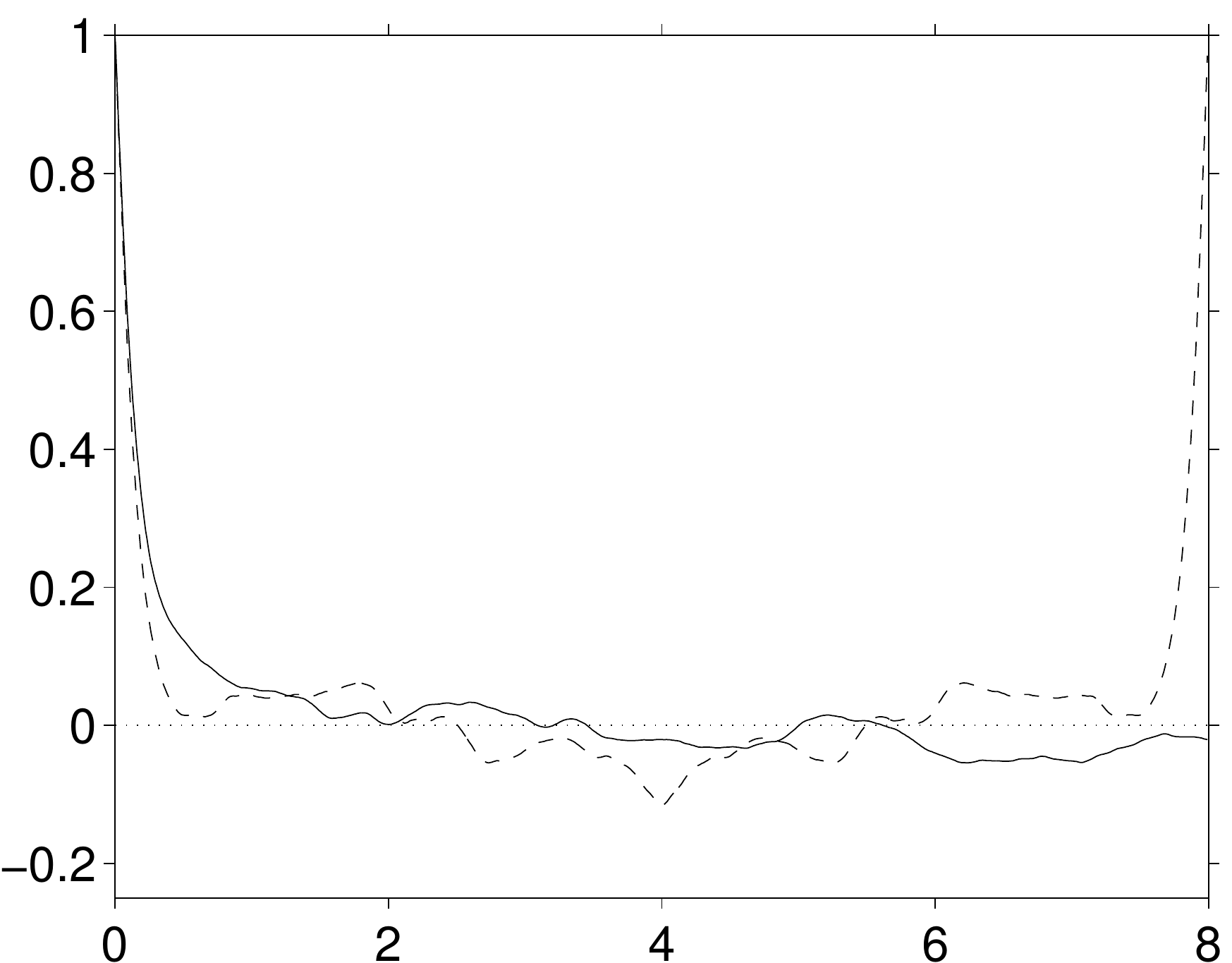}
    \\
    \centerline{$r_x/h$}
  \end{minipage}
  \caption{%
    Two-point correlation functions of fluid phase velocity
    fluctuations for streamwise separations 
    computed from a single snapshot of the flow field in a
    wall-parallel plane at a wall distance of 
    $y/h=0.0938$ (i.e.\ $y^+\approx22$). 
    $(a)$ streamwise velocity component $R_{uu}$, 
    $(b)$ spanwise velocity component $R_{ww}$.  
    The correlation immediately after periodic extension from the small-box
    simulation of \cite{uhlmann:08a} is shown as a dashed line; 
    the correlation at the beginning of the present sampling interval is
    indicated by a solid line.
  }
  \label{fig-corr}
\end{figure}
\subsection{Flow configuration}\label{sec-numa-config}
Figure~\ref{fig-domain} shows the flow configuration under
consideration as well as the coordinate system. 
The plane channel is oriented vertically, $x$ being the streamwise
coordinate direction, $y$ is the wall-normal (with the channel width
equal to $2h$) and $z$ the spanwise direction. 
Fluid flow is directed upwards (in the positive $x$ direction), driven
by a streamwise pressure-gradient. 
The bulk velocity $u_b$ is maintained at a
constant  value, such that the Reynolds number based upon the bulk
velocity, $Re_b=u_bh/\nu$, is imposed (cf.\
table~\ref{tab-particles-channelp-params-phys} for the values of the
principal physical parameters). 
%
%
A large number 
\revision{}{($N_p=8192$)} 
of monodispersed, rigid, spherical particles is
suspended in the flow. 
The nominal terminal velocity of the particles \citep[computed from an
equilibrium of buoyancy force and standard drag force,][]{clift:78} 
is set equal to the bulk velocity of the fluid phase. 
Consequently, the average particle settling velocity obtained in the
actual simulation is roughly zero.
The chosen density ratio $\rho_p/\rho_f=2.2077$ (where $\rho_p$, $\rho_f$ are
the particle and fluid densities) is comparable to the case of glass
particles in water. 
The particle diameter $D$, approximately equal to $11$ wall units, is
comparable to the cross-sectional scales of buffer layer flow
structures.  
Finally, table~\ref{tab-particles-channelp-params-phys} shows that the
suspension is indeed dilute, with less than one half percent of solid
volume fraction. 

As can be seen from table~\ref{tab-particles-channelp-params-elong},
the present simulation is performed in a computational domain which
has twice the streamwise period as compared to 
\cite{uhlmann:08a}, while maintaining an identical small-scale
resolution. The table also shows that an observation interval of
approximately 
$90$ 
bulk time units 
(defined as $T_b=h/u_b$)
has been computed after discarding the initial transient, which will
be discussed next. 
%

\revision{}{%
  The simulations were run on different supercomputing systems,
  typically using between 256 and 1024 processor cores. The total
  number of CPU hours spent was of the order of 5 million. 
}
\subsection{Simulation start-up and initial transient}
\label{sec-numa-transient}
The current simulation was initialized with an exact periodic extension
of a flow field taken at an instant towards the end of the simulation
of \cite{uhlmann:08a}. 
For this purpose, fluid and particle data in the interval
$x\in[0,8h]$ was copied from the reference field. 
In order to obtain data in the interval $x\in[8h,16h]$, 
the shift ${x}=\tilde{x}+8h$ 
was applied 
to the reference field. 
It should be emphasized that no explicit perturbations
whatsoever were added to the initial field. 

Subsequently, the extended simulation was run while different
quantities were monitored in order to determine whether the system has
developed sufficiently such as to ``forget'' the initial state. 
A sensitive measure of independence from the initial condition
is provided by two-point correlations of fluid data.
Therefore,  
correlation functions of fluid velocity components as a function of
streamwise separations have been analyzed. 
As can be seen from figure~\ref{fig-corr}, the field initially
exhibits a periodicity over half the domain length, consistent with
the fact that the simulation was started from an exact ``copy'' of a
field taken from a run in a domain with half the streamwise period. 
Over an interval of approximately $40\,T_b$ 
the artificial periodicity (with period $8h$)
gradually disappears and only the strict periodicity over the
fundamental period of $16h$ remains. 
The transient interval was discarded and statistics have 
then 
been accumulated 
over a sampling interval of $90\,T_b$. 
The plots in figure~\ref{fig-corr} also show the correlation function
computed from a snapshot at the beginning of the interval over which
statistics have been accumulated, 
which clearly lacks any traces of artificial periodicity. 
%
%
%
\begin{figure}
  \centering
  \begin{minipage}{2ex}
    \rotatebox{90}{
      $\langle u_f\rangle_c/u_b$, $\langle u_p\rangle/u_b$}
  \end{minipage}
  \begin{minipage}{.45\linewidth}
    \centerline{$(a)$}
    \includegraphics[width=\linewidth]
    {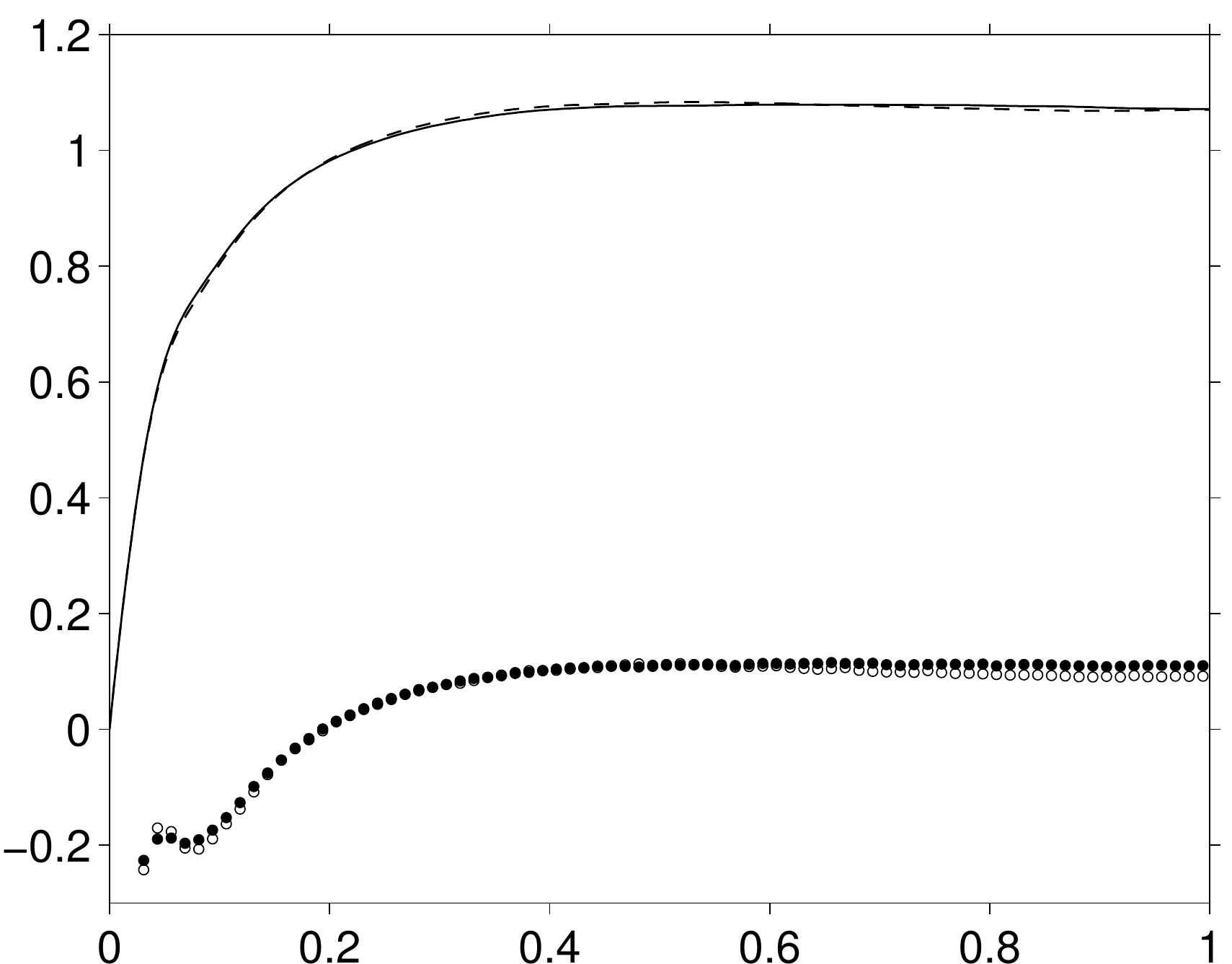}
    \\
    \centerline{$y/h$}
  \end{minipage}
  \begin{minipage}{2ex}
    \rotatebox{90}{
      $(\langle u_p\rangle -\langle u_f\rangle_c)/u_b$}
  \end{minipage}
  \begin{minipage}{.45\linewidth}
    \centerline{$(b)$}
    \includegraphics[width=\linewidth]
    {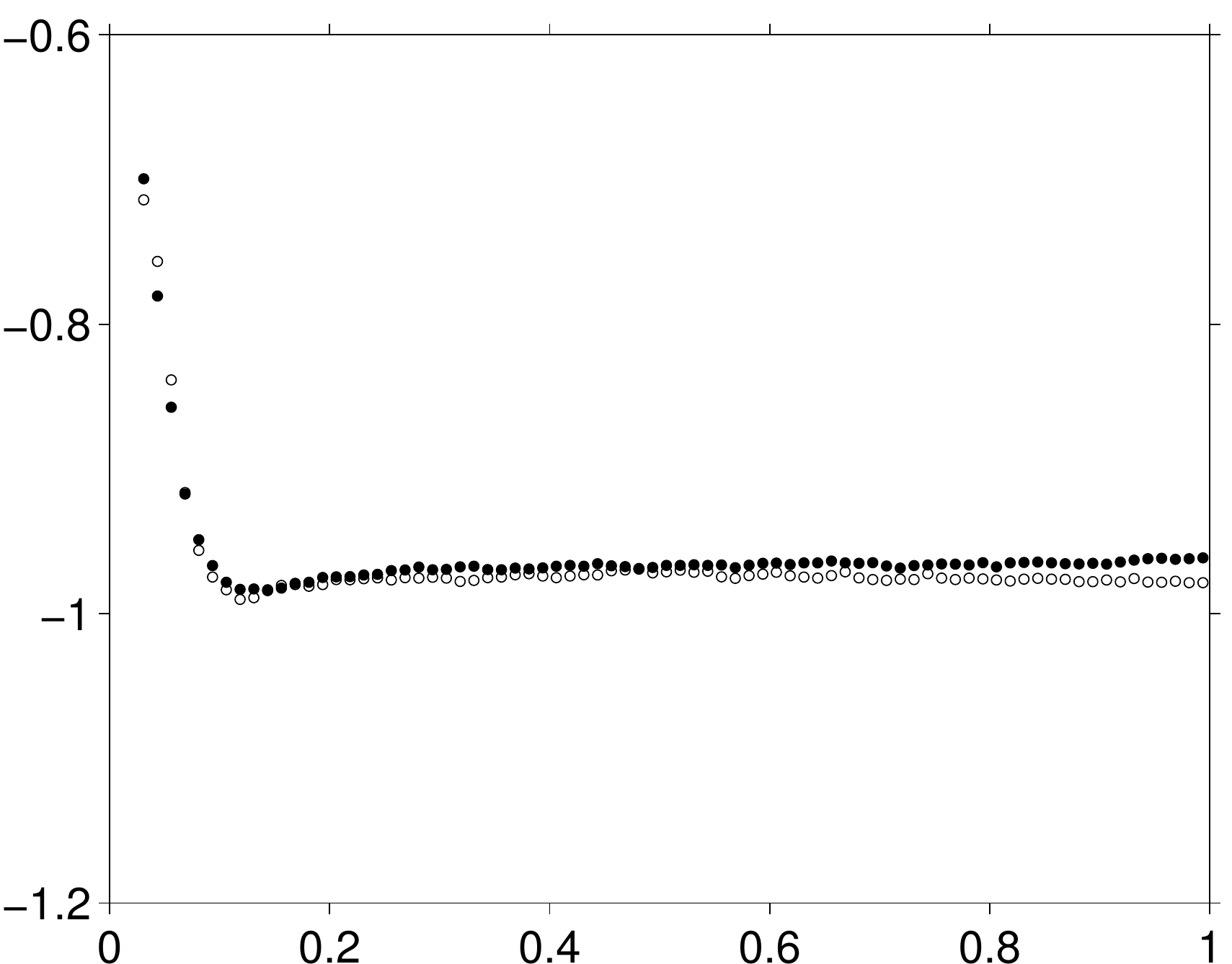}
    \\
    \centerline{$y/h$}
  \end{minipage}
  \caption{%
    $(a)$ Mean velocity profiles of both phases. The fluid phase is
    shown as: \solid, present; \dashed, \cite{uhlmann:08a}. 
    The mean velocity of the particulate phase is shown as: 
    $\bullet$, present; $\circ$, \cite{uhlmann:08a}. 
    $(b)$ Apparent slip velocity between the phases:
    $\bullet$, present; $\circ$, \cite{uhlmann:08a}. 
    Fluid data shown in this figure is 'composite-averaged' (cf.\
    equation~\ref{equ-def-avg-operator-plane-and-time-COMPOSITE}) for
    the purpose of strict comparison; 
    pure fluid averaging data for the present simulation is shown in 
    Appendix~\ref{app-pure-stats}. 
}
  \label{fig-stat-um}
\end{figure}
\subsection{Notation}\label{sec-numa-notation}
Before turning to the results, let us fix the basic notation followed
throughout the present text. 
Velocity vectors
and their 
components 
corresponding to
the fluid and the particle phases 
are distinguished by subscripts ``f'' and ``p'', 
respectively, 
as in
$\mathbf{u}_f=(u_f,v_f,w_f)^T$ and 
$\mathbf{u}_p=(u_p,v_p,w_p)^T$. Similarly, the vector of angular
particle velocity is denoted as
$\boldsymbol{\omega}_p=(\omega_{p1},\omega_{p2},\omega_{p3})^T$ and
linear 
particle acceleration as $\mathbf{a}_p=(a_{p1},a_{p2},a_{p3})^T$, while
fluid acceleration is denoted as
$\mathbf{a}_f=(a_{f1},a_{f2},a_{f3})^T$.  
The equations of motion for an immersed sphere with density $\rho_p$
can be written as: 
\begin{subequations}
\begin{eqnarray}
  \label{equ-numa-newton-rigid-body-linear}
  \rho_p\,V_p\,\mathbf{a}_p&=&
  \mathbf{F}_H+
  \mathbf{F}_B+
  \mathbf{F}_C
  \,,
  \\
  \label{equ-numa-newton-rigid-body-angular}
  I_p\,\dot{\boldsymbol{\omega}}_p&=&
  \mathbf{T}_H
  \,,
\end{eqnarray}
\end{subequations}
where $V_p=\pi D^3/6$ is the volume of a sphere with diameter $D$ and
$I_p=\rho_pD^5\pi/60$ is its moment of inertia, considering a
homogeneous mass density. 
In equation (\ref{equ-numa-newton-rigid-body-linear}) the resulting
force has been separated into a hydrodynamic contribution 
$\mathbf{F}_H=\int_{\cal S}{\boldsymbol{\tau}
    \cdot \mathbf{n}}\,\mathrm{d}\sigma  
  - \int_{\cal S}{ p\mathbf{n} }\,\mathrm{d}\sigma$ 
(${\cal S}$ being the surface of the sphere, $\mathbf{n}$ the outward
pointing normal vector at the surface, 
$\boldsymbol{\tau}= \rho_f\nu \left( \partial u_{fi}/\partial x_j
  +\partial u_{fj}/\partial x_i \right)$ 
the viscous stress tensor 
and $p$ the hydrodynamic pressure), 
a 
relative 
buoyancy force $\mathbf{F}_B=(\rho_p-\rho_f)\mathbf{g}\,V_p$ and a
contribution $\mathbf{F}_C$ from solid-solid contact which is modelled
as discussed in \S~\ref{sec-numa-meth}. 
The angular acceleration in (\ref{equ-numa-newton-rigid-body-angular})
only has a single contribution from viscous hydrodynamic stresses,
viz.\ 
$\mathbf{T}_H=\int_{\cal S}\mathbf{r}_c\times(\boldsymbol{\tau} \cdot
\mathbf{n})\,\mathrm{d}\sigma$ 
($\mathbf{r}_c$ being the distance vector from the particle center),  
since our solid-solid collision treatment is limited to normal forces.  
%

%% file: results_a.tex
\section{Results}\label{sec-results}
\begin{figure} 
  \centering
    \begin{minipage}{4ex}
      $\displaystyle
      \frac{\langle \phi_s\rangle}{\Phi_s}$
    \end{minipage}
    \begin{minipage}{.45\linewidth}
      \centerline{$(a)$}
      \includegraphics[width=\linewidth]
      {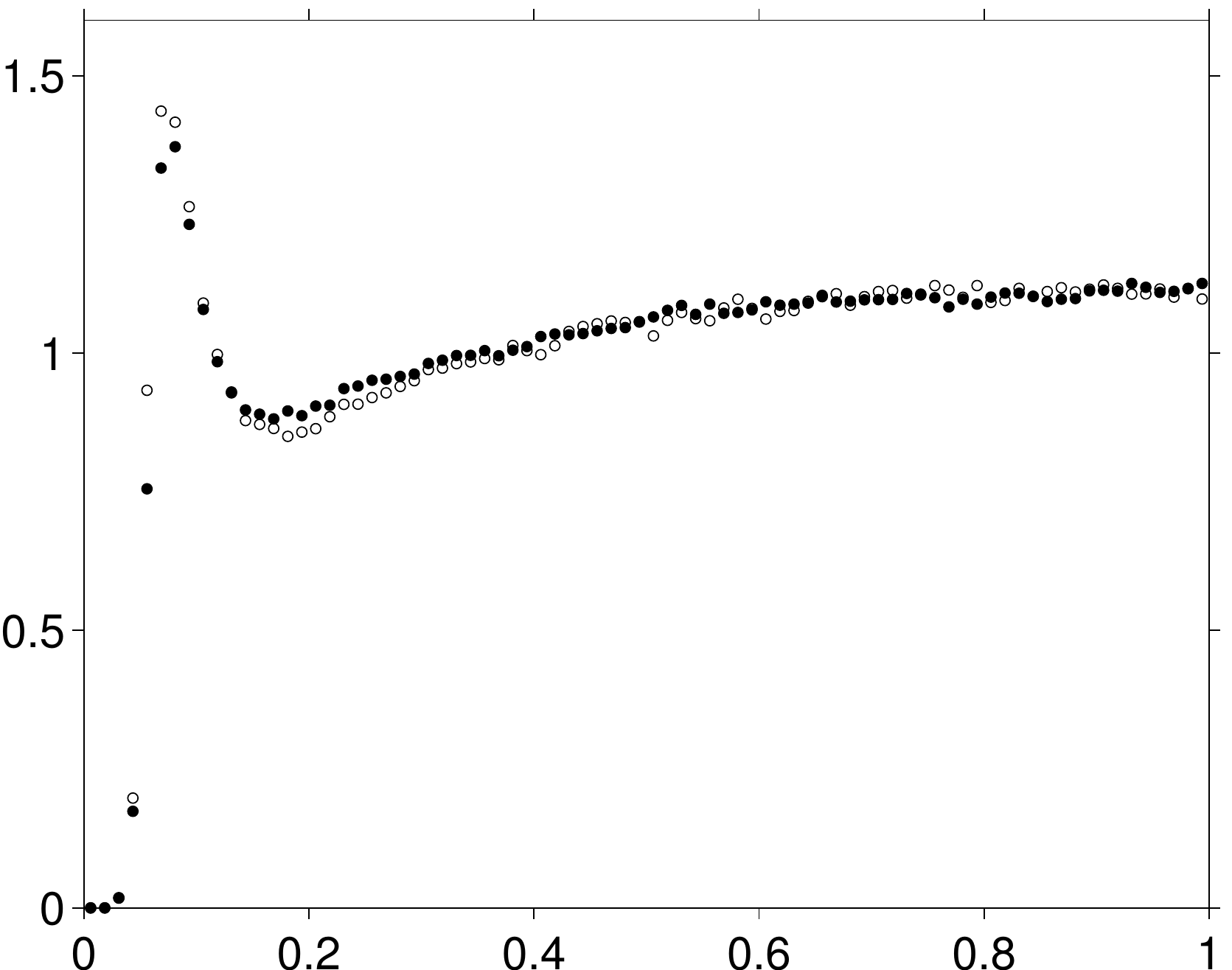}
      \\
      \centerline{$y/h$}
    \end{minipage}
    \begin{minipage}{2ex}
      \rotatebox{90}{$\langle \omega_{p,z}\rangle h/u_b$}
    \end{minipage}
    \begin{minipage}{.45\linewidth}
      \centerline{$(b)$}
      \includegraphics[width=\linewidth]
      {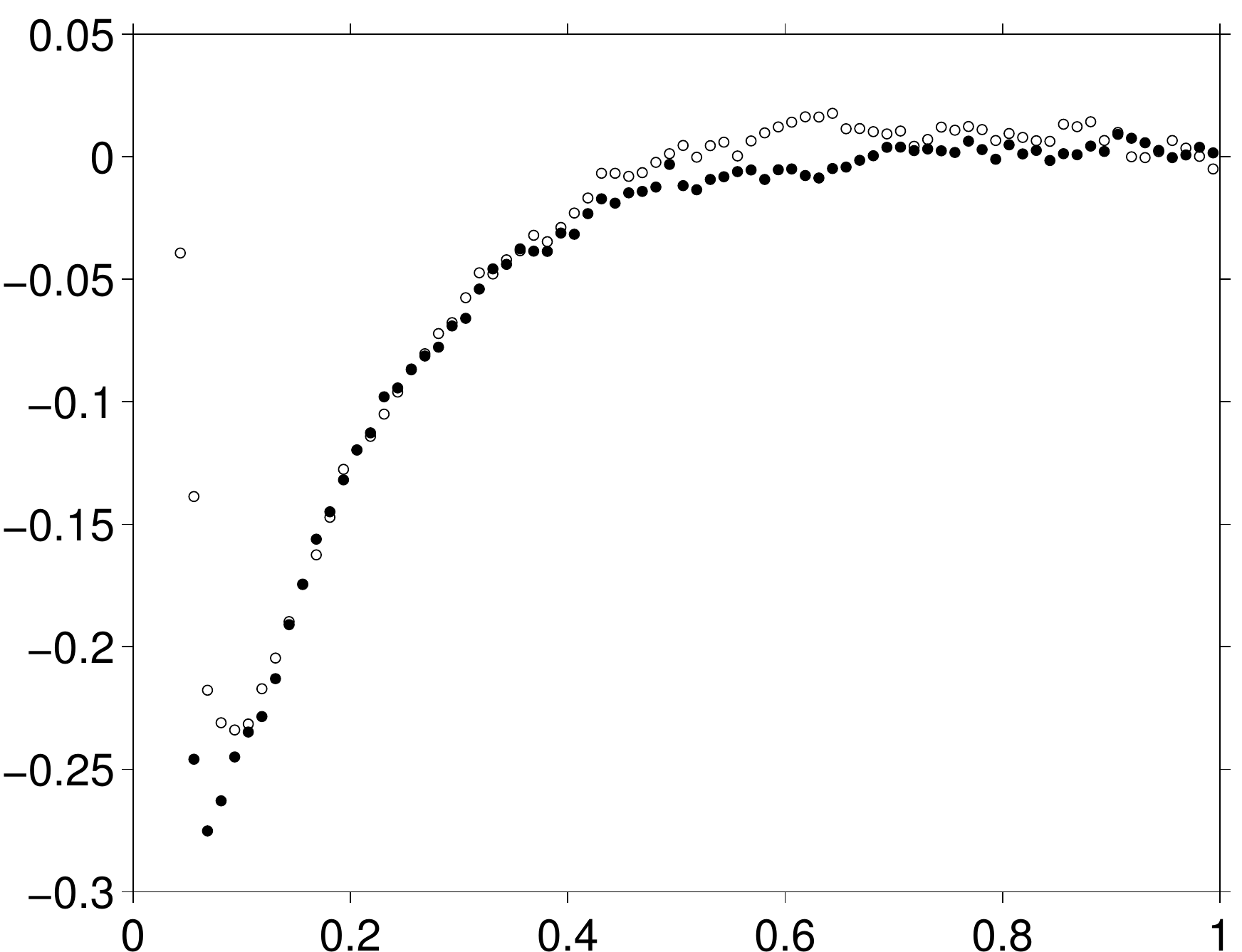}
      \\
      \centerline{$y/h$}
    \end{minipage}
  \caption{%
    Wall normal profiles of: 
    $(a)$ the mean solid volume fraction, 
    %
    $(b)$ the mean value of the spanwise component of angular particle
    velocity.   
    Symbols as in figure~\ref{fig-stat-um}$(b)$. 
  }
  \label{fig-stat-conc-angvel}
\end{figure}
\begin{figure}
  \centering
  \begin{minipage}{3ex}
    \rotatebox{90}{
      $\langle u_{f,\alpha}^\prime u_{f,\alpha}^\prime\rangle_c^{1/2}/u_b$, 
      $\langle u_{p,\alpha}^\prime u_{p,\alpha}^\prime\rangle^{1/2}/u_b$
      }
  \end{minipage}
  \begin{minipage}{.45\linewidth}
    \centerline{$(a)$}
    \includegraphics[width=\linewidth]
    {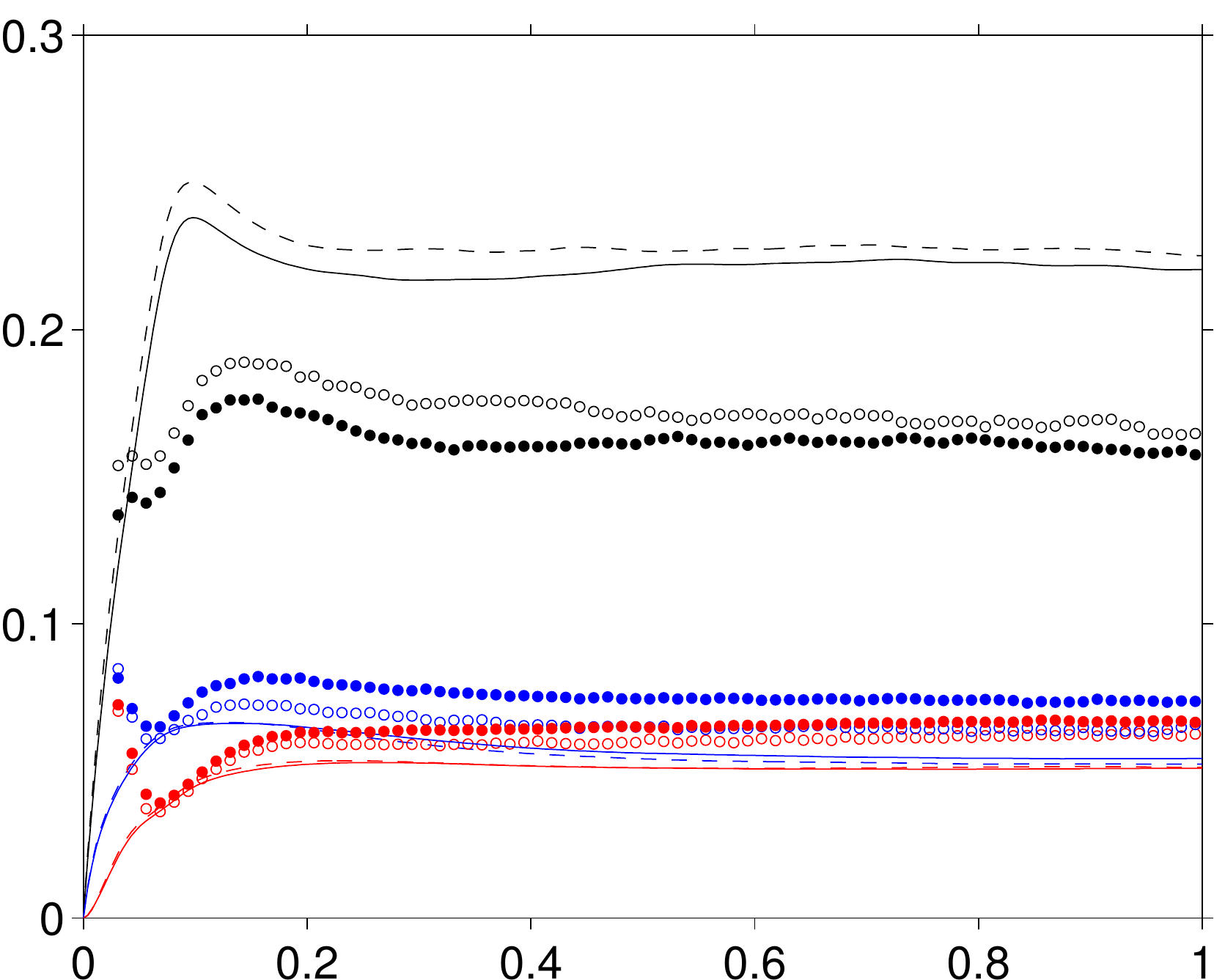}
    \\
    \centerline{$y/h$}
  \end{minipage}
  \begin{minipage}{3ex}
    \rotatebox{90}{
      $\langle u_{f}^\prime v_{f}^\prime\rangle_c/u_b^2$, 
      $\langle u_{p}^\prime v_{p}^\prime\rangle/u_b^2$
    }
  \end{minipage}
  \begin{minipage}{.45\linewidth}
    \centerline{$(b)$}
    \includegraphics[width=\linewidth]
    {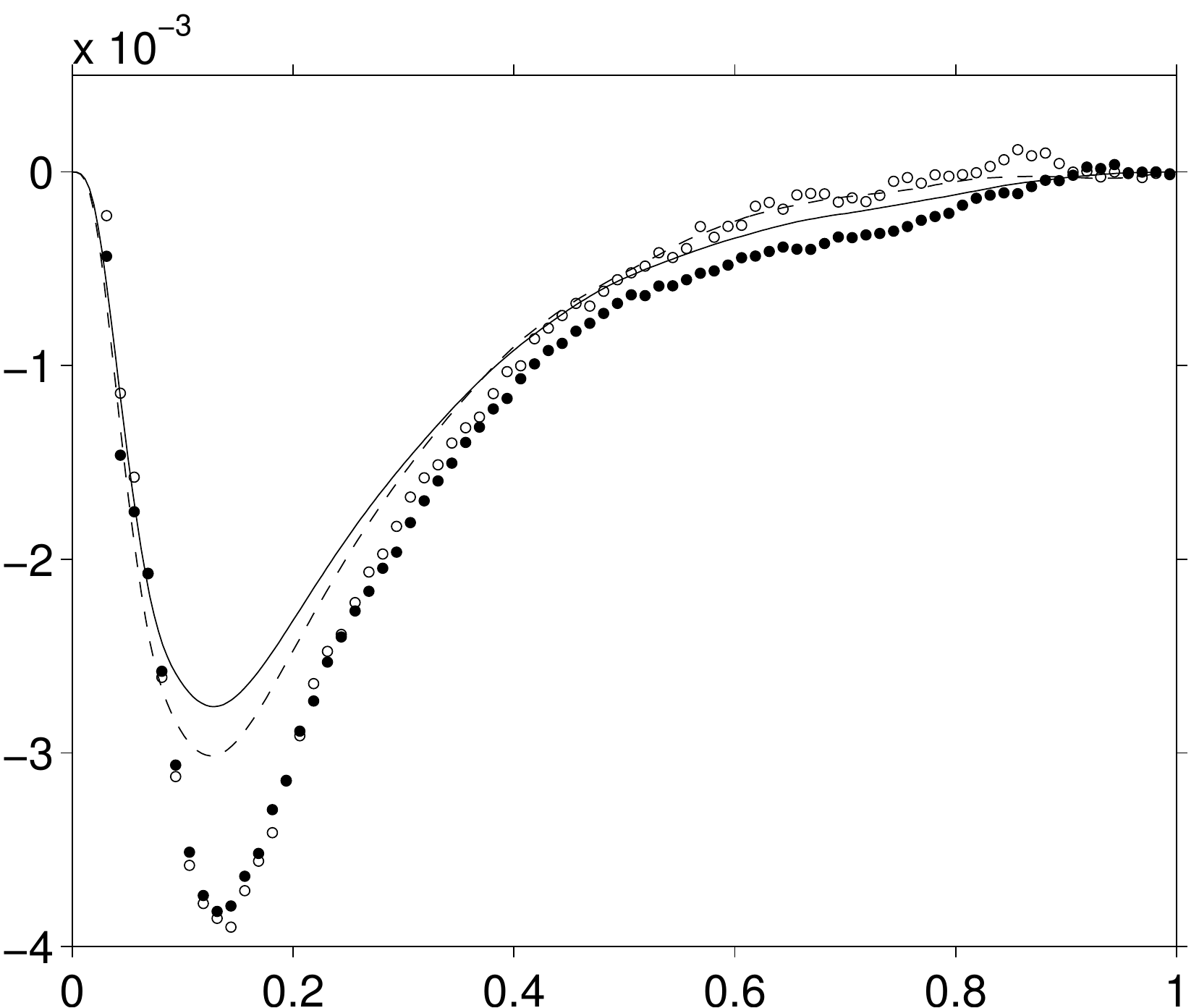}
    \\
    \centerline{$y/h$}
  \end{minipage}
  \caption{%
    $(a)$ R.m.s.\ of velocity fluctuations of both phases, with lines
    corresponding to the fluid phase (\solid, present; \dashed,
    \cite{uhlmann:08a}), 
    and symbols to the particulate phase ($\bullet$, present; $\circ$,
    \cite{uhlmann:08a}), as in figure~\ref{fig-stat-um}$(a)$. 
    The color coding indicates streamwise (black),
    wall-normal (red), spanwise (blue) components.
    $(b)$ Reynolds shear stress of fluid velocity fluctuations as well
    as corresponding velocity correlation of the particle motion.
    All quantities are normalized in bulk units. 
    Fluid data shown in this figure is 'composite-averaged' (cf.\
    equation~\ref{equ-def-avg-operator-plane-and-time-COMPOSITE}) for
    the purpose of strict comparison; 
    pure fluid averaging data for the present simulation is shown in 
    Appendix~\ref{app-pure-stats}. 
}
  \label{fig-stat-uu}
  %
  %
  %
  %
  %
  %
  %
  %
  %
  %
  %
  %
\end{figure}
%
\subsection{Effect of streamwise elongated domain}
The aim of the present section is to examine the effect of the finite
streamwise period imposed upon the fields in the simulation. 
For this purpose we will compare the present results with those
obtained in a shorter domain as presented by \cite{uhlmann:08a}. 

Note that the fluid data shown in figures~\ref{fig-stat-um} and
\ref{fig-stat-uu}  
are based upon `composite' averaging, i.e.\ not distinguishing
between velocity values of those grid nodes which are instantaneously
located in the fluid phase and those inside the solid phase. 
Doing so allows for a strict comparison with results from the
simulation of \cite{uhlmann:08a}, where this distinction was not made  
\cite[cf.\ discussion in Appendix~A of][]{uhlmann:08a}. 
\revision{%
  The correctly evaluated statistics (including only actual grid nodes
  located inside the fluid domain, cf.\ definition in
  equation~\ref{equ-def-avg-operator-plane-and-time-fluid-only})
  corresponding to the fluid quantities shown in figures~\ref{fig-stat-um} and
  \ref{fig-stat-uu} are included in \ref{app-pure-stats} for future
  reference.}{%
  The specific fluid-phase statistics (including only actual grid nodes
  located inside the fluid domain, cf.\ definition in
  equation~\ref{equ-def-avg-operator-plane-and-time-fluid-only})
  corresponding to the fluid quantities shown in figures~\ref{fig-stat-um} and
  \ref{fig-stat-uu} are included in Appendix~\ref{app-pure-stats} for future
  reference.
}

\revision{}{%
  In the present flow it turns out that the average 
  time interval between two collision events experienced by a particle
  is $3.9\,T_b$, showing that indeed collisions are
  relatively infrequent. 
  Furthermore, no particle-wall collisions have been observed during
  the simulation interval. 
}
\subsubsection{Eulerian statistics of both phases}
Mean fluid and particle velocity profiles as well as particle
concentration profiles and the mean spanwise component of the angular
particle velocity are shown in figures~\ref{fig-stat-um} and 
\ref{fig-stat-conc-angvel}. 
It can be seen that these average values are not 
too strongly affected by an increase of the box size from $L_x/h=8$ to
$16$.  
The observed mostly small differences in these quantities can be
attributed to the remaining statistical uncertainty (i.e.\ limited
sampling of the largest flow scales). 
In particular, the wall shear-stress averaged over the observation
interval differs slightly, such that the time-and-wall-averaged
friction Reynolds number $Re_\tau$ takes a value of 
$220.9$ 
\citep[versus $224.4$ in the simulation of][]{uhlmann:08a}. 
Furthermore, the previously observed weak tendency towards a concave
mean velocity profile \citep{uhlmann:08a} is not confirmed by the
present results (cf.\ figure~\ref{fig-stat-um}$a$). It might therefore 
equally be attributed to limited sampling of the large scales in the
previous simulation.
%
The difference between the mean velocities of both phases (cf.\
figure~\ref{fig-stat-um}$b$) is termed
the `apparent slip velocity', which will be denoted by 
$u_{lag}=\langle u_p\rangle -\langle u_f\rangle$ in the following.
The corresponding Reynolds number ($Re_{lag}=|u_{lag}|D/\nu$) measures 
$Re_{lag}\approx132$ in the bulk of the channel, while the value drops
significantly for wall-distances smaller than $y/h\approx0.1$. 

Concerning the mean value of the spanwise component of angular
particle velocity shown in figure~\ref{fig-stat-conc-angvel}$(b)$, it
should be mentioned that the corresponding graph in \cite{uhlmann:08a}
erroneously shows the quantity $\langle\omega_{p,z}\rangle u_b/h$  
\citep[i.e.\ the axis
label of figure~14 in][is incorrect]{uhlmann:08a}. 
The correct non-dimensional quantity $\langle\omega_{p,z}\rangle
h/u_b$ from that reference is shown in the present
figure~\ref{fig-stat-conc-angvel}$(b)$. 
Both datasets exhibit a proportionality as given by
$\langle\omega_{p,z}\rangle=-A\,\mbox{d}\langle u_f\rangle/\mbox{d}y$
with 
$A\approx0.15$,  
%
except for the near-wall region $y^+\leq25$. 
As a consequence, the average shear Reynolds number $Re_s=
|\mbox{d}\langle u_f\rangle/\mbox{d}y|D^2/\nu$ and the average
particle rotation Reynolds number
$Re_\Omega=\langle\omega_{p,z}\rangle D^2/\nu$ are approximately
proportional to each other with the proportionality factor $A$. 
Incidentally, the particle rotation Reynolds number takes values in
the interval $0\leq Re_\Omega\lesssim1.7$ (not shown), with the
maximum occurring at $y/h\approx0.1$.

\begin{figure}
  \centering
  \begin{minipage}{2ex}
     $(a)$\\[5ex]
    \rotatebox{90}{$R_{\alpha\alpha}$}
  \end{minipage}
  \begin{minipage}{.45\linewidth}
    \centerline{$r_x^+$}
    \includegraphics[width=\linewidth]
    {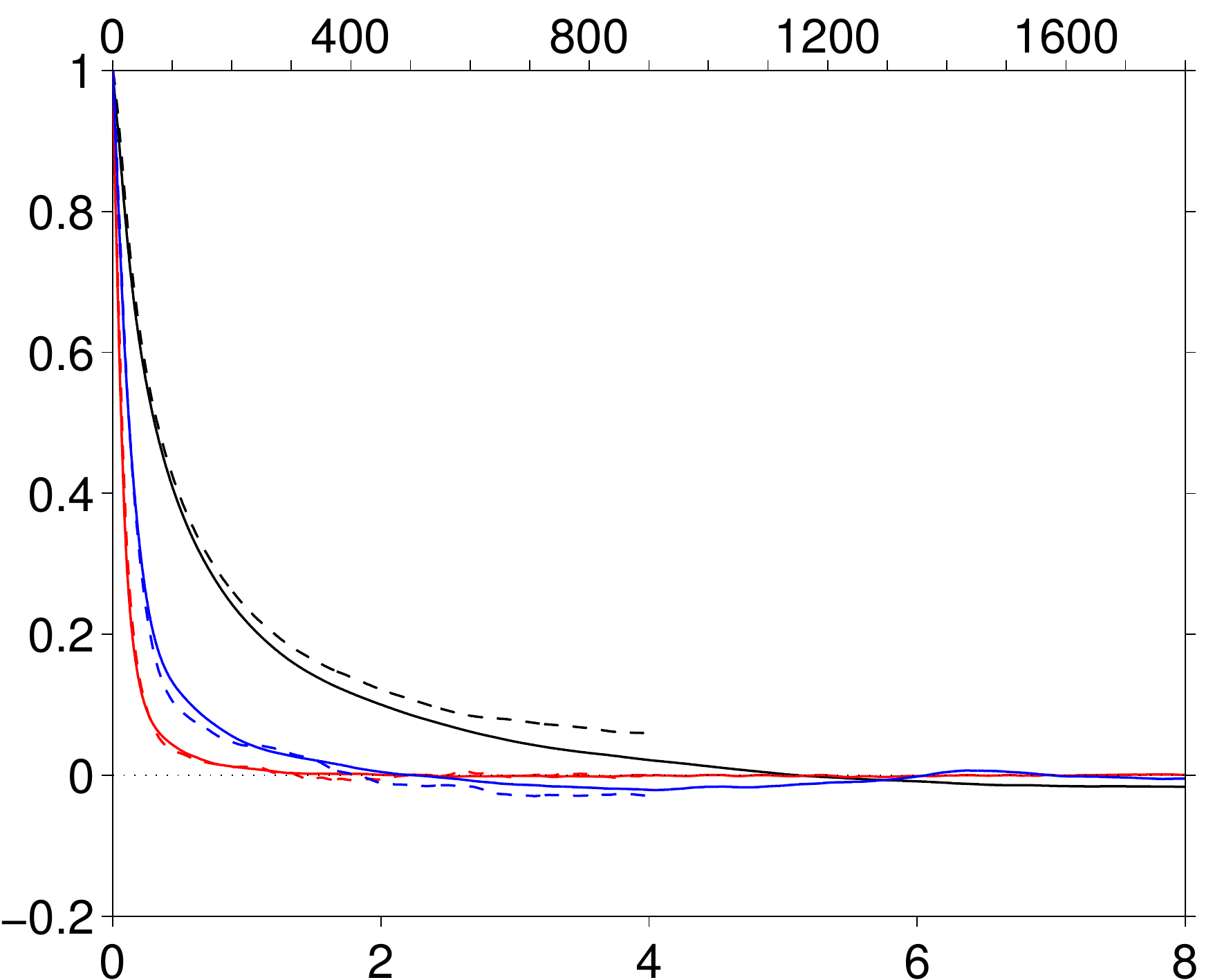}
    \centerline{$r_x/h$}
  \end{minipage}
  \begin{minipage}{2ex}
     $(b)$\\[5ex]
    \rotatebox{90}{$R_{\alpha\alpha}$}
  \end{minipage}
  \begin{minipage}{.45\linewidth}
    \centerline{$r_z^+$}
    \includegraphics[width=\linewidth]
    {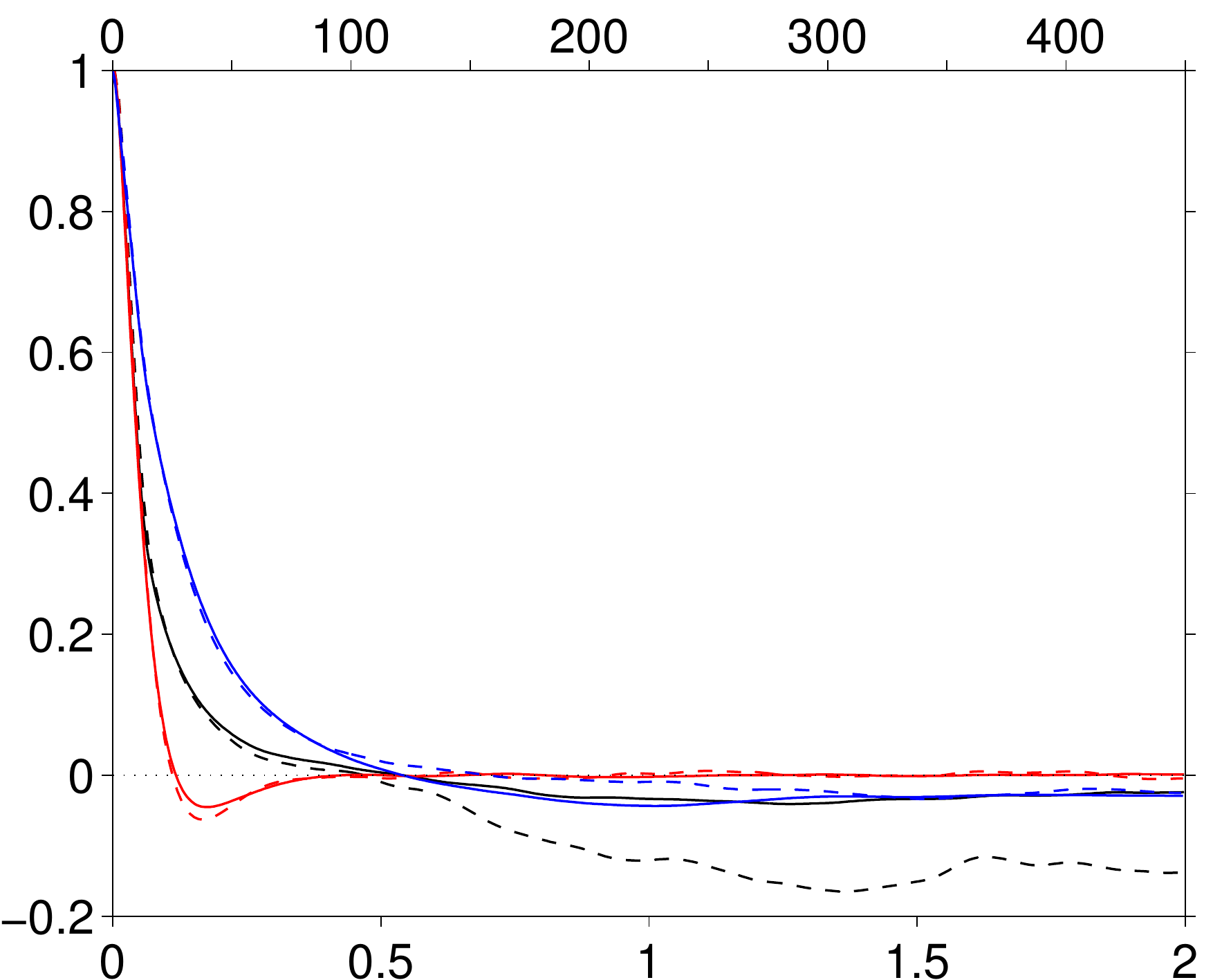}
    \centerline{$r_z/h$}
  \end{minipage}
  \caption{%
    Two-point autocorrelations of fluid velocity fluctuations
    for $(a)$ streamwise separations 
    and $(b)$ spanwise separations. 
    The correlation functions are evaluated from 12  
    instantaneous flow fields in the case of \cite{uhlmann:08a} 
    (dashed lines) and from 
    85 
    flow fields in
    the present simulation (solid lines). In both cases only data
    points in the actual fluid domain are taken into account. 
    {\color{black}\solid} streamwise component ($\alpha=1$), 
    {\color{red}\solid} wall-normal ($\alpha=2$), 
    {\color{blue}\solid} spanwise ($\alpha=3$). 
    The data corresponds to a wall-parallel plane at $y/h=0.0938$
    ($y^+\approx22$).
  }
  \label{fig-2pcorr-12-vs-21}
\end{figure}
%
Velocity covariances for both phases are shown in
figure~\ref{fig-stat-uu}.  
It can be observed that the recorded covariance profiles are of very
similar values in both simulations, exhibiting only relatively weak
differences.  
In absolute terms, the standard deviation values of the streamwise
velocity fluctuations of both phases ($\langle
u_f^\prime u_f^\prime\rangle^{1/2}$ and $\langle u_p^\prime u_p^\prime\rangle^{1/2}$) 
are reduced on average by $0.007\,u_b$ and $0.01\,u_b$, respectively. 
This reduction is consistent with the higher degree of decorrelation
of fluid velocity data in the extended domain, as discussed below. 
The curves of the wall-normal and spanwise fluid velocity fluctuation
intensities in figure~\ref{fig-stat-uu}$(a)$ both nearly collapse with
their counterparts from the simulation of \cite{uhlmann:08a}. 
On the other hand, the r.m.s.\ particle velocity fluctuations in the
wall-normal and spanwise directions are both slightly larger in the
present simulation than in the previous one (on average by
$0.005\,u_b$ and $0.01\,u_b$, respectively). 
The profiles of the fluid Reynolds stress (cf.\ figure~\ref{fig-stat-uu}$b$)
feature a negative peak of slightly smaller amplitude in the present
dataset, while the values of the covariance between streamwise and
wall-normal particle velocity fluctuations $\langle u_{p}^\prime
v_{p}^\prime\rangle$ 
\revision{approximately collapse}{%
are approximately the same in both datasets}  
in the vicinity of the
negative peak at $y/h\approx0.15$. 
Further into the core of the channel (i.e.\ for $y/h\gtrsim0.4$) the
present dataset exhibits slightly 
higher negative 
covariance values, both for
$\langle u_{f}^\prime v_{f}^\prime\rangle$ and $\langle u_{p}^\prime
v_{p}^\prime\rangle$. 
%
%
The observed differences in the covariance values between the two
data-sets is at least in part due to statistical uncertainty,
as the previous dataset is found to exhibit somewhat noisier profiles 
(cf.\ particularly the curve for $\langle u_{p}^\prime
v_{p}^\prime\rangle$ in figure~\ref{fig-stat-uu}$b$). 
However, only when additional simulations with a much longer sampling
interval are available will it be possible to settle the question
about a systematic trend in the fluctuation amplitudes. 
%

\begin{figure}
  \centering
  \begin{minipage}{2ex}
    \rotatebox{90}{$k_x\,E_{\alpha\alpha}(k_x)$}
  \end{minipage}
  \begin{minipage}{.45\linewidth}
    \centerline{$(a)$}
    \includegraphics[width=\linewidth]
    {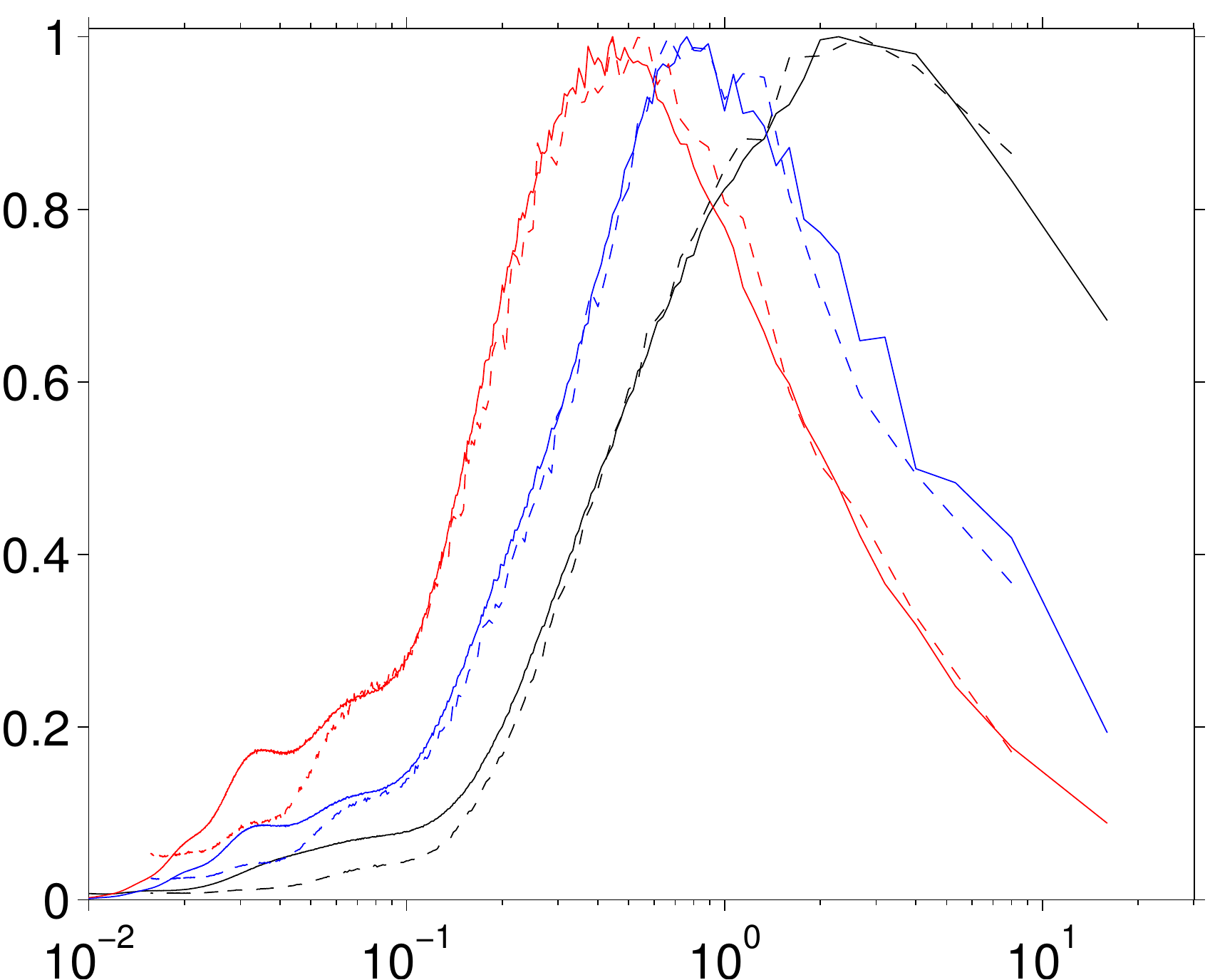}
    \centerline{$\lambda_x/h$}
  \end{minipage}
  \begin{minipage}{2ex}
    \rotatebox{90}{$k_z\,E_{\alpha\alpha}(k_z)$}
  \end{minipage}
  \begin{minipage}{.45\linewidth}
    \centerline{$(b)$}
    \includegraphics[width=\linewidth]
    {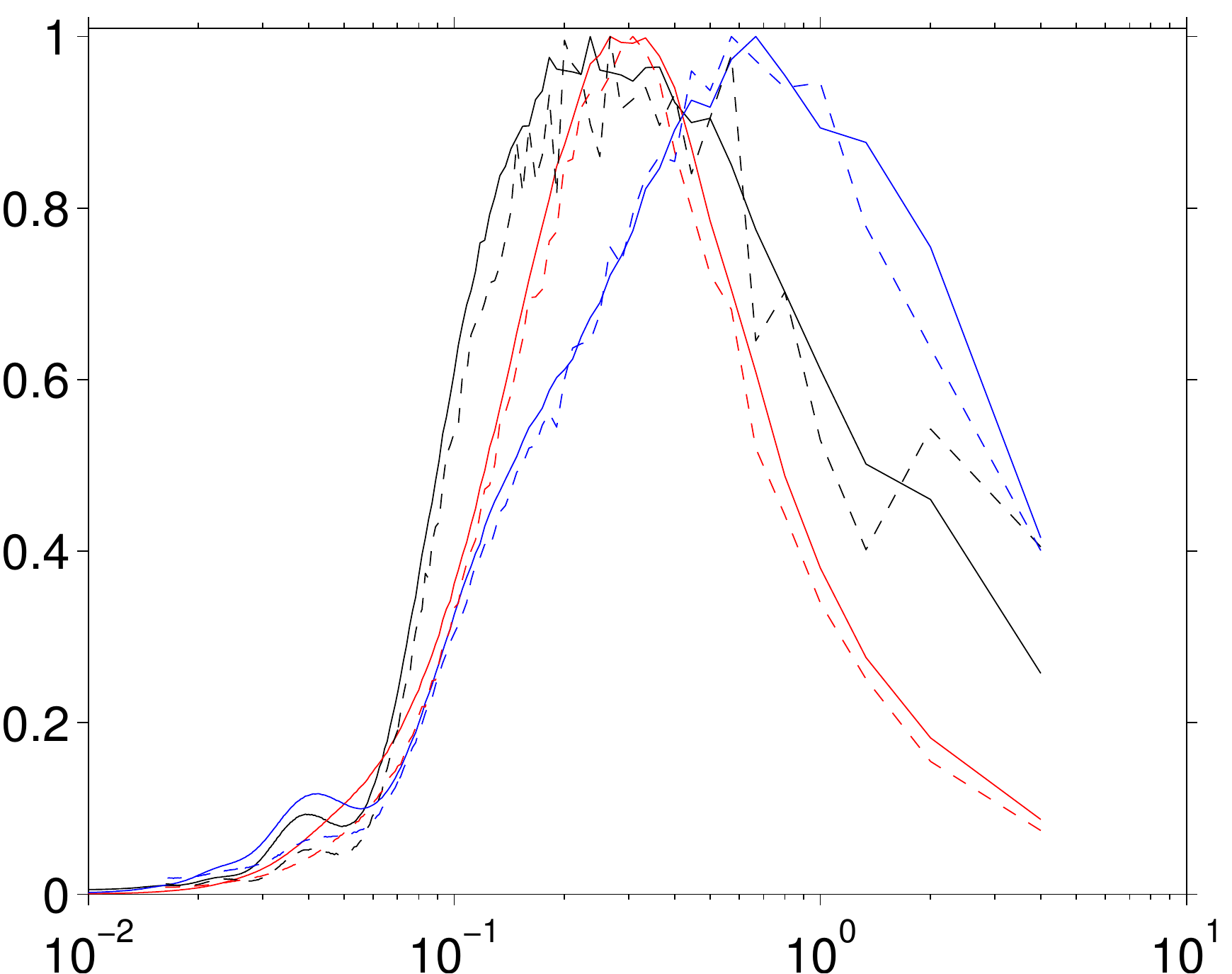}
    \centerline{$\lambda_z/h$}
  \end{minipage}
  \caption{%
    Premultiplied spectra of the 
    three fluid velocity components 
    as function
    of the wavelength, corresponding to the data in
    figure~\ref{fig-2pcorr-12-vs-21} (wall-parallel plane at
    $y^+=22$). Each curve is normalized to a maximum value of unity in
    order to emphasize the frequency content. 
    $(a)$ shows streamwise spectra, 
    and $(b)$ spanwise spectra.
    Line-styles and color-coding as in
    figure~\ref{fig-2pcorr-12-vs-21}. 
    Note that the particle diameter corresponds to
    $D/h=5\cdot10^{-2}$.
    %
  }
  \label{fig-pspec-12-vs-21}
\end{figure}
\begin{figure}
  \begin{minipage}[t]{.47\linewidth}
  \begin{minipage}[c]{3ex}
    $\displaystyle\frac{x}{h}$
  \end{minipage}
  \begin{minipage}[c]{.4\linewidth}
    \centerline{$(a)$}
    \includegraphics*[width=\linewidth,viewport=314 326 888 1474]
    {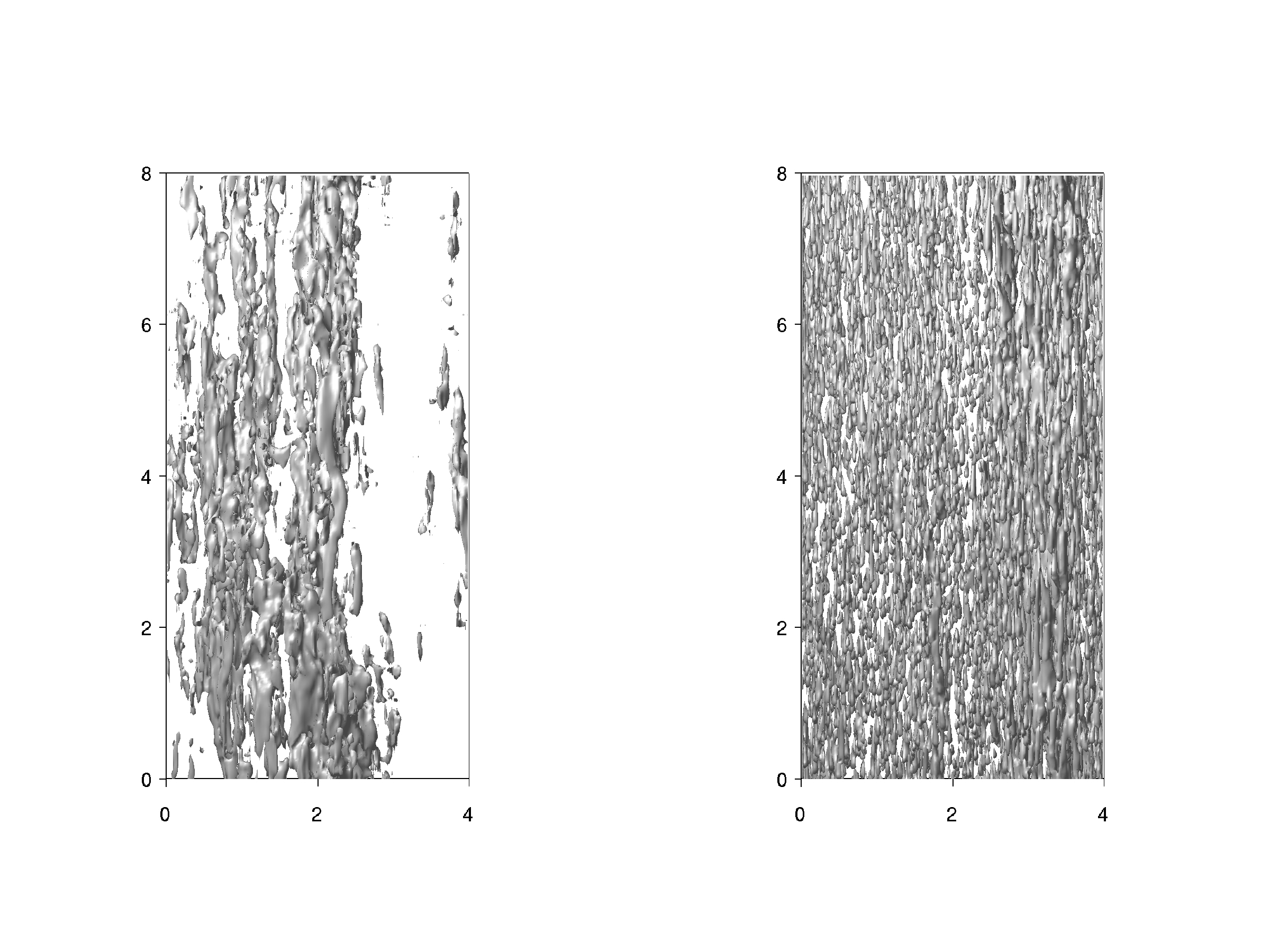}
    \centerline{$z/h$}
  \end{minipage}\hfill
  \begin{minipage}[c]{.4\linewidth}
    \centerline{$(b)$}
    \includegraphics*[width=\linewidth,viewport=375 8 825 1790]
    {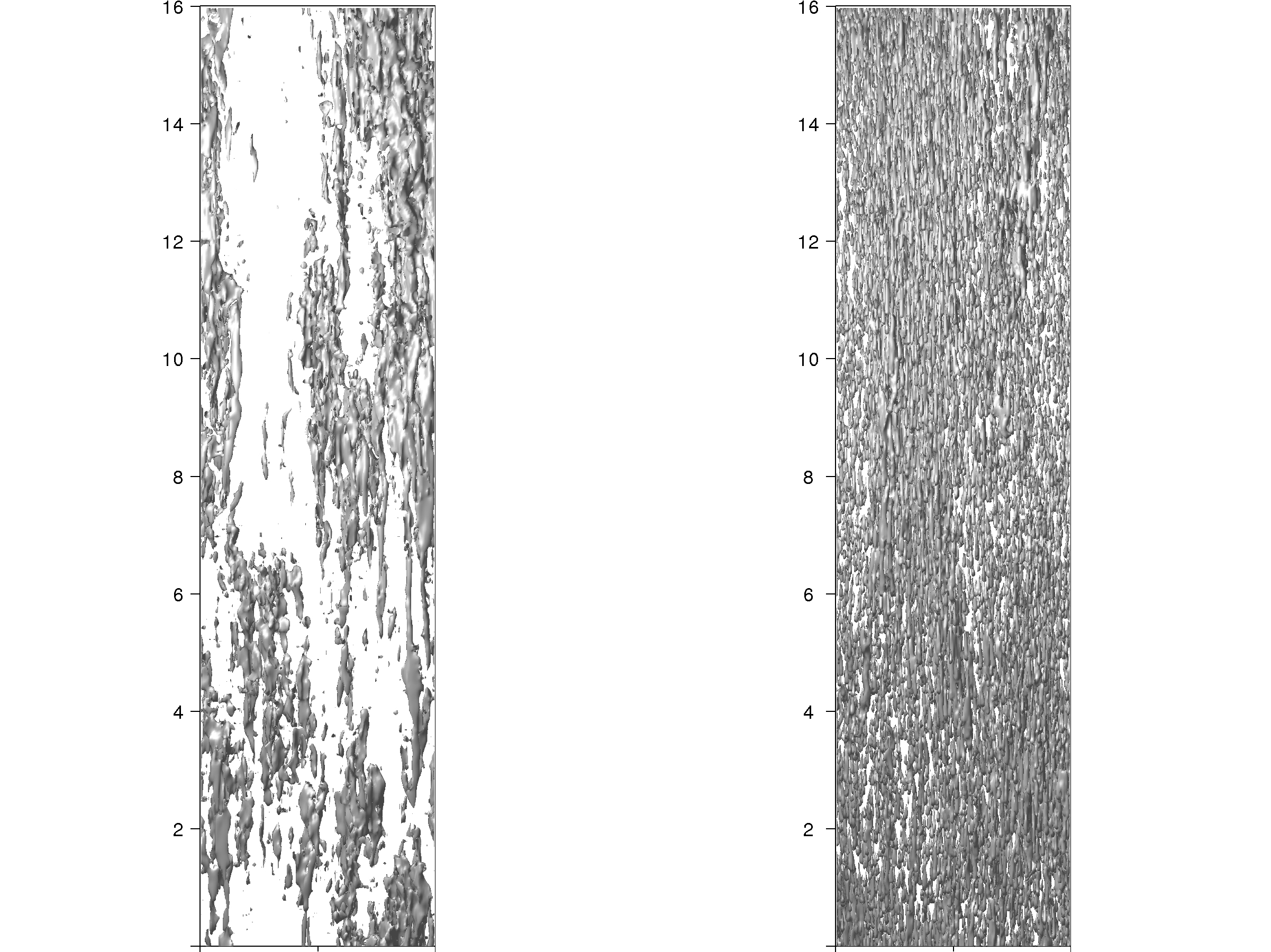}
    \centerline{$z/h$}
  \end{minipage}
  \caption{Instantaneous three-dimensional isosurfaces of streamwise
    velocity fluctuations  $u^\prime=3.6u_\tau$ (equivalent to
    $0.3u_b$). 
    The graph $(a)$ corresponds to the simulation in a shorter box
    \cite{uhlmann:08a},
    $(b)$ is from the present simulation in a streamwise-elongated box. 
    The view 
    is directed into the wall.
    Please note that only every eighth grid point in each direction
    was used. 
  } 
  \label{fig-fluid-snapshot-pos-ufluc}
  \end{minipage}
  \hfill
  \begin{minipage}[t]{.47\linewidth}
  \begin{minipage}[c]{3ex}
    $\displaystyle\frac{x}{h}$
  \end{minipage}
  \begin{minipage}[c]{.4\linewidth}
    \centerline{$(a)$}
    \includegraphics*[width=\linewidth,viewport=1513 326 2089 1474]
    {channelp12_uflucx_xz_full_12_3_6utau.jpg}
    \centerline{$z/h$}
  \end{minipage}\hfill
  \begin{minipage}[c]{.4\linewidth}
    \centerline{$(b)$}
    \includegraphics*[width=\linewidth,viewport=1575 8 2025 1790]
    {channelp210_uflucx_xz_full_post_mi_xc2_field88_3_6utau.jpg}
    \centerline{$z/h$}
  \end{minipage}
  \caption{As figure~\ref{fig-fluid-snapshot-pos-ufluc}, but showing
    negative-valued surfaces at the same magnitude, i.e.\
    $u^\prime=-3.6u_\tau$.   
  } 
  \label{fig-fluid-snapshot-neg-ufluc}
  \end{minipage}
\end{figure}
Two-point correlations of the fluid velocity field evaluated at a
wall-distance of $y/h=0.1$ ($y^+=22$) are shown in
figure~\ref{fig-2pcorr-12-vs-21}. 
Correlation values for the wall-normal and spanwise velocity
components are found to be only very little affected by the
prolongation of the domain as the smaller domain size already yielded 
a reasonable decorrelation at separations of the order of half the
fundamental period. 
On the other hand, fluctuations of the streamwise fluid velocity
component decorrelate somewhat more rapidly with streamwise
separations in the longer box than they do in the shorter one. 
Furthermore, the correlation function $R_{11}$ in the present
simulation reaches values close to zero at the largest separations
$r_x/h\approx8$.  
Therefore, it makes sense to compute streamwise integral length scales 
($L_{\alpha\alpha}^{(x)}=\int_0^{L_x/2}R_{\alpha\alpha}\mbox{d}r_x$) 
which take the following values in the present case:  
$L_{11}^{(x)}/h=0.68$, 
\revision{$L_{33}^{(x)}/h=0.11$}{$L_{22}^{(x)}/h=0.11$}, 
$L_{33}^{(x)}/h=0.19$. 
Concerning spanwise separations, the results from the original and the
streamwise-extended domain are similar, with the exception of $R_{11}$
which is visibly less noisy in the present data-set. 

Pre-multiplied one-dimensional spectra of the fluid velocity field are
shown in figure~\ref{fig-pspec-12-vs-21}, providing an alternative
view of the correlation data of figure~\ref{fig-2pcorr-12-vs-21}.
In order to compute the spectra, velocity values at nodes inside the
particles were set to zero, which leads to the well-known
`step-noise' at wavelengths around and below the particle scale 
\citep{parthasarathy:90a}, i.e.\ at $\lambda/h\approx D/h=0.05$. 
The large-wavelength end of the spectrum, however, is not affected by
the discontinuities at the phase-interfaces. 
The spectra are normalized by their respective maximum value, in order
to enable a comparison of the frequency contents. 
Concerning the streamwise component of the fluid velocity field, it
can be seen in figure~\ref{fig-pspec-12-vs-21} that the peak of the
energy spectrum is captured both in the longer and shorter
boxes. However, the decay with wavelength is more complete in the
larger domain.  
Nevertheless, the data suggest that still a considerably larger
streamwise period would be needed in order to allow for a complete
energy decay at the large-scale end of the spectrum. 
Unfortunately, a quantitative estimate of the required length
cannot be deduced from the present data.

Figures~\ref{fig-fluid-snapshot-pos-ufluc} and
\ref{fig-fluid-snapshot-neg-ufluc} show streamwise velocity
fluctuations from instantaneous fields of simulations in the original
domain of \cite{uhlmann:08a} and the extended computational domain of
the present simulation. 
Visual inspection of these and other snapshots corroborates the above
finding that the large-scale streak-like structures \citep[first
revealed in][]{uhlmann:08a} 
still exist in the present case with a prolonged domain. However, 
they do not appear to span the entire box length of $L_x/h=16$.
%
\begin{figure}
  \begin{center}
    \begin{minipage}{5ex}
      $(a)$\\[5ex]
      $R_{Lp,\alpha}$
    \end{minipage}
    \begin{minipage}{.43\linewidth}
      \includegraphics[width=\linewidth]
      {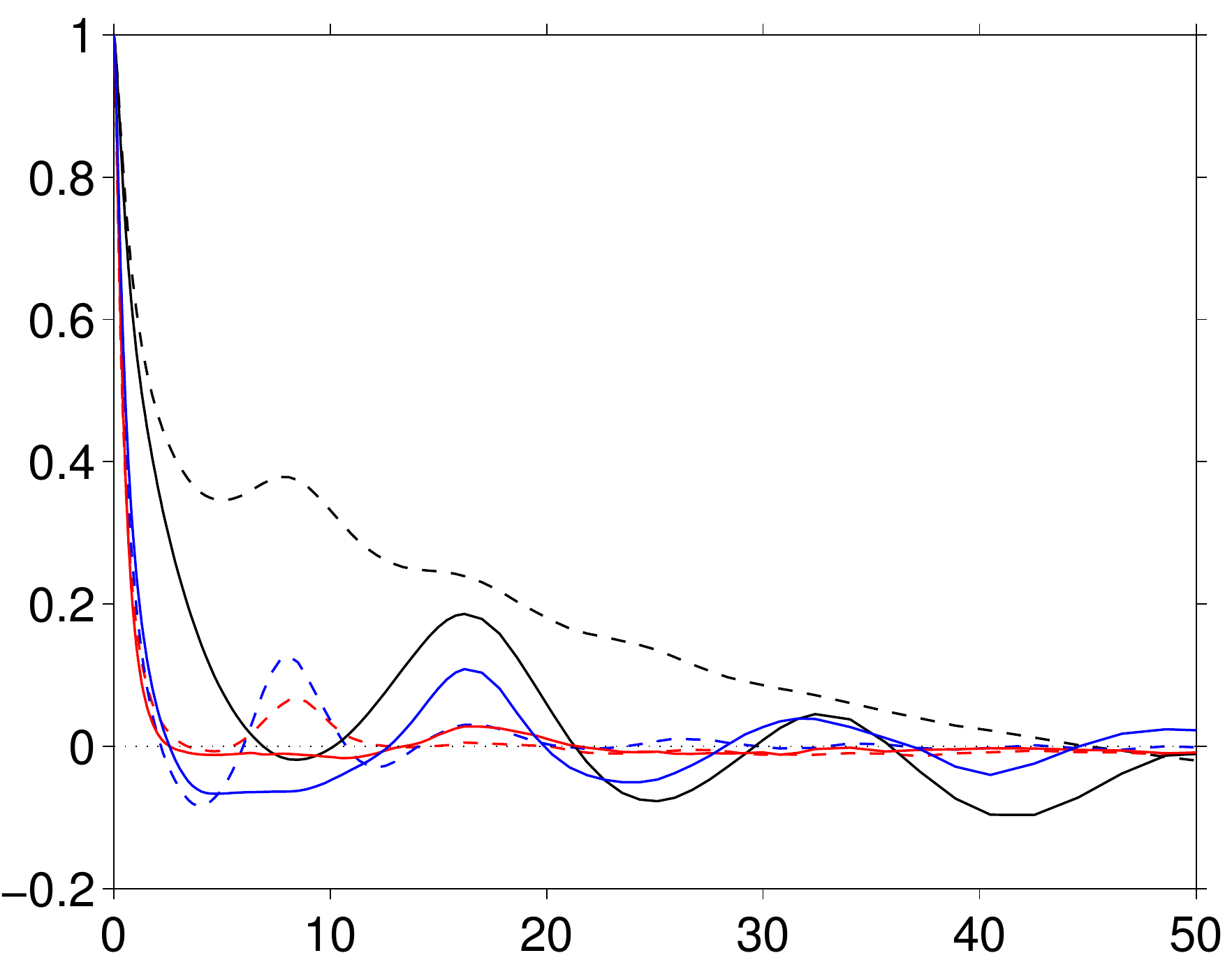}
      \\
      \centerline{$\tau/T_b$}
    \end{minipage}
    \begin{minipage}{5ex}
      $(b)$\\[5ex]
      $R_{Lp,\alpha}$
    \end{minipage}
    \begin{minipage}{.43\linewidth}
      \includegraphics[width=\linewidth]
      {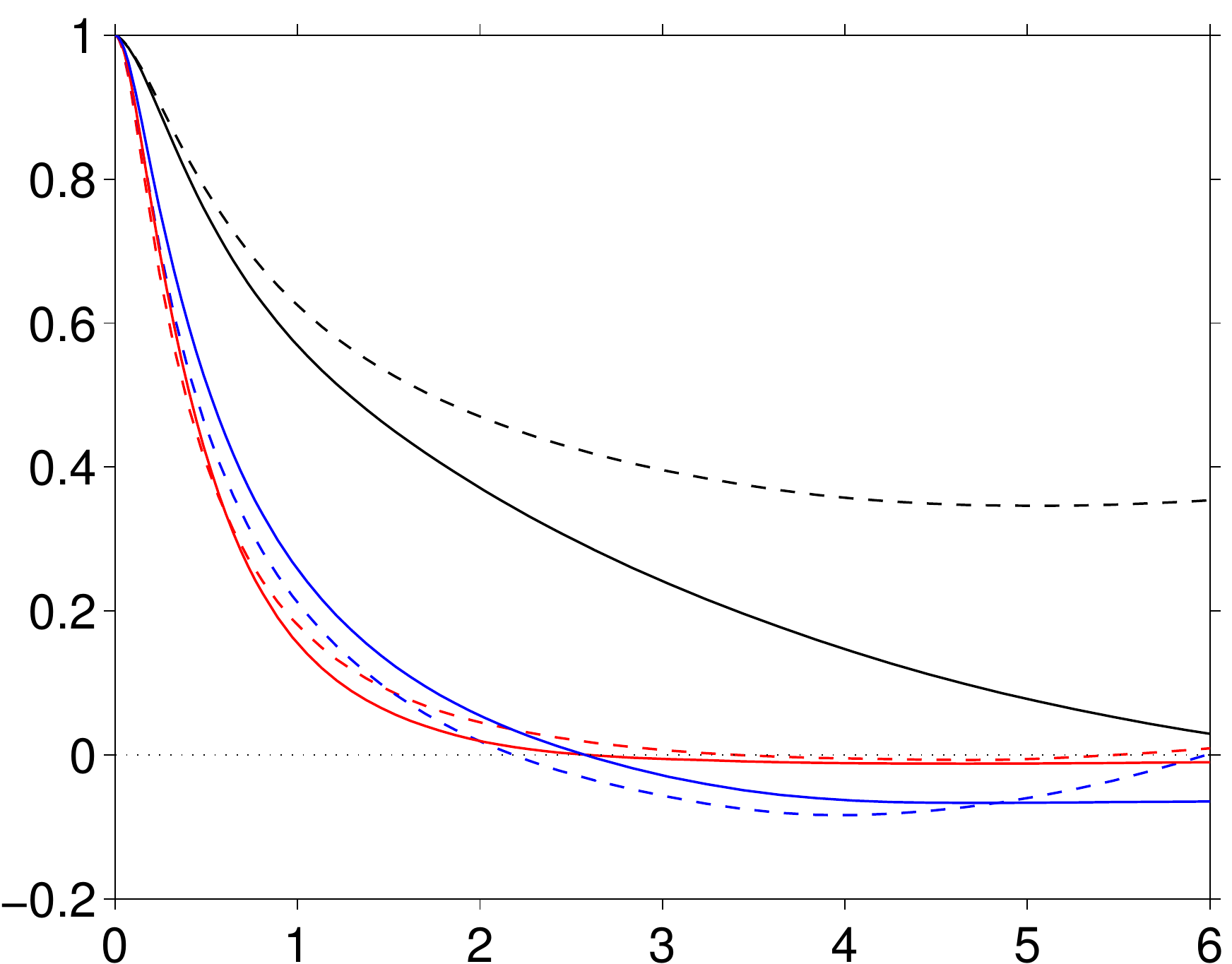}
      \\
      \centerline{$\tau/T_b$}
    \end{minipage}
  \end{center}
  \caption{%
    Lagrangian particle velocity autocorrelation as a function of the
    separation time $\tau$: 
    {\color{black}\solid}~streamwise velocity component ($\alpha=1$); 
    {\color{red}\solid}~wall-normal ($\alpha=2$); 
    {\color{blue}\solid}~spanwise ($\alpha=3$). 
    The results previously obtained in a shorter domain
    \citep{uhlmann:08a} are indicated by dashed lines.  
    The graph in $(b)$ shows a close-up of the same data as $(a)$ for
    small separation times. 
    %
  }
  \label{fig-lagcorr}
\end{figure}
\subsubsection{Lagrangian particle velocity correlations}
\label{sec-results-lag-vel}
\cite{uhlmann:08a} found that Lagrangian velocity
correlations were strongly affected by the flow structures with the
largest streamwise extension. 
For the present case, figure~\ref{fig-lagcorr} shows the
auto-correlation of particle velocity components along the
trajectories.  
Please note that the correlation data is averaged over all particles
and integrated over the full observation interval 
\citep[cf.\ definition in equation~7 of][]{uhlmann:08a}, which means
that no distinction is made concerning the particles' instantaneous
wall-distance.  
It had been observed by \cite{uhlmann:08a} that the finite (periodic)
domain in conjunction with the presence of very long-lived flow
structures leads to a statistical bias manifesting itself in the form
of successive peaks in the Lagrangian particle velocity
autocorrelation curves at intervals corresponding to an average return
time (i.e.\ equivalent to the domain length $L_x$ divided by the
apparent velocity lag $u_{rel}\approx u_{b}$). 
In the present case we still observe repeated peaks in the correlation
functions shown in figure~\ref{fig-lagcorr}, but now at intervals of
approximately $16\,T_{b}$ -- again consistent with the expected
return time ($L_x/u_{b}$). 
Furthermore, a faster initial decorrelation of the streamwise particle
velocity component is observed when using the enlarged box. This
finding is consistent with the fact that the spatial decorrelation of
the corresponding fluid velocity field is faster in the longer box
(cf.\ figure~\ref{fig-pspec-12-vs-21}), although it is still not
complete in the present spatial domain. 
%

%
Despite the statistical bias due to the finite streamwise period, the
short-time behavior of the auto-correlation in the present case is
expected to be well represented. 
\revision{%
  Therefore, we have computed the 
  Taylor micro-scale, i.e.\ the intercept with the horizontal axis
  of the osculating parabola at zero separation: it measures $0.72\,T_b$,
  $0.35\,T_b$ and $0.40\,T_b$ for the streamwise, wall-normal and
  spanwise velocity components, respectively.
}{%
  As shown in figure~\ref{fig-lagcorr}$(b)$, 
  the initial decay of the correlation function is fastest for the
  wall-normal component, slightly slower for the spanwise component
  and significantly slower for the streamwise component of the
  particle velocity. 
  The corresponding time scale (Taylor micro-scale, i.e.\ the
  intercept with the horizontal axis of the osculating parabola at
  zero separation) measures $0.72\,T_b$,
  $0.35\,T_b$ and $0.40\,T_b$ for the streamwise, wall-normal and
  spanwise velocity components, respectively.
  This result is in qualitative agreement with the auto-correlation of
  fluid particles in single-phase turbulent channel flow
  \citep{choi:04}. The explanation for the directional differences put
  forth by these authors is based upon the anisotropy of the near-wall
  coherent structures (velocity streaks and quasi-streamwise
  vortices). 
  Despite the presence of particle wakes and additional large scales,
  the flow structure in the present case is similar to single-phase
  channel flow. Therefore, we expect a similar cause to be responsible
  for the anisotropic autocorrelation of the particle velocity in the
  present case.  
  }
%

%% file: results_b.tex
\begin{figure}
  \centering
  \begin{minipage}{.2\linewidth}
    \includegraphics[width=\linewidth,clip=true,
    viewport=365 100 840 1700]
    {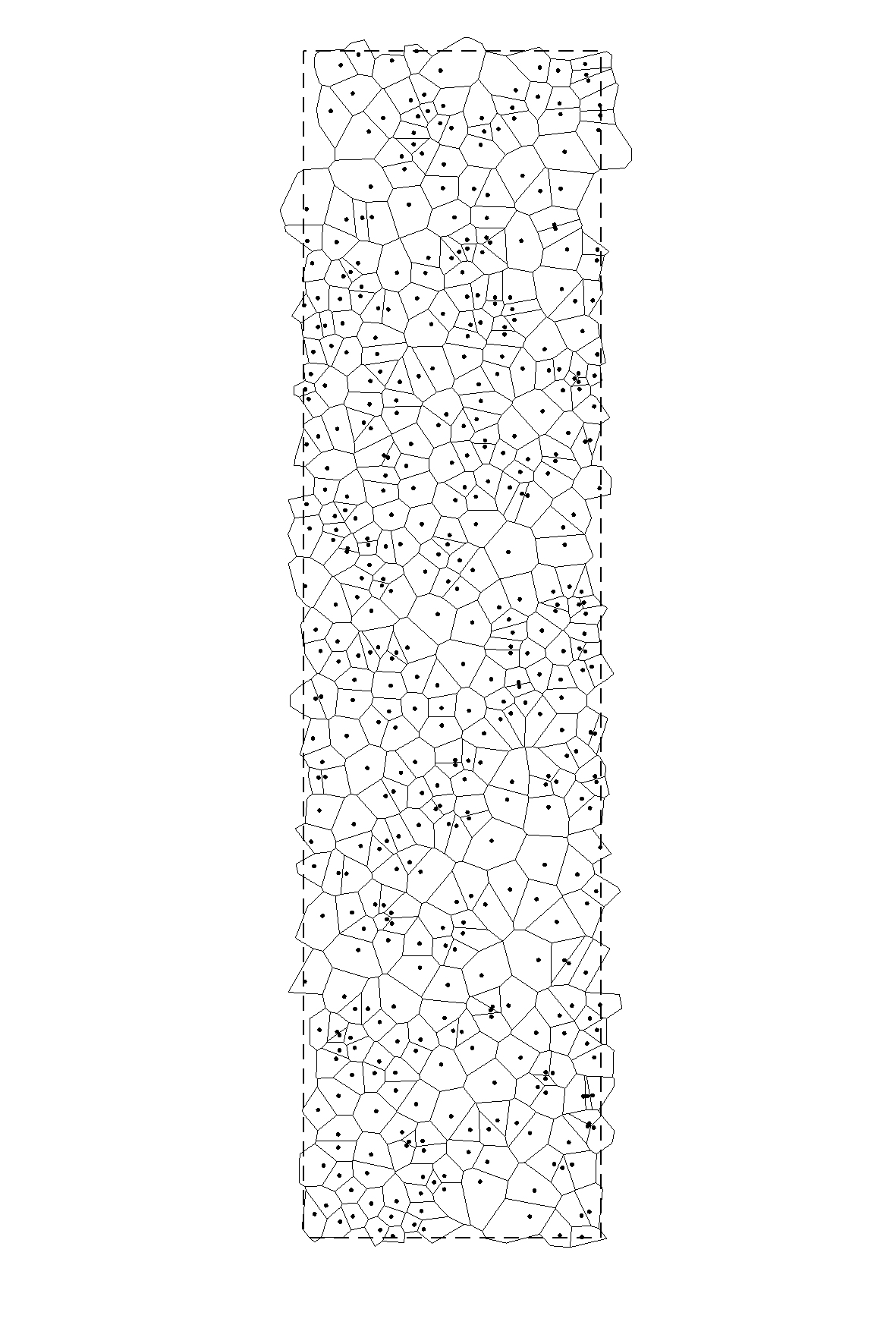}
  \end{minipage}
  \hspace*{.1\linewidth}
  \begin{minipage}{.4\linewidth}
    \includegraphics[height=\linewidth,clip=true]
    {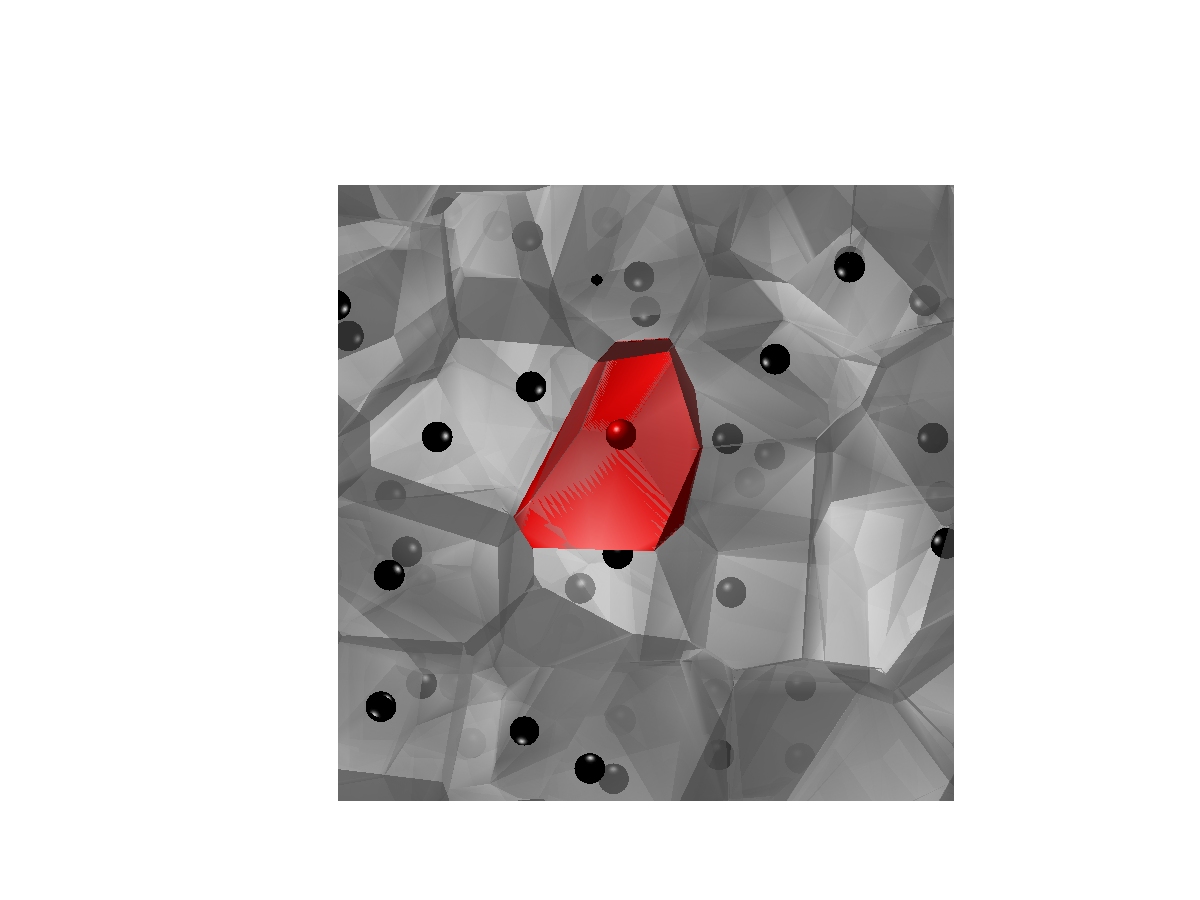}
  \end{minipage}
  \caption{
    $(a)$ Example of a Voronoi tesselation of a bi-periodic (wall-parallel)
    plane, performed with respect to the locations indicated as black dots. 
    Note that the cells are continued periodically in order to
    guarantee a space-filling and non-redundant tesselation. The
    periodic boundaries are indicated by dashed lines.
    $(b)$~Close-up of a Voronoi tesselation in three-dimensions, as
    performed in the present case. 
  } 
  \label{fig-voronoi-meth}
\end{figure}
\subsection{Voronoi analysis of spatial particle distribution}
A number of techniques have been established for the purpose of
characterizing the spatial structure of the dispersed phase. In the
precursor study \citep{uhlmann:08a} conventional box-counting
\citep{fessler:94}, nearest-neighbor statistics \citep{kajishima:04},  
as well as genuine clustering detection algorithms
\citep{wylie:00,melheim:05} were employed. 
Based upon these measures, it was concluded by \cite{uhlmann:08a} that 
no significant instantaneous accumulation of particles takes place. 

A new technique based upon Voronoi tessellation has recently been
proposed in the context of particulate flows \citep{monchaux:10b,monchaux:12}. 
Figure~\ref{fig-voronoi-meth} shows an example in two
dimensions, where starting with given particle center positions (the
'sites') the space is covered with cells which have the property that
each point inside the cell is closer to the cell's site than to any other
cell's site. As a consequence, the inverse of a Voronoi cell's
volume is an indicator of the local particle concentration. 
A statistical analysis of Voronoi tesselations can be performed by
computing the p.d.f.\ of cell volumes, followed by a comparison with
reference data for random particle positions \citep{monchaux:10b,monchaux:12}.
Compared to previous approaches for characterizing the spatial
distribution of the dispersed phase, Voronoi analysis offers two 
key advantages: 
(i) richness of geometrical data, offering the possibility to 
compute various derived diagnostic quantities; 
(ii) computational efficiency due to the availability of a fast
algorithm for the tessellation.
In the following, we present the results from an application of this
technique to our present data-set.  

\begin{figure}
  \centering
  \begin{minipage}{2ex}
    \rotatebox{90}{$pdf$}
  \end{minipage}
  \begin{minipage}{.45\linewidth}
    \centerline{$(a)$}
    \includegraphics[width=\linewidth]
    {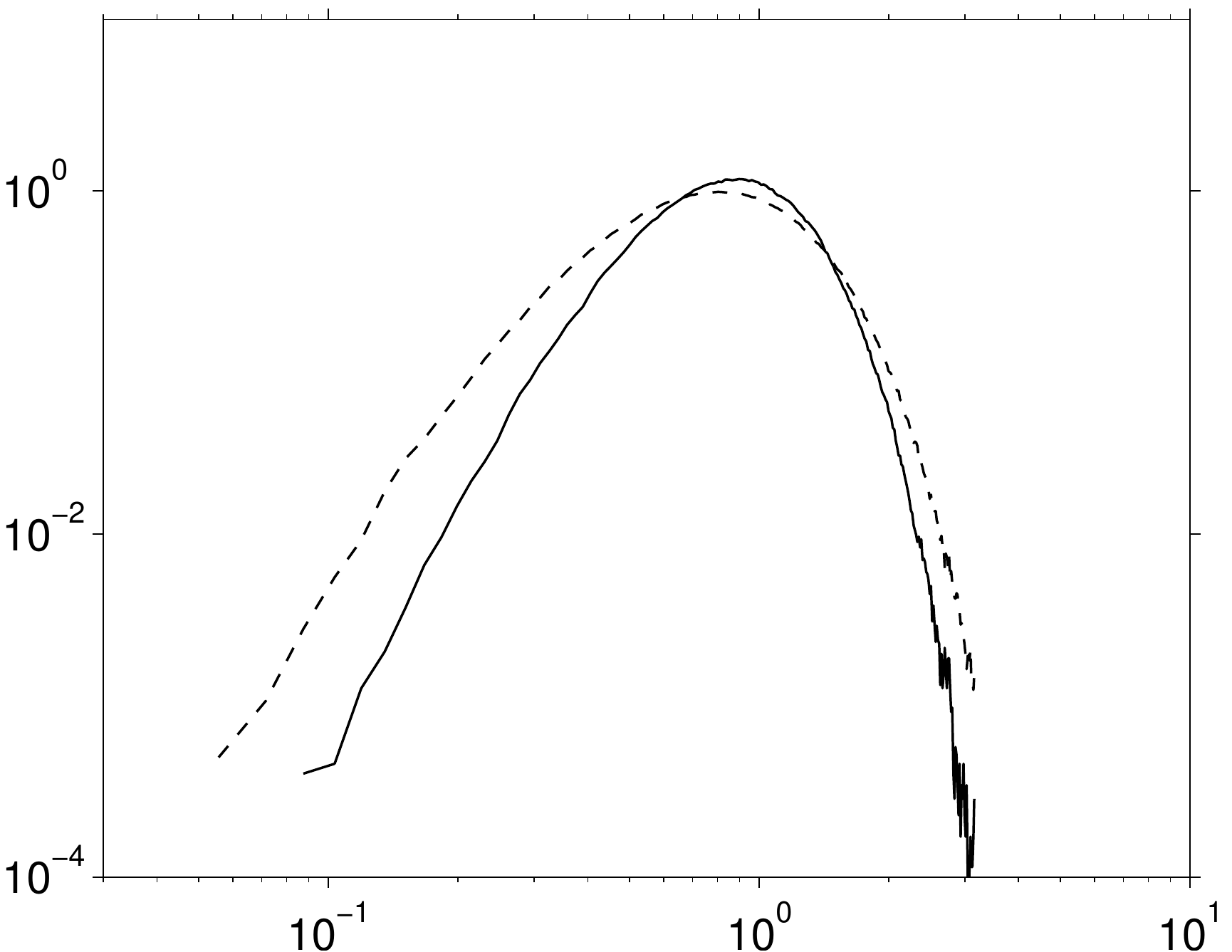}
    \\
    \centerline{$V/\langle V\rangle$}
  \end{minipage}
  \begin{minipage}{2ex}
    \rotatebox{90}{$pdf$}
  \end{minipage}
  \begin{minipage}{.45\linewidth}
    \centerline{$(b)$}
    \includegraphics[width=\linewidth]
    {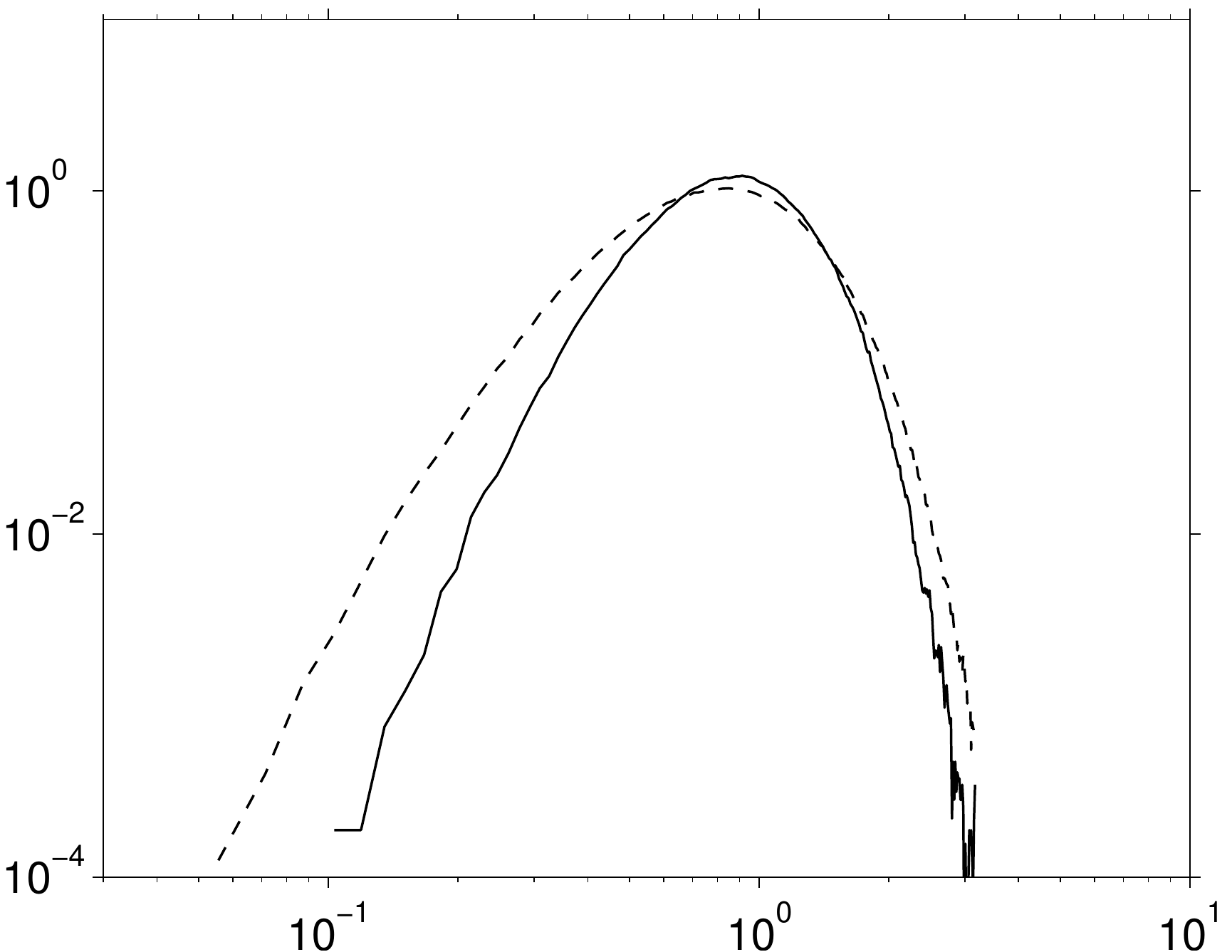}
    \\
    \centerline{$V/\langle V\rangle$}
  \end{minipage}
  \caption{
    Pdf of Voronoi cell volumes. 
    After three-dimensional Voronoi tesselation, pdfs of the cell
    volumes are computed in 20 uniform bins across the channel width. 
    The graph in $(a)$ shows the pdf corresponding to the first bin
    near the wall, $(b)$ is for the bin adjacent to the channel
    centerline. 
    The present DNS data is shown by a solid black curve; the dashed 
    curve corresponds to randomly placed (non-overlapping) particles. 
    Data is averaged over 
    2200 
    instantaneous snapshots. 
  } 
  \label{fig-voronoi-vol}
\end{figure}
A random distribution of points in unbounded space yields a Gamma
distribution for the Voronoi volumes \citep{ferenc:07}. However, since
our particles are of finite size, a random positioning of particle
centers without additional constraints can lead to non-physical
overlap of particle boundaries. Furthermore, the presence of domain
boundaries (walls) can further modify the p.d.f.\ of Voronoi cell volumes. 
Therefore, we have numerically determined the Voronoi tessellation of
particle positions which have been generated randomly, with the
additional constraint that no (particle/particle, particle/wall)
overlap is obtained. The corresponding 
p.d.f.'s (which are indeed close to a Gamma distribution) are added to the
graphs for the purpose of comparison with the actual DNS data. 

An additional issue arises in the present context due to the
spatial inhomogeneity of the average particle concentration. In order
to properly account for this variation (i.e.\ non-constant average
volume of Voronoi cells) in the wall-normal direction, we have divided
the channel height into 20 intervals of equal width. Statistics of
Voronoi cells were then performed individually for each `bin', i.e.\
computed over all Voronoi cells whose site is located inside the
corresponding wall-normal interval. 

A graph of the p.d.f.\ of Voronoi cell volumes based upon an ensemble of
particle positions at 
2200 
different times in our present simulation is
depicted in figure~\ref{fig-voronoi-vol}, for the interval adjacent to
the wall (figure~\ref{fig-voronoi-vol}$a$) and for the centerline 
(figure~\ref{fig-voronoi-vol}$b$). 
In studies of particulate flows which exhibit significant preferential
concentration, an appreciable deviation of the Voronoi cell area p.d.f.
from the random case is observed \citep{monchaux:10b}, characterized by
larger-than-random probabilities of finding very small and very large
Voronoi cells while roughly maintaining the overall shape of the
p.d.f.. 
Contrarily, our data in figure~\ref{fig-voronoi-vol} shows the
opposite behavior:  
very small Voronoi cells and very large Voronoi cells are less probable
than in the case of randomly placed particles.
In other words, large deviations from the average cell volume are
found to be less probable than in a case of randomly placed
particles. Therefore, the state of the dispersed phase can be
characterized as more ordered than random, slightly (but
significantly) tending towards a homogeneous spatial distribution.  
It should be noted that the results of \cite{uhlmann:08a} for the
average distance to the nearest particle (cf.\ figure~23 therein) did
indicate a slight deviation from a fully random particle ensemble,
tending towards the value for a homogeneous particle array. 
Box-counting and direct cluster identification, however, did not pick
up noticeable differences with respect to randomness. 

The deviation from a random state as presently observed from Voronoi
cell volume statistics is consistent with particles having a weak 
tendency 
to form a regular pattern. 
While the extreme case of particles forming a uniform cubical lattice
would yield a Dirac distribution (with a pulse at the average cell volume), 
the present data shown in figure~\ref{fig-voronoi-vol}
only weakly deviates from the purely random reference curve. 
On the other hand, preferential particle concentration outside of
vortical structures can indeed be ruled out in the present case,
confirming the previous findings of the precursor
study \citep{uhlmann:08a}. 

\begin{figure}
  \centering
  \begin{minipage}{2ex}
    \rotatebox{90}{$pdf$}
  \end{minipage}
  \begin{minipage}{.45\linewidth}
    \centerline{$(a)$}
    \includegraphics[width=\linewidth]
    {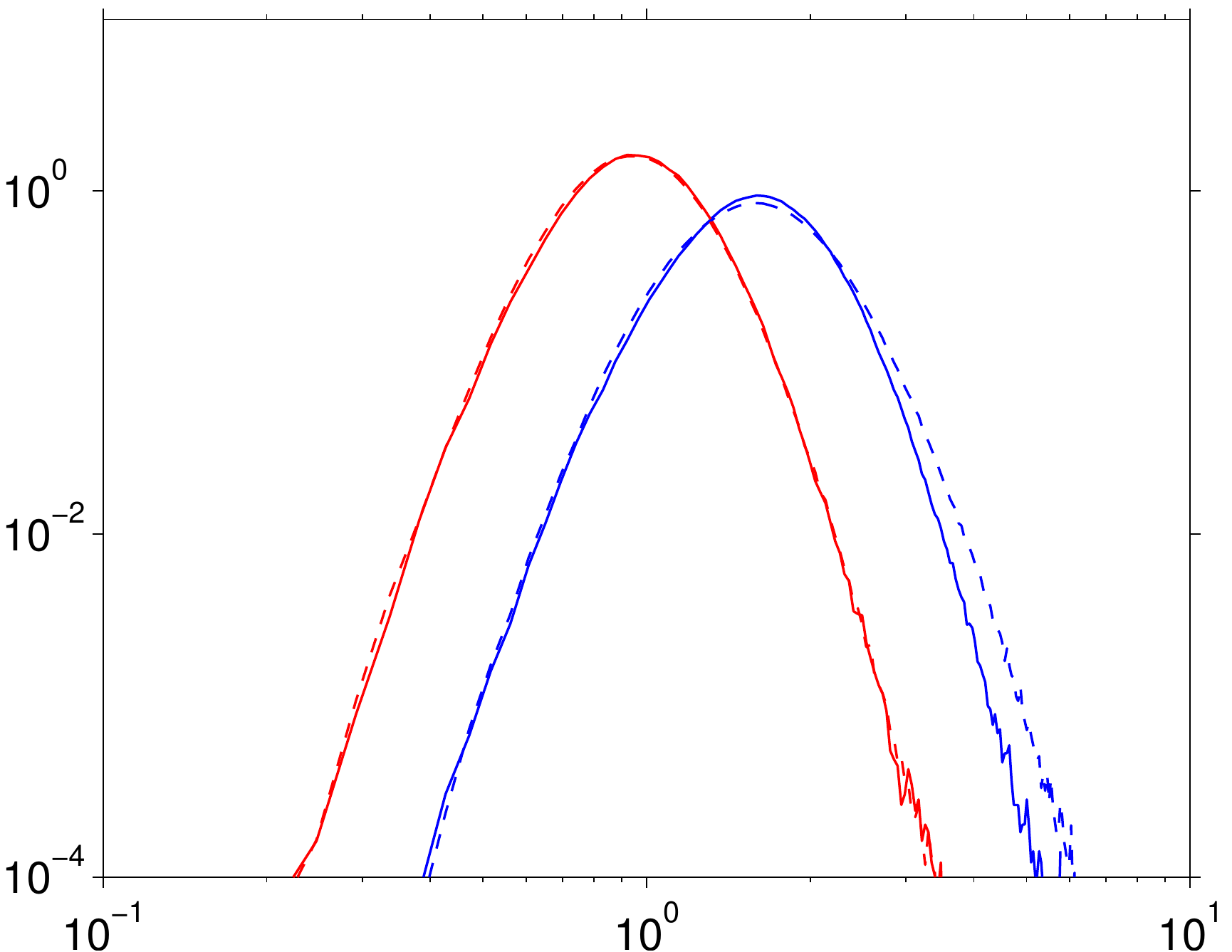}
    \\
    \centerline{$A^{V}_\alpha$}
  \end{minipage}
  \begin{minipage}{2ex}
    \rotatebox{90}{$pdf$}
  \end{minipage}
  \begin{minipage}{.45\linewidth}
    \centerline{$(b)$}
    \includegraphics[width=\linewidth]
    {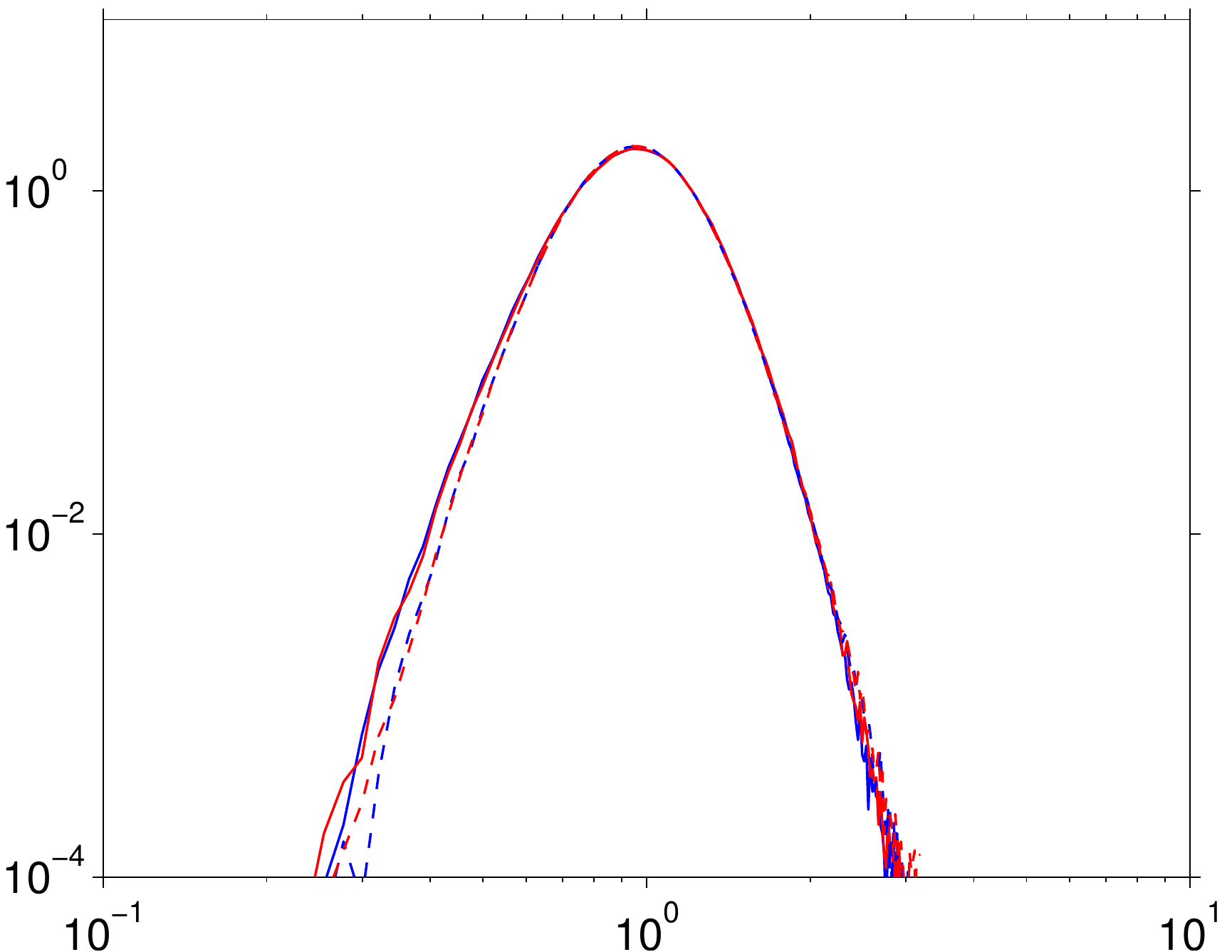}
    \\
    \centerline{$A^{V}_\alpha$}
  \end{minipage}
  \caption{
    Pdf of Voronoi cell aspect ratios, defined as the maximum
    extension of a given cell in the respective coordinate
    directions: 
    streamwise/wall-normal ($A^{V}_y$) are shown in blue, 
    streamwise/spanwise ($A^{V}_z$) are shown in red.
    As in figure~\ref{fig-voronoi-vol}, graph $(a)$ shows the pdf
    corresponding to the first bin (out of 20) near the wall, 
    $(b)$ is for the bin adjacent to the channel centerline; 
    solid lines correspond to DNS data, dashed lines to random particle
    positions. 
  } 
  \label{fig-voronoi-aspect}
\end{figure}
Additional quantities of interest can be deduced from the Voronoi
tessellation. In our case, where there exist preferred spatial
directions due to gravity and the general non-isotropy of channel
flow, particle accumulation (if any) is not expected to take place in
an isotropic manner. In particular, possible elongated particle chains
with a preferential orientation can be expected to lead to
non-isotropic statistics of Voronoi cells.  
Therefore, we have computed the aspect ratio of Voronoi cells 
dividing the largest streamwise extension $l_x$ by the largest
wall-normal $l_y$ and spanwise extensions $l_z$ of each Voronoi cell,
viz. 
\begin{equation}\label{equ-def-voronoi-aspect}
  A^{V}_\alpha=\frac{l_x}{l_\alpha}
  \quad\forall\,\alpha=2,3
  \,.
\end{equation}
The corresponding p.d.f.'s are shown in figure~\ref{fig-voronoi-aspect}
(again for two different wall-normal intervals).
Adjacent to the solid surface, the streamwise/spanwise and the
streamwise/wall-normal aspect ratios are clearly distinct: the latter
having a mean of $1.74$ due to the constraint by the
wall-boundary. 
%
In the center of the channel the constraint by the wall is no longer
felt and the curves for the streamwise/spanwise and the
streamwise/wall-normal aspect ratios coincide.  
Except for the first interval adjacent to the walls, deviations of the
DNS data from the curves for random particle arrangements are only
observed as an increased probability of very small values $A^{V}_y$
and $A^{V}_z$, i.e.\ Voronoi cells which are squeezed in the
streamwise direction are found more frequently than in a random
arrangement. This small but systematic difference (cf.\
figure~\ref{fig-voronoi-aspect}$b$) implies that particles have indeed
a weak tendency to align along the streamwise direction.  
It can be speculated that such an alignment is induced by the
sheltering effect of particle wakes, which is known to cause trailing
particles to approach leading particles \citep{wu:98,fortes:87}.
%

%% file: results_c.tex
\begin{figure}
  \begin{minipage}[t]{.47\linewidth}
  \begin{minipage}{10ex}
    $\langle a_{p,i}\rangle^+$,
    \\[2ex]
    $\displaystyle\frac{\mbox{d}\langle u_i^\prime v^\prime\rangle^+}{\mbox{d}y^+}$
  \end{minipage}
  \begin{minipage}{.8\linewidth}
    \centerline{$y/D$}
    \includegraphics[width=\linewidth]
    {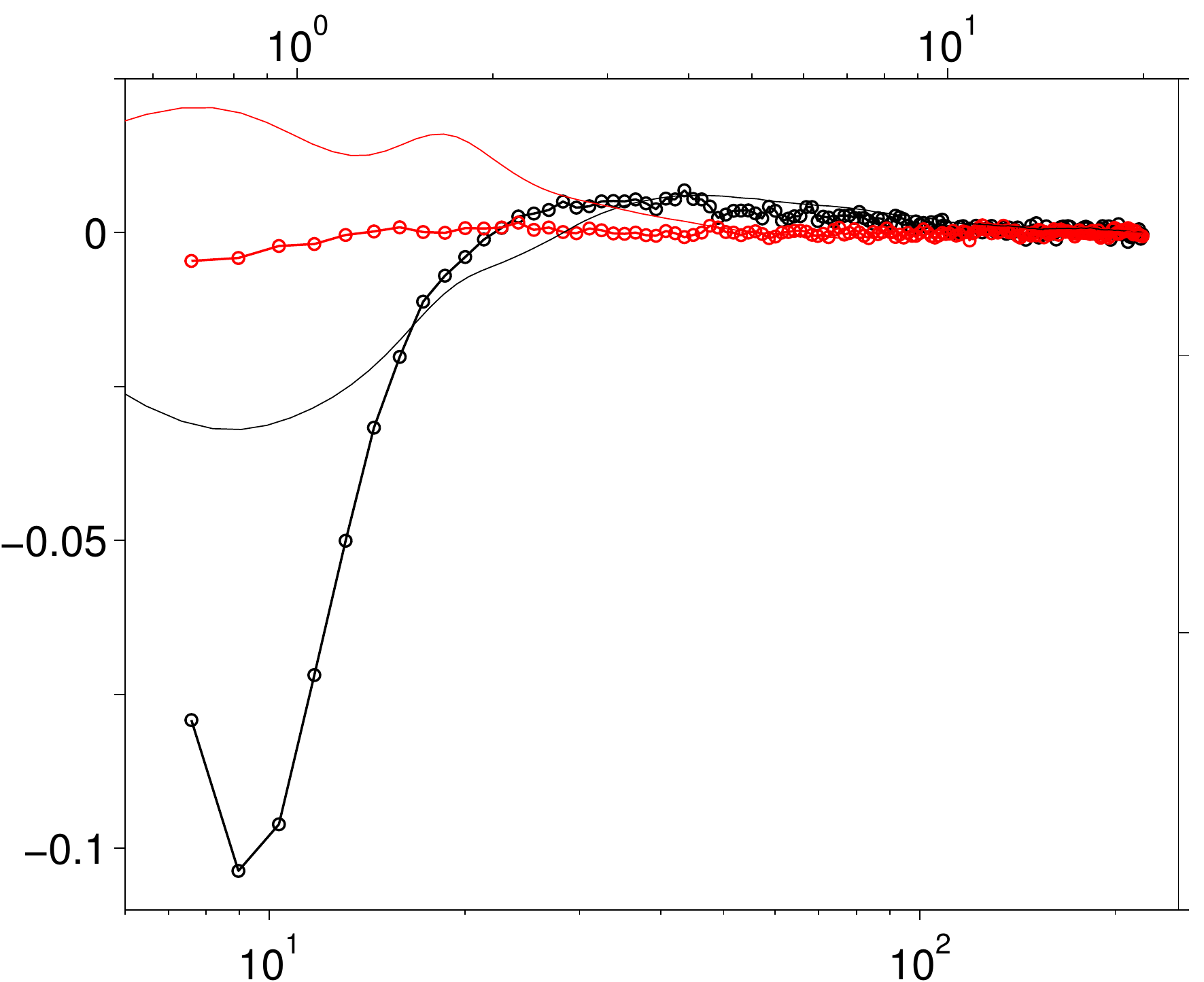}
    \\
    \centerline{$y^+$}
  \end{minipage}
  \caption{
    Mean particle acceleration, as a function of the
    wall distance. 
    {\color{black}$-\!\!\circ\!\!-$},~streamwise component $\langle
    a_{p,1}\rangle$;  
    {\color{red}$-\!\!\circ\!\!-$},~wall-normal component $\langle
    a_{p,2}\rangle$.
    The lines without symbols show the wall-normal gradients of
    Reynolds stresses:
    {\color{black}\solid},~$\mbox{d}\langle u^\prime v^\prime\rangle/\mbox{d}y$; 
    {\color{red}\solid},~$\mbox{d}\langle v^\prime v^\prime\rangle/\mbox{d}y$.
    Wall units have been used as the normalization scale ($a_{ref}=u_\tau^3/\nu$).
  }
  \label{fig-force-mean}
  \end{minipage}
  \hfill
  \begin{minipage}[t]{.47\linewidth}
  \begin{minipage}{3ex}
    \rotatebox{90}{$
      (\langle a_{p,\alpha}^\prime a_{p,\alpha}^\prime\rangle^+)^{1/2}$
    }
  \end{minipage}
  \begin{minipage}{.8\linewidth}
    \centerline{$y/D$}
    \includegraphics[width=\linewidth]
    {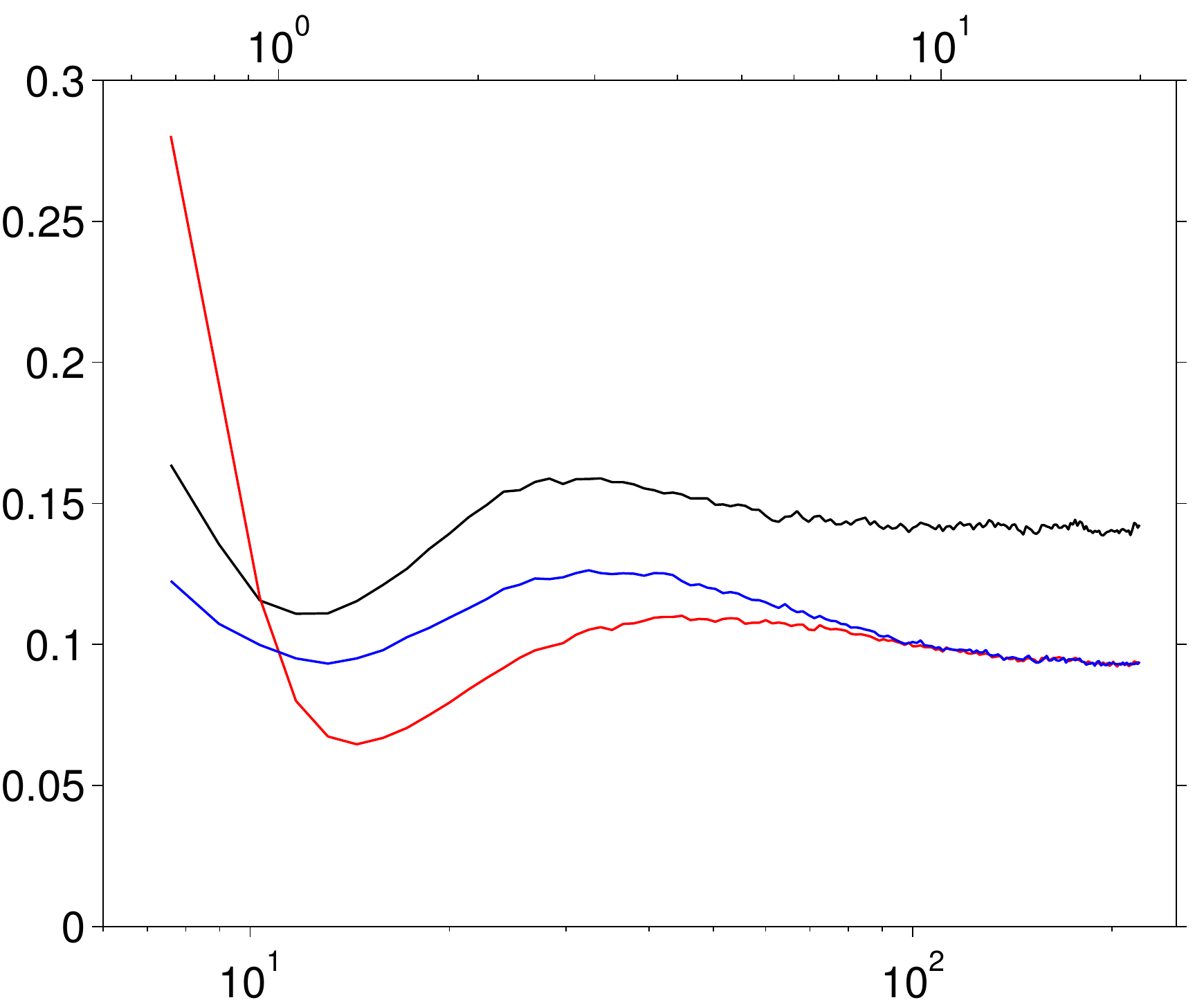}
    \\
    \centerline{$y^+$}
  \end{minipage}
  \caption{
    Standard deviation of particle acceleration as a function of the wall
    distance. 
    %
    %
    {\color{black}\solid},~streamwise component; 
    {\color{red}\solid},~wall-normal component; 
    {\color{blue}\solid},~spanwise component. 
  }
  \label{fig-force-rms}
  \end{minipage}
\end{figure}
\subsection{
  Particle acceleration statistics
}
%
In this section, we study the Lagrangian acceleration statistics of
the solid particles in the present flow.  
In recent years much attention has been given to the Lagrangian
acceleration statistics of fluid and solid particles, see for example
the review by \cite{toschi:09}. Apart from being  
relevant to a number of applications,
the growing interest has been mainly due to the appearance of new experimental methods
and the abundance of direct numerical simulations employing the
point-particle approach. 
Most previous studies have been performed for homogeneous flows. 
Exceptions are, for example,  the study of Lagrangian acceleration
statistics of 
sub-Kolmogorov size
inertial particles in a turbulent boundary layer, 
by means of experiments \citep{gerashenko:08} and DNS
\citep{lavezzo:10}. 
%
Finite size effects on Lagrangian particle acceleration statistics
have 
\revision{only}{%
  mainly}
been studied experimentally, using 
both neutrally buoyant and heavy particles 
\citep{qureshi:07,qureshi:08,xu:08,brown:09}. 
\revision{}{%
  Two contributions based upon DNS studies should be mentioned. 
  \cite{calzavarini:09} have extended the point-particle approach by
  including Fax\'en corrections, while \cite{homann:10} have
  simulated the motion of a single, resolved, finite-size 
  particle, both in forced homogeneous-isotropic turbulence. 
  In both cases, however, there was no apparent slip
  velocity due to the exclusion of gravity in the former case and due
  to matched density in the latter. 
}
%

Before turning to the \revision{}{present} results, it should be
pointed out that we have 
filtered from the particle force data all records corresponding to
particle collisions (i.e.\ those where there is a force contribution
$\mathbf{F}_C$ in equation~\ref{equ-numa-newton-rigid-body-linear}
stemming from the artificial repulsion force).  

\begin{figure}
    \begin{minipage}{3ex}
      \rotatebox{90}{
      $(\langle F_{H\,i}\rangle\pm\langle
        F_{H\,i}^{\prime\,2}\rangle^{1/2})/F_{ref}^{(i)}$
    }
    \end{minipage}
    \begin{minipage}{.45\linewidth}
      \centerline{$(a)$}
      \includegraphics[width=\linewidth]
      {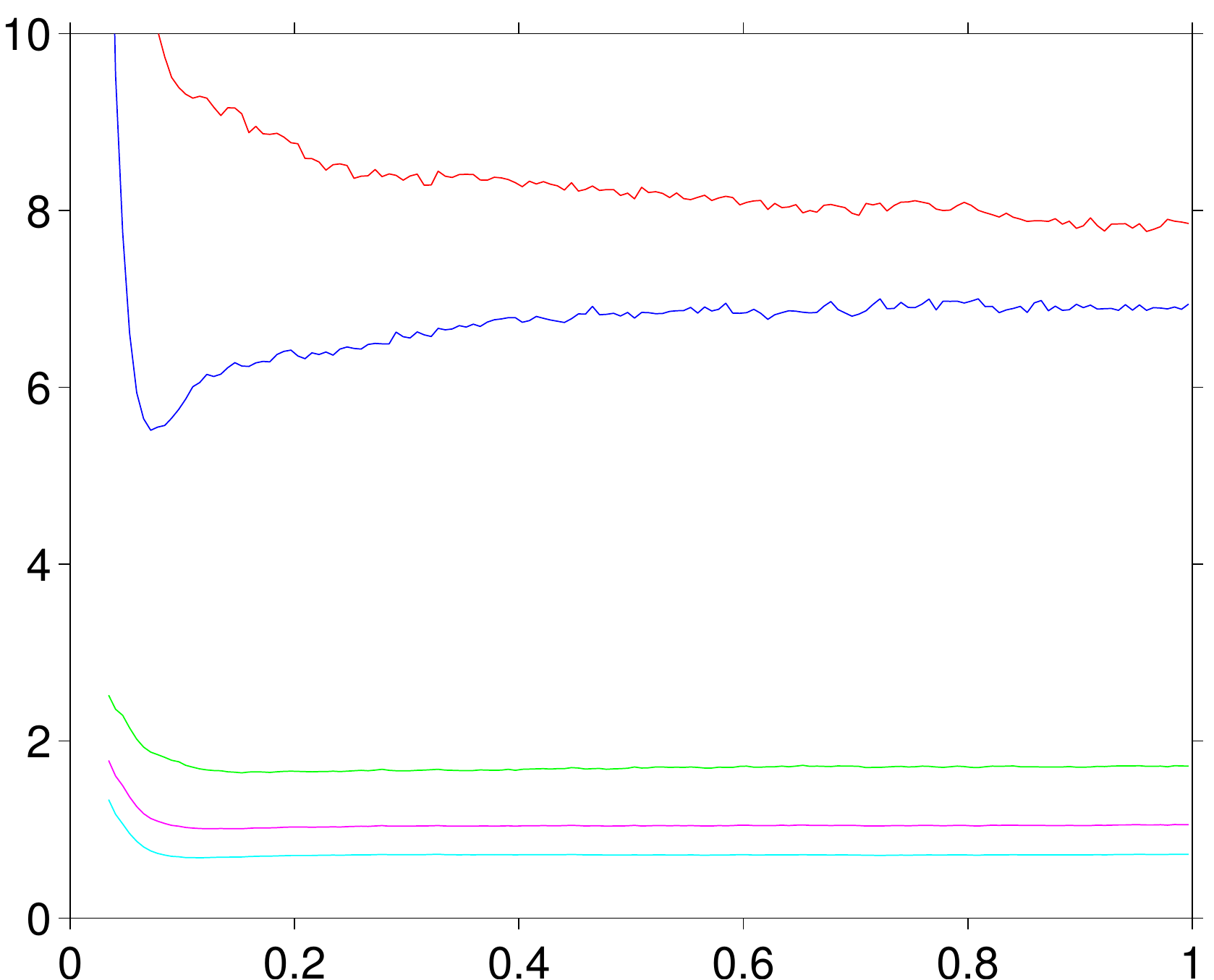}
      \\
      \centerline{$y/h$}
    \end{minipage}
    \begin{minipage}{3ex}
      \rotatebox{90}{
        $(\langle F_{H\,i}\rangle\pm\langle
        F_{H\,i}^{\prime\,2}\rangle^{1/2})/F_{ref}^{(i)}$
        \,,\quad
        $c_D$
      }
    \end{minipage}
    \begin{minipage}{.45\linewidth}
      \centerline{$(b)$}
      \includegraphics[width=\linewidth]
      {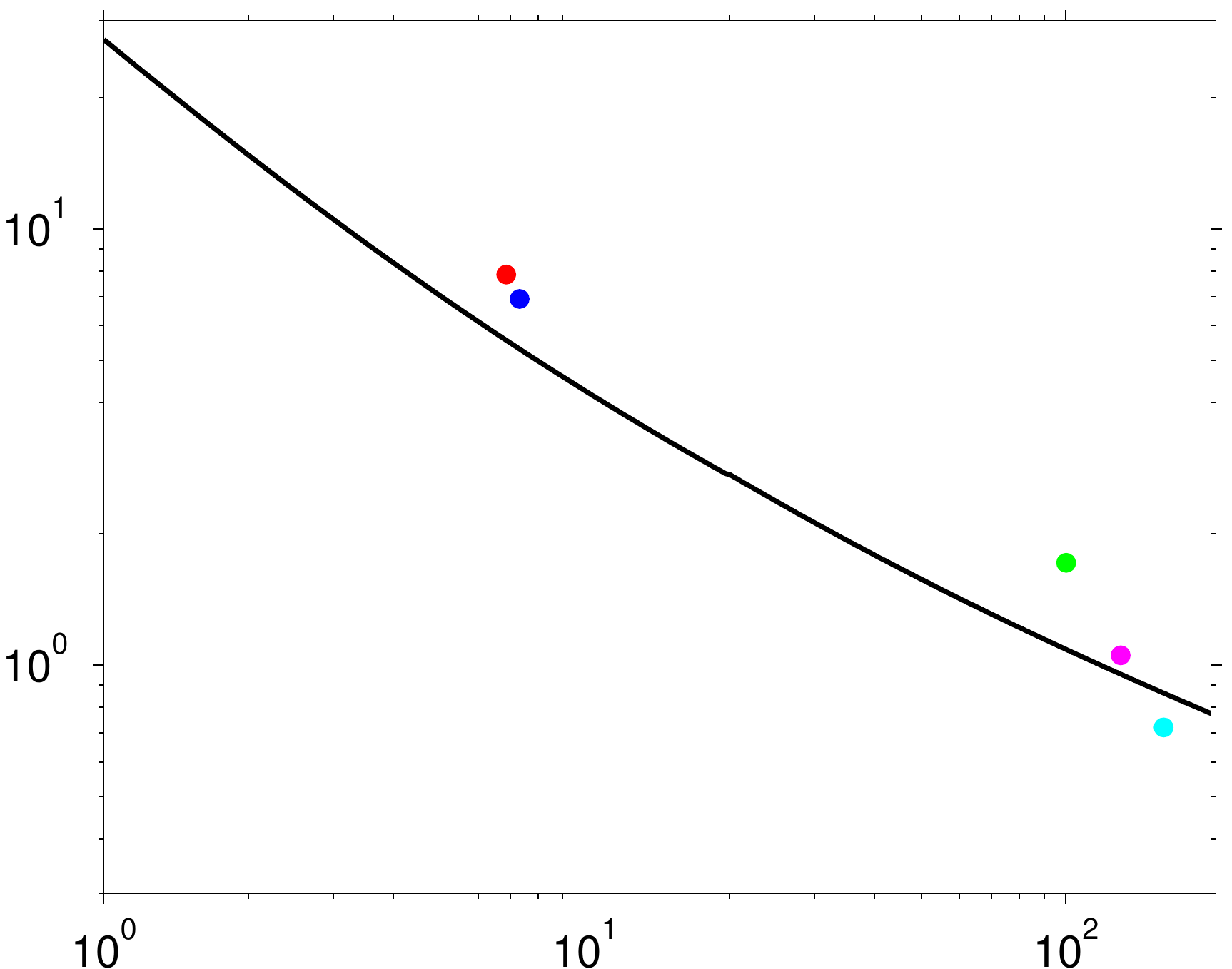}
      \\
      \centerline{$Re_D^{(i)}$\,,\quad
        $Re_D$}
    \end{minipage}
   \caption{%
    Mean and rms hydrodynamic particle force components
    normalized by a reference force
    $F_{ref}^{(i)}=\frac{1}{2}\rho_fA_pu_{rel}^{(i)2}$ (where $A_p=\pi
    D^2/4$). 
    $(a)$ shows wall-normal profiles, while in 
    $(b)$ the data is presented as a function of the local Reynolds
    number based on an approximation for the relative velocity.
    The red (blue) line is for the rms particle force in the
    wall-normal (spanwise) direction, the relative velocity in the
    reference force being taken as the rms fluid velocity, i.e.\ 
    $u_{rel}^{(2)}=\langle v_f^\prime v_f^\prime\rangle^{1/2}$ 
    ($u_{rel}^{(3)}=\langle w_f^\prime w_f^\prime\rangle^{1/2}$ ). 
    The magenta-colored line indicates the average particle force in
    the streamwise direction, the relative velocity being taken as the
    mean apparent slip velocity $u_{rel}^{(1)}=\langle
    u_p\rangle-\langle u_f\rangle$. 
    The cyan (green) colored lines indicate the mean plus (minus) the
    rms particle force in the streamwise direction, normalized by a
    reference force based upon the apparent slip velocity plus (minus)
    the rms streamwise fluid velocity. 
    Concerning the graph in $(b)$:  
    the horizontal axis represents the Reynolds number based upon the
    particle diameter and the respective relative velocity $u_{rel}^{(i)}$. 
    %
    %
    The black solid line indicates the standard drag law for spheres in
    uniform flow \citep[][table~5.2]{clift:78}. 
    The symbols correspond to the data in $(a)$ on the channel
    centerline, using the same color coding.  
    %
  }
  \label{fig-force-cd}
\end{figure}
Since the present flow is inhomogeneous in the wall-normal direction,
we start by considering  
the mean Lagrangian acceleration of the particles, a quantity which is
zero in 
statistically stationary 
homogeneous flows.
Recall that in 
turbulent channel flow
the
mean
 streamwise (wall-normal) Lagrangian acceleration of fluid particles turns out to be 
equal to the wall-normal gradient of $\langle u_f'v_f' \rangle$ ($\langle
v_f'v_f' \rangle$) \citep{yeo:10}.  
Fig.~\ref{fig-force-mean} displays the mean streamwise $\langle a_{p1}
\rangle$ and wall-normal  $\langle a_{p2} \rangle$ acceleration of the
solid  particles. Also for comparison the mean streamwise and wall-normal
Lagrangian fluid particle accelerations  
$\langle a_{f1} \rangle=\partial_y \langle u'_f v'_f \rangle$
and $\langle a_{f2} \rangle=\partial_y \langle v'_f v'_f  \rangle$
 are shown. 
The streamwise and wall-normal acceleration components 
are non-zero near the wall, while the mean spanwise acceleration is zero everywhere (not shown).
In the central region, $\langle a_{p1} \rangle$ is roughly zero, as
can be expected 
from an equilibrium between buoyancy and drag forces. 
The presence of the wall alters this equilibrium and a
negative 
peak in $\langle a_{p1} \rangle$ appears
at $y^+\approx10$, which is 
significantly 
larger in magnitude than the corresponding
peak observed for $\langle a_{f1} \rangle$.  
Furthermore, the mean particle acceleration $\langle a_{p1} \rangle$
presents a milder positive peak at $y^+\approx30$. 
Concerning the wall-normal acceleration, $\langle a_{p2} \rangle$ 
exhibits slightly 
positive values in the interval $15\lesssim y^+\lesssim40$ (equivalent
to a mean lift 
acting towards the channel center) 
which are smaller in magnitude than those presented by 
the fluid counterpart
$\langle a_{f2}\rangle$. Closer to the wall ($y^+\lesssim15$) the mean
wall-normal particle acceleration takes slightly negative values. 
%
The peak in $\langle a_{p1} \rangle$ might be explained by the following mechanism. 
%
Similar to turbulent fluid motion, inertial particle motion 
near the wall  
exhibits a preference for Q2 and Q4 events 
\revision{(in the terminology of quadrant analysis of the streamwise
  and wall-normal velocity fluctuations)}{%
  (in the terminology of quadrant analysis of the streamwise and
  wall-normal velocity fluctuations, where Q2 refers to ejections of
  low-speed fluid away from the wall and Q4 to an inrush -- or sweep
  -- of high-speed fluid)}, 
as shown by \cite{uhlmann:08a}.  
%
%
As a consequence, significant average values of the cross-correlation 
(i.e.\ `Reynolds stress') $\langle u_p^\prime v_p^\prime\rangle$ arise. 
Since the particles have larger inertia than a corresponding blob of
fluid, it can be expected that it takes longer for the particles to
adjust to the surrounding fluid conditions, and consequently the
correlation values can be expected to exceed those of the fluid
counterpart, as is indeed the case (cf.\
figure~\ref{fig-stat-pure-uu}$b$).   
Now, since particles arriving in the near-wall region from larger
wall-distances carry on average an excess axial velocity value (Q4
events), these particles will tend to 
\revision{experience 
  a negative acceleration
  (decrease in drag force).}{%
  experience a smaller amount of positive streamwise force (drag). 
  As a consequence, the (negative) submerged weight will exceed the
  positive drag and a negative particle acceleration in the vertical
  direction will result.
}
The opposite is expected for particles being
ejected away from the wall (Q2 events). 
The location where the average acceleration changes sign from negative
values to positive ones ($y/h\approx0.1$,
$y^+\approx20$) 
coincides approximately with
the location where the gradient of the Reynolds shear stress changes
sign; it is also an upper bound for the near-wall region in which
appreciable wall-normal gradients of the wall-normal particle velocity
fluctuation energy $\langle v_{p}^\prime v_{p}^\prime\rangle$ exist
(cf.\ figure~\ref{fig-stat-pure-uu}).    
As a consequence, the turbophoretic effect
\citep{caporaloni:75,reeks:83,uhlmann:08a} can be expected to lead to
a preponderance of negative particle acceleration for
$y/h\lesssim0.1$, as observed in figure~\ref{fig-force-mean}.   
%


\begin{figure}
  \begin{minipage}{3ex}
    \rotatebox{90}{$\sigma\cdot pdf$}
  \end{minipage}
  \begin{minipage}{.45\linewidth}
    \centerline{$(a)$}
    \includegraphics[width=\linewidth]
    {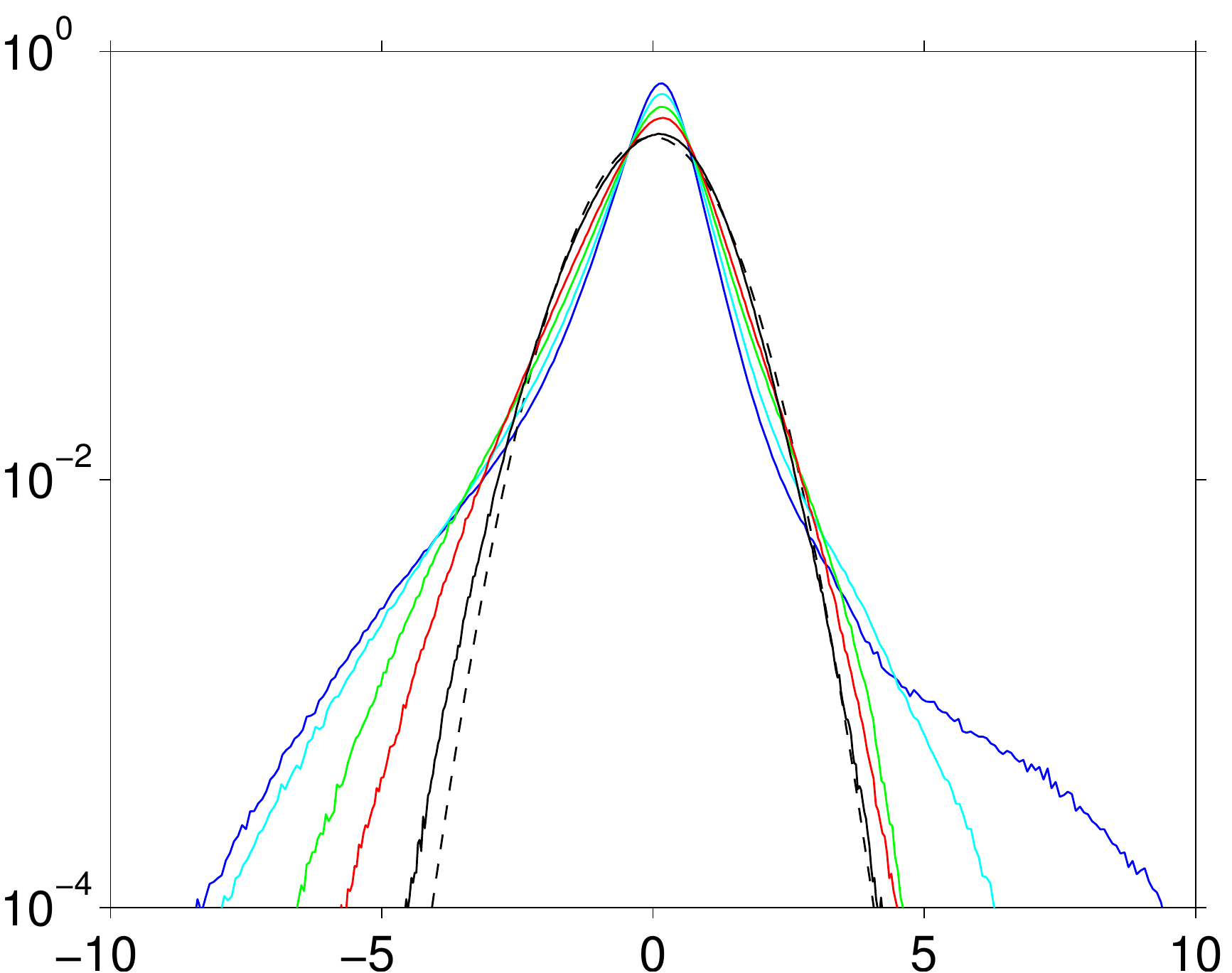}
    \\
    \centerline{$\delta_{u_i^{(p)}}/\sigma$}
  \end{minipage}
  \begin{minipage}{3ex}
    \rotatebox{90}{$\sigma\cdot pdf$}
  \end{minipage}
  \begin{minipage}{.45\linewidth}
    \centerline{$(b)$}
    \includegraphics[width=\linewidth]
    {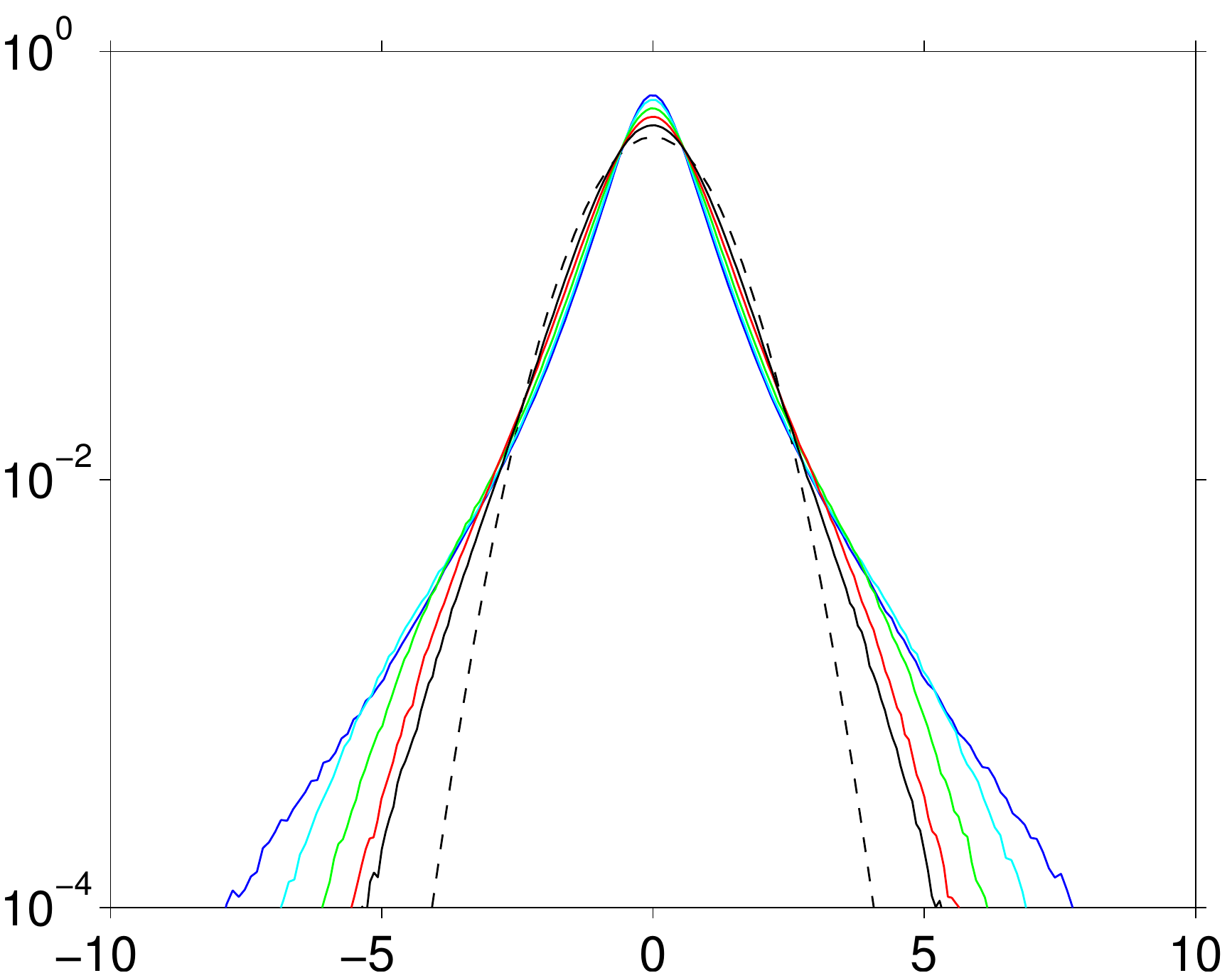}
    \\
    \centerline{$\delta_{u_i^{(p)}}/\sigma$}
  \end{minipage}
  \caption{
    Normalized probability density function of Lagrangian particle
    velocity increments
    $\delta_{u_{p,i}}=u_{p,i}^{\prime}(t+\tau)-u_{p,i}^{\prime}(t)$
    in 
    $(a)$ the streamwise direction  
    $(b)$ the wall-normal direction. 
    The different curves correspond to various values of the time lag 
    {\color{blue}\solid}, $\tau^+=1.4$; 
    {\color{cyan}\solid}, $\tau^+=2.9$; 
    {\color{green}\solid}, $\tau^+=7.2$; 
    {\color{red}\solid}, $\tau^+=14.3$; 
    {\color{black}\solid}, $\tau^+=70.8$.
  }
  \label{fig-force-pdfs-increment}
\end{figure}
Fig.~\ref{fig-force-rms} shows the r.m.s.\ values of the three components of the particle acceleration
scaled in wall units. The values obtained are 
comparable in magnitude to the r.m.s.\ fluid particle 
acceleration in single-phase channel flow  \citep{yeo:10} 
and to the r.m.s.\ acceleration of inertial particles in a turbulent
boundary layer \citep{gerashenko:08}.
The three components present a similar trend: a peak near the wall, a dip in the buffer layer
 and a gradual increase
towards a mild maximum and finally a plateau in the central region.
The r.m.s.\ of streamwise acceleration is larger than the other two components by about 50\%
except near the wall ($y^+<10$), where the r.m.s. of the wall-normal component presents maximum values.

\begin{figure}
  \centering
  \begin{minipage}{2ex}
    $K$
  \end{minipage}
  \begin{minipage}{.45\linewidth}
    \includegraphics[width=\linewidth]
    {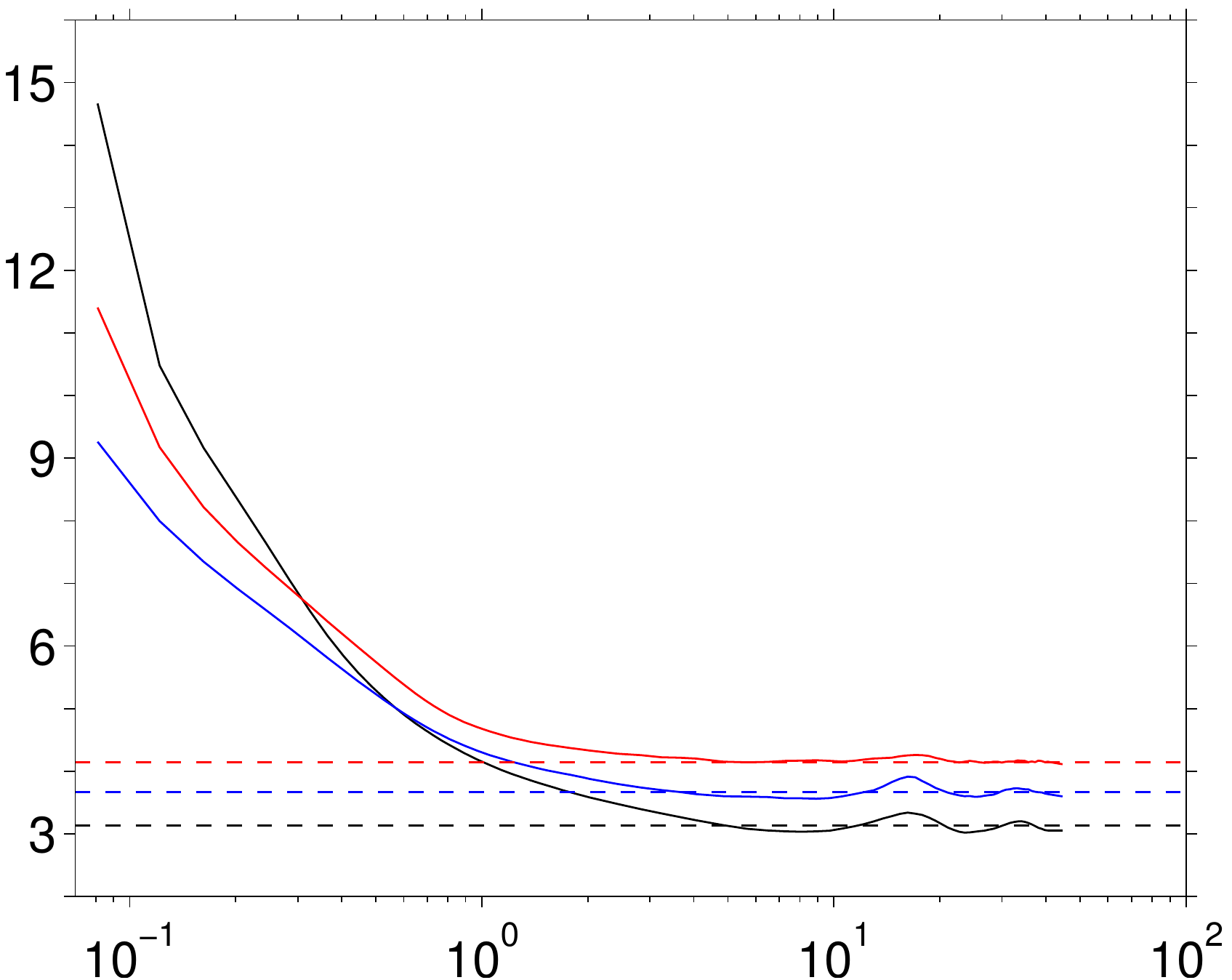}
    \\
    \centerline{$\tau/T_b$}
  \end{minipage}
  \caption{
    \revision{
    }{%
      Kurtosis $K(\delta_{u_{p,i}})$ of the Lagrangian velocity
      increments of solid particle 
      velocities
      $\delta_{u_{p,i}}(\tau)={u^\prime_{p,i}}(t+\tau)-{u^\prime_{p,i}}(t)$,
      plotted as a function of the separation time $\tau$.  
      The three coordinate directions are color-coded as in
      figure~\ref{fig-force-rms}.  
      For each component the dashed line indicates the asymptotic
      value $K(u^\prime_{p,i})/2+3/2$. 
    }
  } 
  \label{fig-lag-vel-incr-kurtosis}
\end{figure}
\begin{figure}
  \begin{minipage}{6ex}
    $(a)$\\[5ex]
    $R_{Lp,F_\alpha}$
  \end{minipage}
  \begin{minipage}{.45\linewidth}
    \includegraphics[width=\linewidth]
    {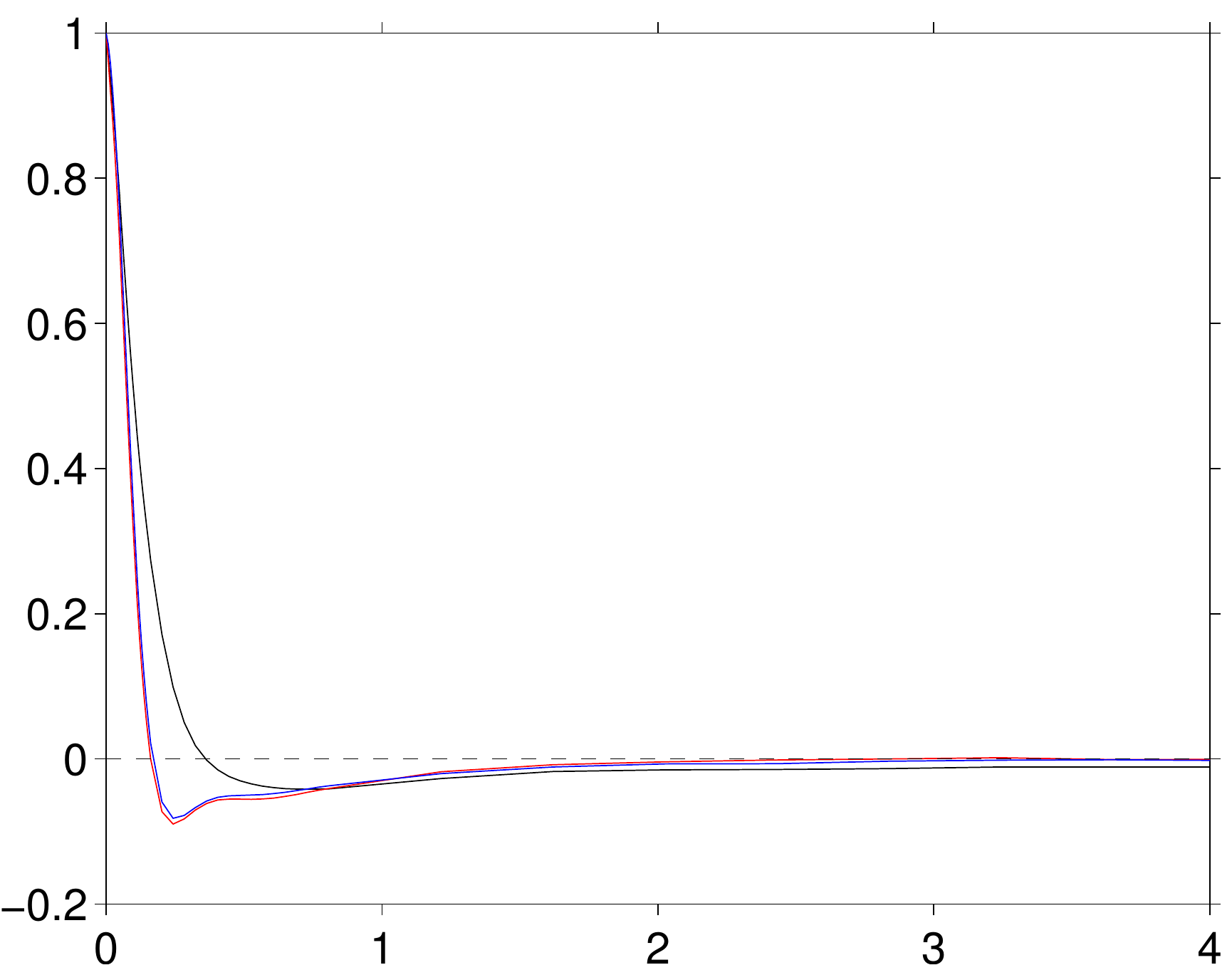}
    \\
    \centerline{$\tau/T_b$}
  \end{minipage}
  \begin{minipage}{6ex}
    $(b)$\\[5ex]
    $R_{Lp,F_\alpha}$
  \end{minipage}
  \begin{minipage}{.45\linewidth}
    \centerline{$\tau^+$}
    \includegraphics[width=\linewidth]
    {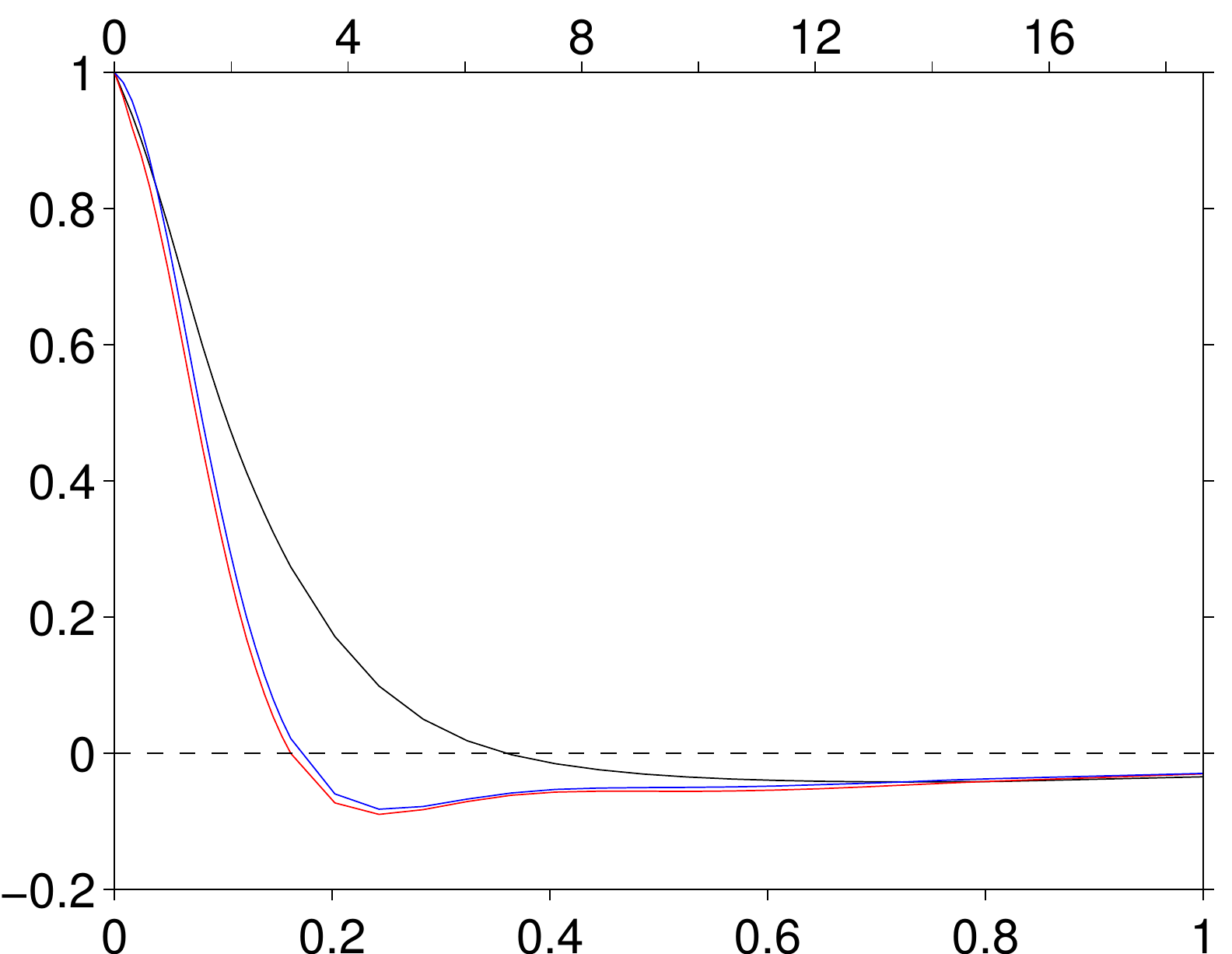}
    \\
    \centerline{$\tau/T_b$}
  \end{minipage}
   \caption{%
     Lagrangian autocorrelation of hydrodynamic force
     acting upon the particles, plotted as a function of
     the separation time $\tau$:  
     {\color{black}\solid},~streamwise ($\alpha=1$), 
     {\color{red}\solid},~wall-normal ($\alpha=2$), 
     {\color{blue}\solid},~spanwise ($\alpha=3$) components. 
     The graph in $(b)$ shows a close-up of the same data as $(a)$ for
     small separation times, including scaling of the abscissa in
     bulk time units and in wall units. 
     %
   }
   \label{fig-lagcorr-force}
 \end{figure}
%
Further insight can be obtained by studying the r.m.s.\ of the
hydrodynamic forces on the particles, since by Newton's law studying
the force is equivalent  
(up to a constant coefficient)
to studying the acceleration. We distinguish here 
the streamwise direction and the transverse directions. For the
streamwise direction the mean drag force 
is approximately balanced by the gravity force, while for the
transverse directions the mean lift and spanwise forces are zero,
except for the lift very near the wall, as discussed above.  
We consider, therefore, for the transverse directions
the r.m.s.\ of the 
hydrodynamic 
force and for the streamwise direction we consider
first, the mean 
hydrodynamic 
force, and second
the mean 
hydrodynamic 
force plus (or minus) the r.m.s.\ of the 
hydrodynamic 
force. 
For each direction, we use as a reference force 
$F_{ref}^{(i)}=\frac{1}{2}\rho_fA_pu_{rel}^{(i)2}$ where $A_p=\pi
D^2/4$
and $u_{rel}^{(i)}$ is a reference velocity characteristic for the
relative flow in the $i$th coordinate direction.
%
We estimate the relative velocity scale as follows: for the transverse
directions the r.m.s.\ fluid velocity is employed, viz.\ $u_{rel}^{(2)}=\langle
v_f^\prime v_f^\prime\rangle^{1/2}$  and $u_{rel}^{(3)}=\langle
w_f^\prime w_f^\prime\rangle^{1/2}$; 
in the streamwise direction, the relative velocity
is taken as the mean apparent slip velocity (plus or minus the r.m.s.\
fluid velocity), $u_{rel}^{(1)}=u_{lag}$ (or $u_{rel}^{(1)}=u_{lag}\pm\langle
u_f^\prime u_f^\prime\rangle^{1/2}$), according to whether the mean or
the mean plus/minus the r.m.s.\ value of the hydrodynamic force are
considered. 
Using this normalization, the mean and r.m.s.\ hydrodynamic forces are shown in Fig.~\ref{fig-force-cd}$(a)$.
All curves are rather flat for $y/h>0.2$, and they can be interpreted
as force coefficients. Defining a particle Reynolds number
$Re_D^{(i)}$ in each direction based on the respective relative
velocity $u_{rel}^{(i)}$,  
and plotting the force coefficients
from the central region of the channel as a function of  $Re_D^{(i)}$,
as shown in Fig. \ref{fig-force-cd}$(b)$, it turns out that the force coefficients are consistent 
with the standard drag law for spheres in uniform flow  \citep[][table~5.2]{clift:78}. 
We have checked that the scaling used for the transverse direction leads in the case 
of the streamwise direction to a force coefficient which is not consistent with the standard drag law.
%
As might be expected, the interpretation of the normalized
hydrodynamic force in terms of a drag coefficient (as shown in
Fig.~\ref{fig-force-cd}$b$) does provide an explanation for the large
differences in r.m.s.\ aceleration/force observed when simply
normalizing all components in wall units (cf.\ Fig.~\ref{fig-force-rms}). 
However, given the rather crude assumptions it is indeed remarkable
that the mean streamwise force coefficient in
Fig.~\ref{fig-force-cd}$b$ turns out relatively close to the value
given by the standard drag law valid for a fixed sphere in uniform flow.  

\begin{figure}
  \centering
  \begin{minipage}{3ex}
    \rotatebox{90}{$\sigma\cdot pdf$}
  \end{minipage}
  \begin{minipage}{.45\linewidth}
    \includegraphics[width=\linewidth]
    {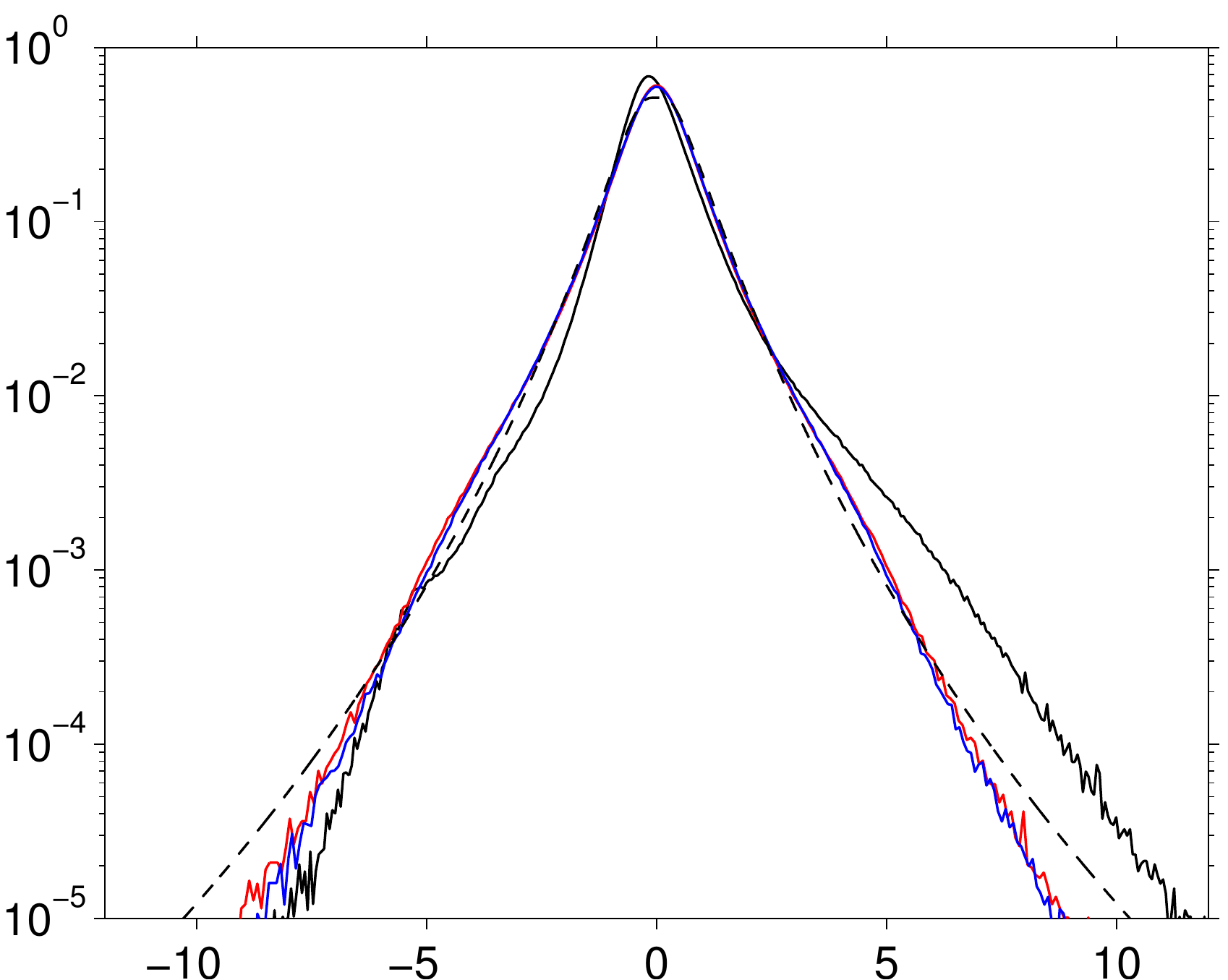}
    \\
    \centerline{$F_{i}^{(p)\prime}/\sigma$}
  \end{minipage}
  \caption{
    Normalized probability density function of hydrodynamic
    forces acting on the particles, 
    averaged over the full channel height, 
    {\color{black}\solid},~streamwise force component; 
    {\color{red}\solid},~wall-normal component; 
    {\color{blue}\solid},~spanwise component; 
    {\color{black}\dashed}, log-normal fit proposed in 
    \cite{qureshi:08}. 
  }
  \label{fig-force-pdfs}
\end{figure}
Next we consider 
the p.d.f.'s  of Lagrangian particle velocity increments
$\delta_{u_{p,i}}(\tau)={u^\prime_{p,i}}(t+\tau)-{u^\prime_{p,i}}(t)$. 
Fig. \ref{fig-force-pdfs-increment} displays the p.d.f.'s of streamwise and wall-normal
increments for various time lags.
Similar 
graphs 
have been reported by \cite{mordant:02}
and \cite{qureshi:07}, among others. 
\revision{The figure shows that for increasing time increments the
  p.d.f.'s tend to a Gaussian distribution.
}{%
The figure shows that for increasing time increments the
  p.d.f.'s become more like Gaussian distributions.
}
 For short time increments, of the order of the viscous time scale,  pronounced tails are present,
 which are the signature of intermittent Lagrangian dynamics. 
It is interesting to note the skewness that appears in the p.d.f.\ corresponding to the streamwise
velocity increments, a feature that has not been observed in previous investigations
in other flow geometries 
\citep{mordant:02,qureshi:07}. 
This 
point 
will be further discussed below. 
\revision{}{
  The decrease of the tails of the velocity increment p.d.f.'s with
  increasing separation time $\tau$ can be gauged by examining the
  temporal evolution of the kurtosis of $\delta_{u_{p,i}}(\tau)$, as
  shown in figure~\ref{fig-lag-vel-incr-kurtosis}. It can be seen that
  the initial decrease is fastest for the streamwise direction, while
  the curves for the two horizontal directions are similar to each
  other, but with a slight shift in the value of the kurtosis. 
  The three components exhibit slight `bumps' at multiples
  of $16T_b$ corresponding to the finite-box-size bias (cf.\
  discussion in \S~\ref{sec-results-lag-vel}), the curves reaching
  asymptotic 
  values which are close to Gaussian for the streamwise direction 
  ($K(\delta_{u_{p}}(\tau=40T_b)=3.0$) and somewhat larger for
  the two horizontal directions 
  ($K(\delta_{v_{p}}(\tau=40T_b)=4.2$ and 
  $K(\delta_{w_{p}}(\tau=40T_b)=3.6$). 
  It can be shown that if the signals at time $t$ and $t+\tau$ are
  completely decorrelated, the kurtosis of the increment will have a
  value of $K(u^\prime_{p,i})/2+3/2$ (note that this limit 
  is indeed approached by our data at long times as can be seen in
  figure~\ref{fig-lag-vel-incr-kurtosis}). 
  As a consequence, the different long-time limits of
  $K(\delta_{u_{p,i}})$ reflect the 
  differences with respect to intermittency of the three components  
  of the particle velocity \citep[cf.\ the one-time p.d.f.'s shown in
  figure~17 of][]{uhlmann:08a}.
}
%

\begin{figure}
  \begin{minipage}{3ex}
    $S$
  \end{minipage}
  \begin{minipage}{.45\linewidth}
    \centerline{$(a)$}
    \centerline{$y/D$}
    \includegraphics[width=\linewidth]
    {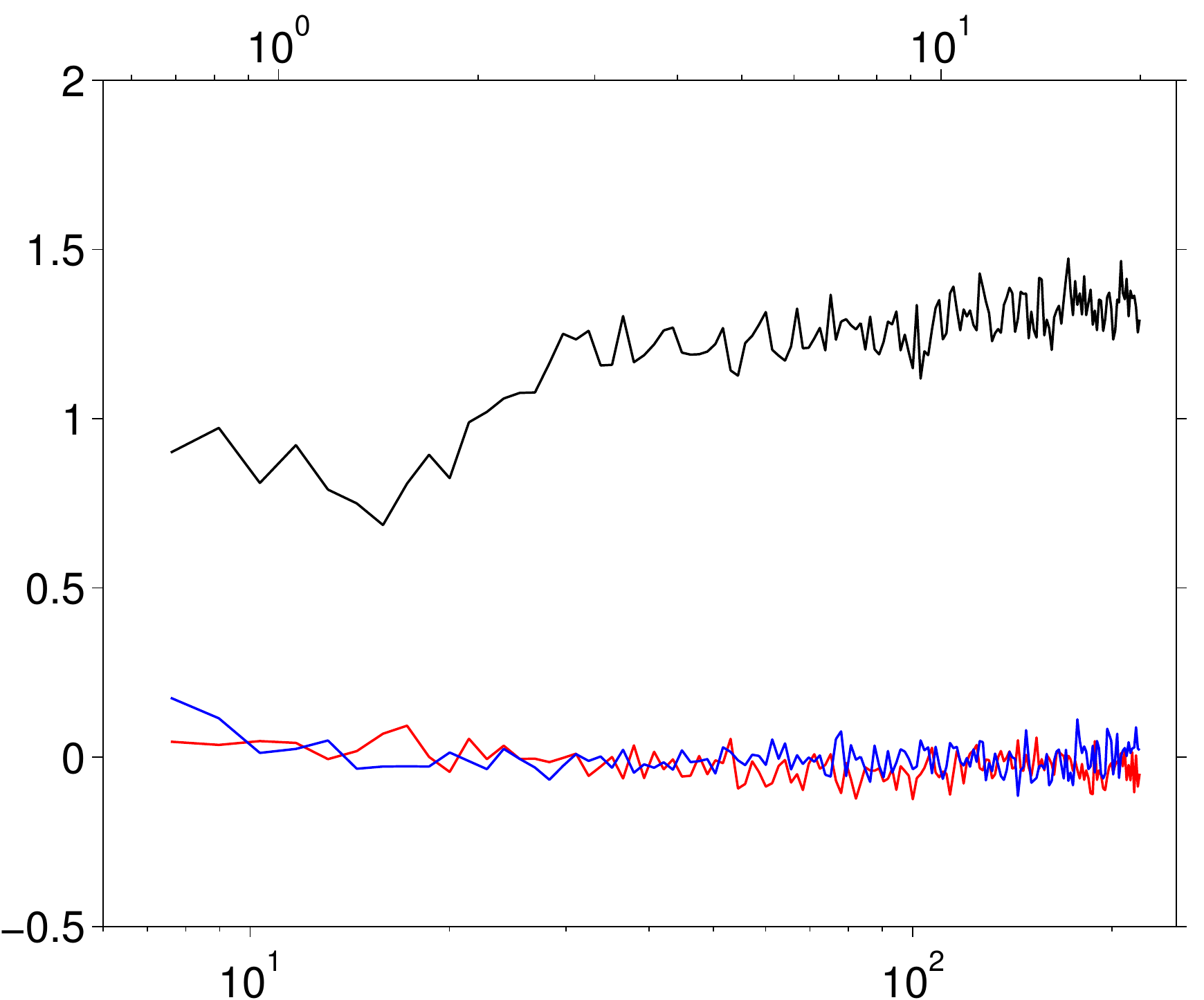}
    \\
    \centerline{$y^+$}
  \end{minipage}
  \begin{minipage}{3ex}
    $K$
  \end{minipage}
  \begin{minipage}{.45\linewidth}
    \centerline{$(b)$}
    \centerline{$y/D$}
    \includegraphics[width=\linewidth]
    {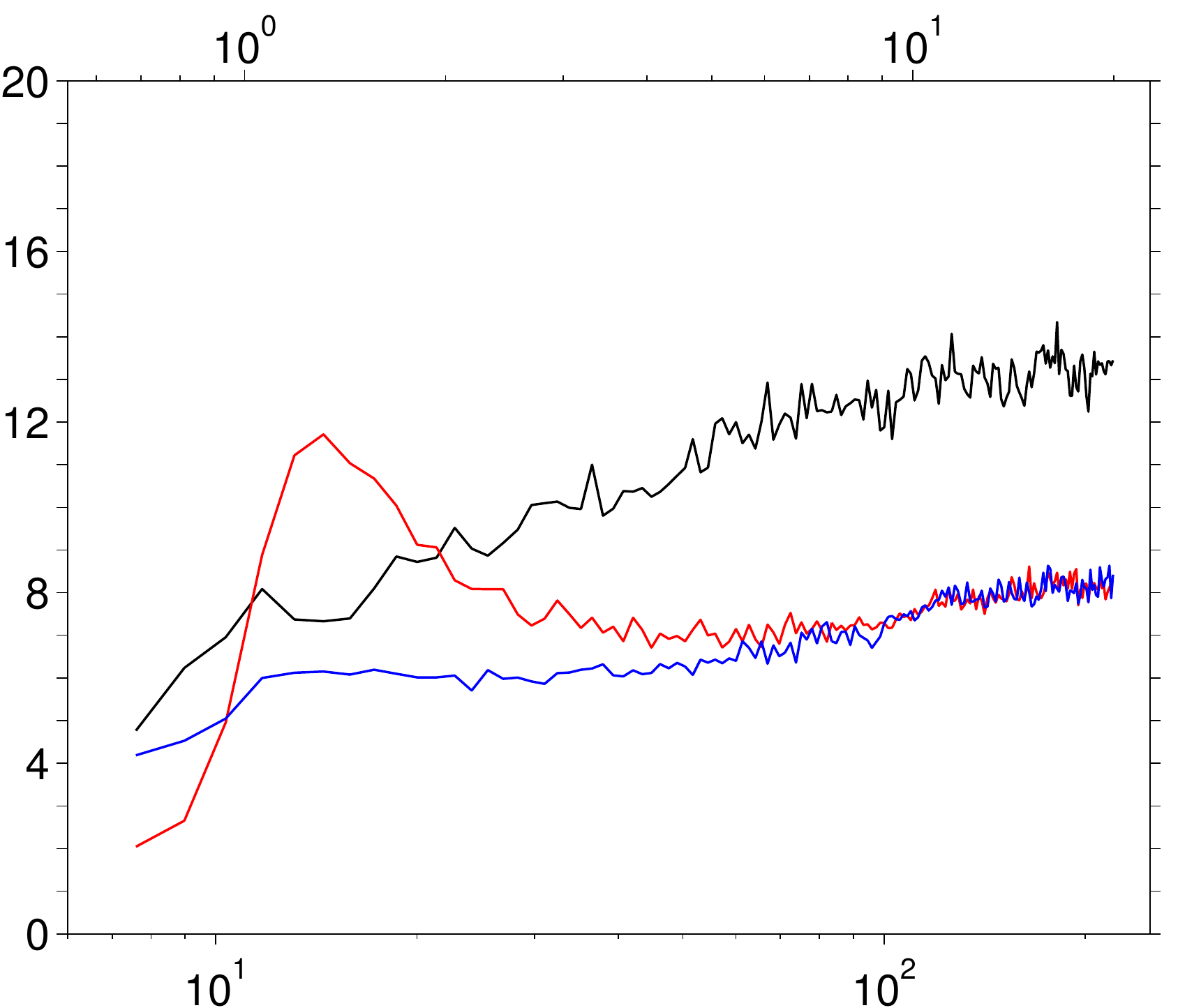}
    \\
    \centerline{$y^+$}
  \end{minipage}
  \caption{
    Higher-order moments of particle acceleration:  
    $(a)$ skewness $S$, 
    $(b)$ kurtosis $K$. 
    Lines as in figure~\ref{fig-force-rms}. 
    %
  }
  \label{fig-force-skew-kurto}
\end{figure}
\revision{The time scale at which the p.d.f.'s become Gaussian is
  closely related to the decorrelation of the Lagrangian
  statistics. Therefore, it can be obtained in a more precise way by
  using the Lagrangian autocorrelation of the hydrodynamic forces
  acting on the particles (which 
  is equivalent to the 
  autocorrelation of 
  particle acceleration)
  shown in Fig.~\ref{fig-lagcorr-force}.
}{%
  Let us now turn to the analysis of correlation times of the
  Lagrangian statistics. 
  Fig.~\ref{fig-lagcorr-force} shows the Lagrangian autocorrelation of
  the hydrodynamic forces acting on the particles (which  
  is equivalent to the autocorrelation of particle acceleration). 
}
It can be observed that the autocorrelation functions first cross the 
zero-axis after several viscous time units (the streamwise component
after $7\,\nu/u_\tau^2$, the wall-normal and spanwise components after
approx.\ $3\,\nu/u_\tau^2$), then taking small negative values which
return to zero on a longer time scale of the order of $2\,T_b$. 
The initial decay can be characterized through the Taylor micro-scale,
which is found to measure $2.95$, $0.92$ and $0.80$ viscous time
units for the streamwise, wall-normal and spanwise force components,
respectively. 
The figure shows that although the initial decorrelation strongly
differs among the three force components, their long-time behavior
(for separation times $\tau/T_b\gtrsim0.7$) is roughly equivalent. 
%
\revision{%
  A similar directional dependence has been found for Lagrangian
  autocorrelations of fluid particle acceleration by \cite{choi:04}, who
  gave an explanation in terms of the predominant coherent structures. 
  Therefore, the presently observed difference in the initial
  decorrelation time scale (i.e.\ the first zero-crossing in
  figure~\ref{fig-lagcorr-force}) 
  between streamwise and cross-stream components of particle 
  acceleration can probably be attributed to the anisotropic nature of
  the wall-bounded turbulent flow scales.
}{%
  A very similar directional dependence of the Lagrangian
  autocorrelation at short separation times has also been observed for
  the acceleration of fluid particles in single-phase channel flow
  \citep{choi:04}.  
  As in the case of Lagrangian velocity autocorrelations (discussed in
  \S~\ref{sec-results-lag-vel}), these authors attribute the effect to
  the anisotropy of the near-wall coherent structures. 
  In view of the striking similarity between the present data
  (figure~\ref{fig-lagcorr-force}$b$) and the fluid particle data 
  \citep[figure~13,][]{choi:04}, we believe that the presently
  observed difference in the
  initial decorrelation time scale can be attributed to the
  anisotropic nature of the wall-bounded turbulent flow scales. 
}
%
 
In the limit of very short time increments, the p.d.f.'s  of Lagrangian
particle velocity increments, 
discussed above,
become the p.d.f.'s of Lagrangian particle
acceleration. \cite{qureshi:08} have shown that acceleration
statistics of finite size inertial particles are found very robust to
size and density variations, the influence of which is mostly carried
by the acceleration variance. 
They found that the shape of the normalized p.d.f.\ remained unchanged over the whole range of
sizes and density ratios explored. The p.d.f.'s shown in
Fig. \ref{fig-force-pdfs} confirm this result in the present case for
the transverse directions (wall-normal and spanwise). Both curves
follow 
the lognormal fit proposed by \cite{qureshi:08} relatively closely
over several decades. 
On the other hand, the streamwise 
acceleration deviates from this curve, showing significantly higher
probabilities for positive 
(upward) acceleration fluctuations. 
It is not clear why this p.d.f.\ is positively skewed. 
One possible mechanism that provides positive skewness is as
follows. Assume that a non-linear drag law holds instantaneously,
i.e.\ $C_d=f(Re_{inst})$, where $Re_{inst}$ is based on the
instantaneous relative velocity felt by the particle,
and that the drag coefficient decays more slowly with Reynolds than
$Re_{inst}^{-1}$.  
In such a case, positive velocity fluctuations would lead to a smaller
decrease in the drag coefficient than the 
corresponding increase due to negative velocity fluctuations
of the same magnitude. 
As a consequence the drag 
force (as well as the corresponding acceleration) can become positively skewed 
even when the velocity fluctuations are symmetric w.r.t.\ the mean. 
%
The data in Fig.~\ref{fig-force-cd}$(a)$ is consistent with this
model: 
it can be seen that the streamwise force coefficient obtained using the mean drag
plus the r.m.s.\ drag is closer to the force coefficient of the mean
drag than the force coefficient obtained using the mean drag minus the
r.m.s.\ drag. 
%
We have 
further 
tested the hypothesis 
based upon non-linear drag by using a Gaussian relative velocity p.d.f.\ 
as input to the standard drag law. 
Although this 
simple model indeed 
yields 
positively skewed hydrodynamic forces (figure omitted), 
the magnitude of the skewness obtained 
is smaller
than the one obtained in the simulation. Therefore, this issue deserves further investigation. 

%

The force/acceleration p.d.f.'s shown in Fig. \ref{fig-force-pdfs} have been computed 
without
 taking into account
the inhomogeneous nature of the flow. In order to illustrate how the wall affects the
p.d.f.'s we have also computed them 
in $160$ wall-normal slabs of uniform thickness.
It suffices to discuss the wall-normal variation of the p.d.f.'s in terms
of their higher order moments, of which skewness and kurtosis
are shown in Fig.~\ref{fig-force-skew-kurto}. 
Only the streamwise component presents 
significant 
positive skewness which varies from $S\approx0.8$ 
close to the wall to $S\approx1.3$ in the central region. The
wall-normal and spanwise 
force/acceleration components are roughly symmetrically distributed,
with the exception of the region near the wall where a small positive
skewness is obtained. 
Concerning the kurtosis, all three components tend to present higher
values in the central 
region of the channel where  the streamwise component is larger than
the other two components by approximately $60\%$. This indicates that the
tails of the p.d.f.\ of the 
streamwise component are broader as can be appreciated in
Fig.~\ref{fig-force-pdfs}. 
The kurtosis of the wall-normal and spanwise components are equal in
the central region of the channel, 
indicating that in this zone these two directions are equivalent as
mentioned above. When approaching 
the wall, they deviate from each other. The kurtosis of the spanwise
component remains more or less constant at a value of $K\approx6$ up
to the immediate near-wall region where it decreases mildly. 
The kurtosis of the wall-normal component, on the other hand, presents
a remarkable peak in the buffer  
layer (with $K\approx12$) which is even higher than the kurtosis of
the streamwise component in this region. Closer to the wall 
the kurtosis of the wall-normal component tends towards a Gaussian
value. 
%

%% file: results_d.tex
 \begin{figure}
   \begin{center}
     \begin{minipage}{5ex}
       $\displaystyle I_r$, 
       \\[2ex]
       $\displaystyle \tilde{I}_r$
     \end{minipage}
     \begin{minipage}{.45\linewidth}
       \includegraphics[width=\linewidth]
       {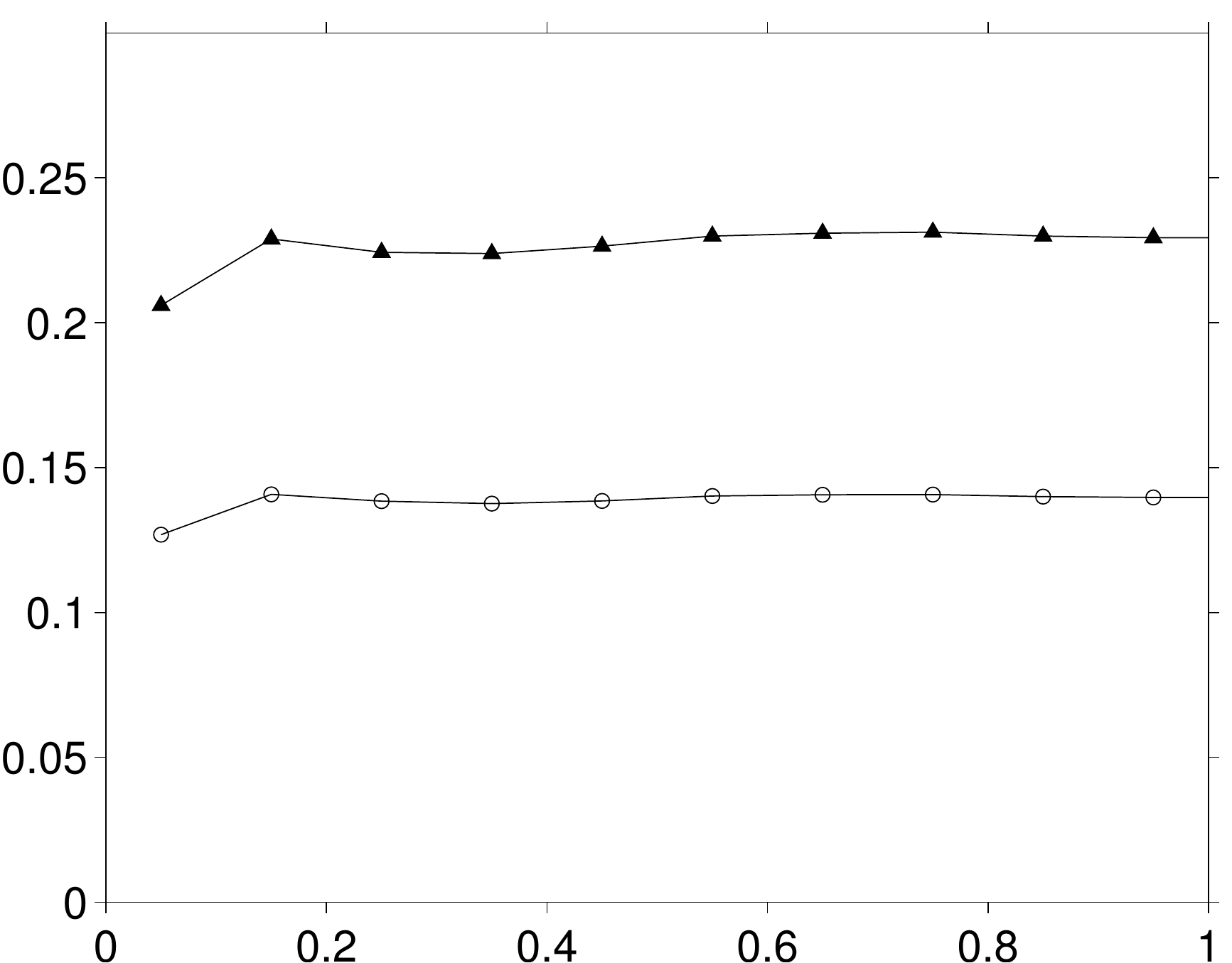}
       \\
       \centerline{$y/h$}
     \end{minipage}
   \end{center}
   \caption{
     Relative turbulence intensity as a function of
     the wall-distance: 
     {\color{black}$-\!\!\circ\!\!-$},~$\tilde{I}_r=(\langle
     u_i^\prime u_i^\prime\rangle/3)^{1/2}/u_{lag}$;  
     {\color{black}$-\!\!\blacktriangle\!\!-$},~$I_r=\langle
     u^\prime u^\prime\rangle^{1/2}/u_{lag}$;  
   }
   \label{fig-wake-relative-turb-intensity}
 \end{figure}
\subsection{%
Particle-conditioned averaging
}
\label{sec-wake}
%
%
In this section we turn our attention to the statistics of the flow in
the neighborhood of the particles. For this purpose we have carried
out averaging of the flow field conditioned upon the presence of
particles. Before 
presenting 
a brief description of the literature on
this topic (\S~\ref{sec-wake-lit}) 
as well as explaining the averaging procedure
(\S~\ref{sec-wake-notation}),  
let us introduce the parameters which are believed to
characterize the local flow field. 

In addition to the Reynolds number based upon the average relative
flow velocity and the sphere diameter ($Re_{lag}$), the
most 
prominent 
parameter describing turbulent flow 
around spheres is the relative turbulence intensity, i.e.\ the
ratio between the intensity of the incoming fluid flow fluctuations
and the apparent slip velocity.  
Whereas in homogeneous flows the definition of a relative turbulence
intensity $\tilde{I}_r$ is often based upon the three-component turbulence
intensity, viz.\ 
$\tilde{I}_r=(\langle u_i^\prime u_i^\prime\rangle/3)^{1/2}/u_{lag}$, 
in cases with unidirectional mean flow (in the $x$-coordinate
direction) the following definition is commonly employed: 
$I_r=\langle u^\prime u^\prime\rangle^{1/2}/u_{lag}$. 
In order to facilitate a relation of our current results to previous
and future work, wall-normal profiles of the relative turbulence
intensity according to both definitions are shown in 
figure~\ref{fig-wake-relative-turb-intensity}. 
These profiles are found to be practically uniform across the channel, 
with values of $\tilde{I}_r\approx0.14$ and $I_r\approx0.23$. 
 \begin{figure}
   \begin{center}
     \begin{minipage}{3ex}
       $(a)$
     \end{minipage}
     \begin{minipage}{3ex}
       $\displaystyle\frac{\tilde{z}}{R}$
     \end{minipage}
     \begin{minipage}{.6\linewidth}
       \includegraphics[width=\linewidth]
       {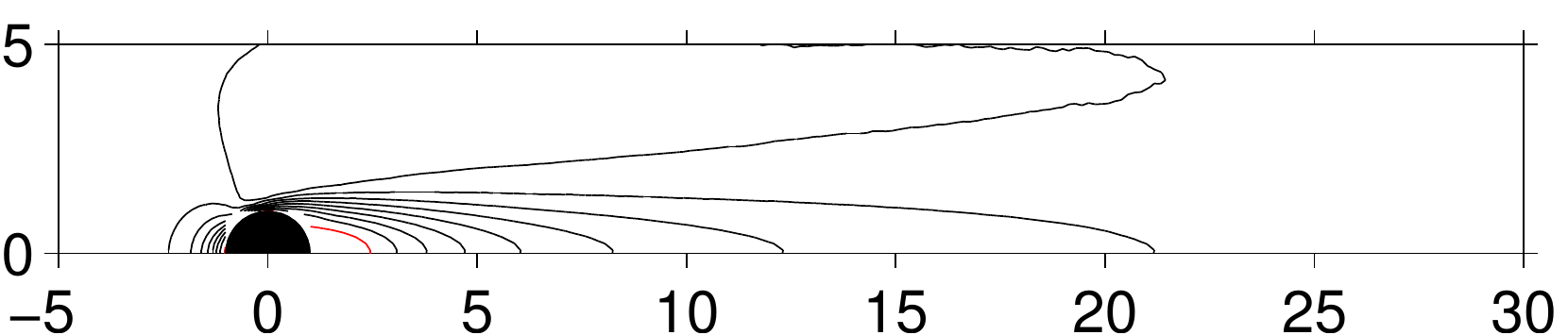}
     \end{minipage}     
     \hspace*{3ex}
     \begin{minipage}{15ex}
       $y^{(s)}/h=0.05$,\\
       $y^{(s)+}=11$,\\
       $y^{(s)}/D=1$
     \end{minipage}
     \\
     \begin{minipage}{3ex}
       $(b)$
     \end{minipage}
     \begin{minipage}{3ex}
       $\displaystyle\frac{\tilde{z}}{R}$
     \end{minipage}
     \begin{minipage}{.6\linewidth}
       \includegraphics[width=\linewidth]
       {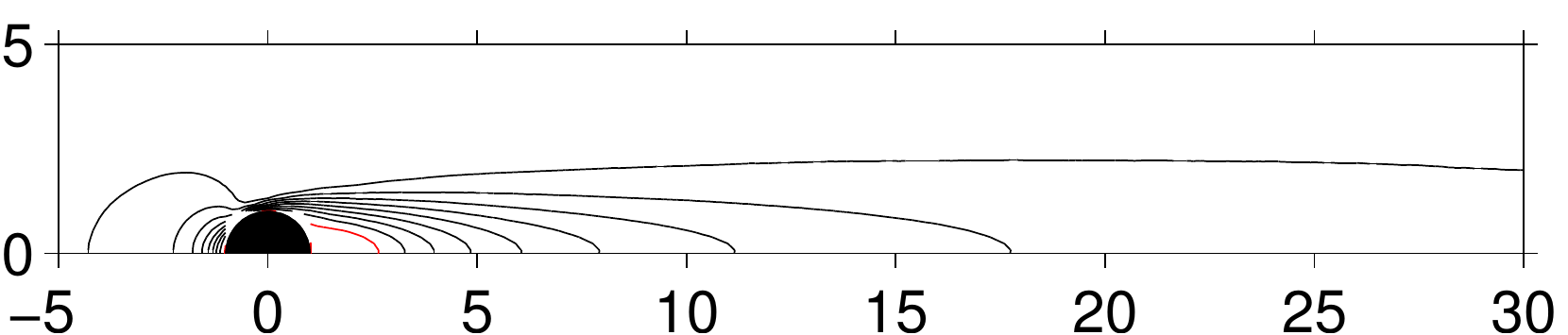}
     \end{minipage}     
     \hspace*{3ex}
     \begin{minipage}{15ex}
       $y^{(s)}/h=0.15$,\\
       $y^{(s)+}=33$,\\
       $y^{(s)}/D=3$
     \end{minipage}\\
     \begin{minipage}{3ex}
       $(c)$
     \end{minipage}
     \begin{minipage}{3ex}
       $\displaystyle\frac{\tilde{z}}{R}$
     \end{minipage}
     \begin{minipage}{.6\linewidth}
       \includegraphics[width=\linewidth]
       {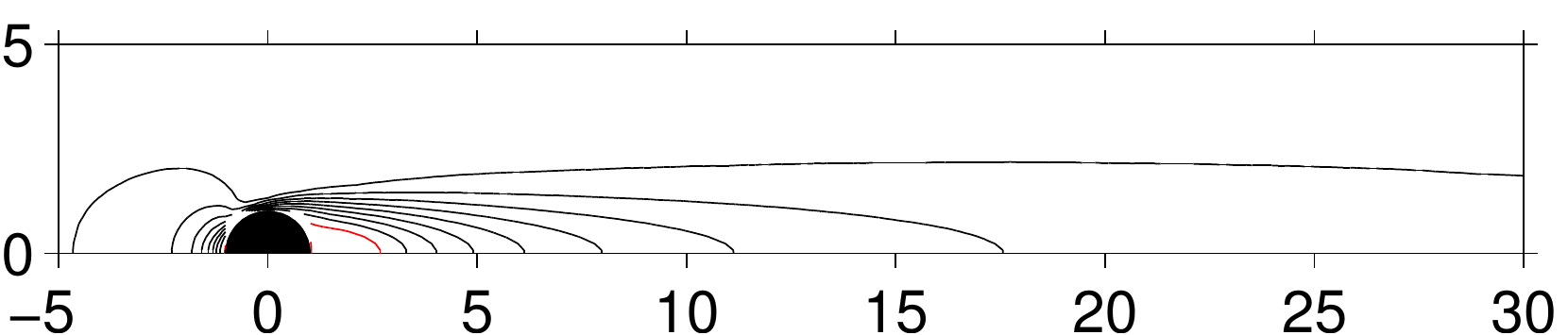}
     \end{minipage}     
     \hspace*{3ex}
     \begin{minipage}{15ex}
       $y^{(s)}/h=0.35$,\\
       $y^{(s)+}=77$,\\
       $y^{(s)}/D=7$
     \end{minipage}\\
     \begin{minipage}{3ex}
       $(d)$
     \end{minipage}
     \begin{minipage}{3ex}
       $\displaystyle\frac{\tilde{z}}{R}$
     \end{minipage}
     \begin{minipage}{.6\linewidth}
       \includegraphics[width=\linewidth]
       {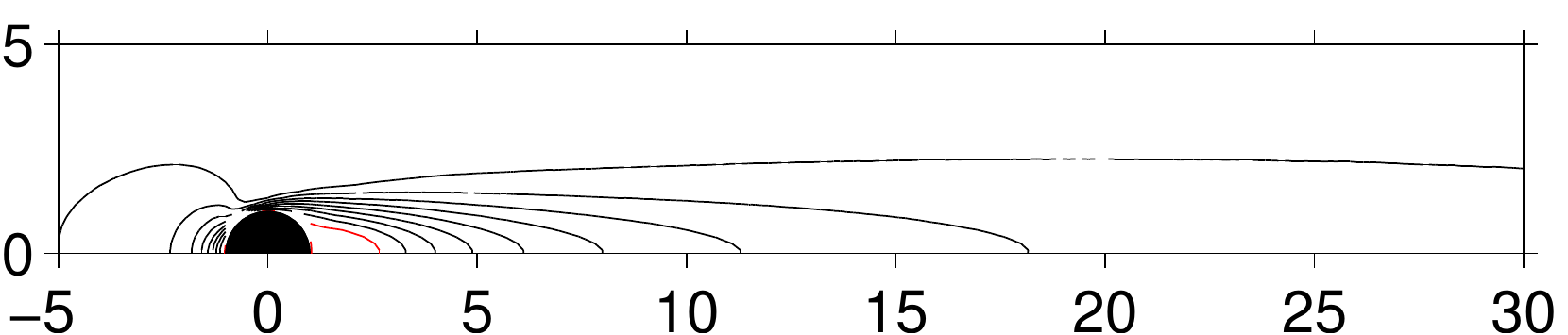}
       \\
       \centerline{$\tilde{x}/R$}
     \end{minipage}     
     \hspace*{3ex}
     \begin{minipage}{15ex}
       $y^{(s)}/h=0.95$,\\
       $y^{(s)+}=208$,\\
       $y^{(s)}/D=19$
     \end{minipage}
   \end{center}
   \caption{
     Contours of the streamwise component of the average relative
     velocity $\tilde{u}$ in wall-parallel planes through the center
     of the 
     particles, plotted for slabs at different wall-distances $y^{(s)}$.  
     Contourlines are at uniformly distributed values
     $(0:.1:0.9)$ times the maximum value in each plane. 
     The zero-valued contour is plotted in red color. 
  }
  \label{fig-wake-avg-cont}
\end{figure}
\subsubsection{Previous work}
\label{sec-wake-lit}
%
A number of previous studies provide useful reference data for the
following discussion of our present results.  
%
\cite{bagchi:04} have simulated the flow around
a single fixed particle in an unbounded domain, swept by
homogeneous-isotropic turbulence (at $Re_\lambda=164$) with a
superposed uniform mean flow.  
They considered two values of relative turbulence intensity,
$\tilde{I}_r=0.1$ and $0.25$, while varying the particle Reynolds
number based upon the average relative flow from $Re_{lag}=58$ to
$610$. 
Although structurally different, homogeneous-isotropic flow is still
relevant to plane channel flow if the central region of the latter
configuration is considered, where a far less anisotropic state
than in the near-wall region can be found.
Therefore, the data of \cite{bagchi:04} will be compared to our
present results in the vicinity of the centerplane of the channel. 

%
\cite{amoura:10} have investigated the flow around a 
fixed sphere swept by higher intensity incident turbulence (which was 
approximately homogeneous-isotropic) with experimental
techniques. They have considered particle Reynolds numbers based upon the
average relative flow from $Re_{lag}=100$ to $1000$, while the
relative turbulence intensity was varied from $I_r=0.26$ to $0.45$. 
At the same time, the ratio of particle diameter to integral length
scale of the incident turbulence was significantly larger than in most
previous studies. 
%

%
\cite{wu:94b} have performed experimental measurements of
the flow around fixed spheres on the centerline of turbulent pipe flow
($Re_b=17000$ up to $50000$, based upon pipe radius). The particle
Reynolds number based upon the average relative velocity was varied from
$Re_{lag}=135$ to $1560$.

%
\cite{legendre:06} have simulated the
flow around a stationary particle on the centerline of turbulent pipe
flow ($Re_b=3000$ and $Re_ \tau=200$, based upon the pipe radius)
by means of LES.  
Their particle has a diameter of $D^+=10.23$ and a particle Reynolds 
number based upon the average relative velocity of $Re_{lag}=200$, 
providing a parameter point which is not far removed from our present
study.  
The prime difference is their relative turbulence intensity
$\tilde{I}_r=0.037$, which is significantly lower than in the present
case. 

%
\cite{zeng:10} have simulated turbulent channel flow
($Re_\tau\approx180$) with a single fixed particle, either located in
the buffer layer ($y^+\approx18$) or at the center of the
channel. They also varied the size of the particle in the range $D^+=3.5$
to $25$, such that a total of six different cases were considered.
Out of that data-set, their case 2 is directly comparable to our data,
featuring a particle with diameter $D^+=10.7$ placed in the buffer
layer.  

Please note that all of the studies mentioned above consider the flow
around single particles which were fixed in space. 
Consequently, effects of particle mobility and collective effects were
absent in those cases. 
We are not aware of previous work involving particle-conditioned
averaging of turbulent flow-fields at the scale of the particles. 
\revision{}{%
  One exception is the experimental work of \cite{poelma:07} 
  who did measure the conditionally-sampled flow field
  around mobile particles in decaying grid-generated turbulence. 
  However, the near-field results presented in that reference are not
  sufficiently detailed in order to serve for the purpose of
  comparison in the present context. 
}
%
%
\begin{figure}
  \centering
  \begin{minipage}{3ex}
    $\displaystyle\frac{L_e}{R}$
  \end{minipage}
  \begin{minipage}{.45\linewidth}
    \includegraphics[width=\linewidth]
    {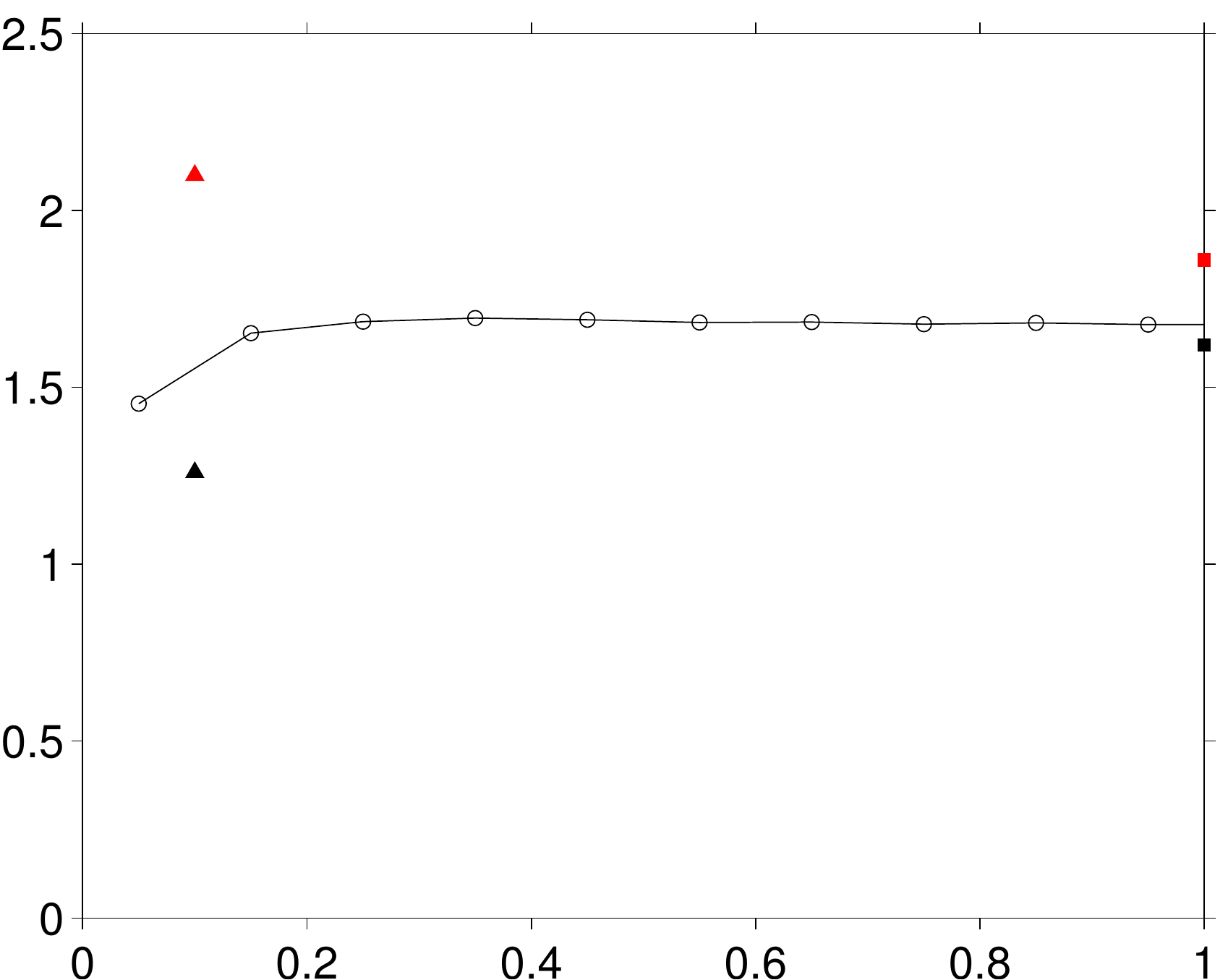}
    \\
    \centerline{$y/h$}
  \end{minipage}
  \caption{
    Average streamwise recirculation length $L_e$ in the wake of the
    particles, shown as a function of the wall-distance.
    Results for fixed single particles from the literature are
    indicated by the following symbols: 
    $\blacktriangle$, particle in buffer layer of turbulent channel
    flow \citep[case 2 of][]{zeng:10}; 
    {\color{red}$\blacktriangle$}, particle in buffer layer of
    laminar channel flow~\citep{zeng:10};  
    {\scriptsize$\blacksquare$}, particle in isotropic turbulence with
    mean relative-flow Reynolds number of $Re_{lag}=114$ and
    relative turbulence intensity $\tilde{I}_r=0.25$ 
    \citep[case 4 of][]{bagchi:04}; 
    {\color{red}\scriptsize$\blacksquare$}, particle in uniform flow at mean
    relative-flow Reynolds number of $Re_{lag}=114$ \citep{bagchi:04}.
    %
    %
    %
  }
  \label{fig-wake-avg-recirc}
\end{figure}
\subsubsection{Preliminaries}
\label{sec-wake-notation}
In the following we present an analysis of the local relative velocity
field in the vicinity of the particles. For this purpose we have
averaged the Eulerian velocity field in a frame of reference attached
to the particle centers, summing over particles located in 20
wall-normal slabs (width of the slabs equal to $2D$) as well as over a
number of 
$97$   
instantaneous fields (yielding more than 
70000 
samples per
slab). 
%
%
Before turning to the results, let us first make the averaging
process more precise.  

The instantaneous velocity difference between the two phases, computed
with respect to a given particle, is defined as follows: 
\begin{equation}\label{equ-def-relative-velocity}
  \mathbf{u}_{r}^{(i)}(\tilde{\mathbf{x}},t)=
  \mathbf{u}_f(\tilde{\mathbf{x}},t)-\mathbf{u}_p^{(i)}(t)
  \,,
\end{equation}
where 
the instantaneous coordinate relative to the $i$th particle's center
position $\mathbf{x}_p^{(i)}(t)$ is given by
\begin{equation}\label{equ-def-relative-position}
  \tilde{\mathbf{x}}^{(i)}(t)=\mathbf{x}-\mathbf{x}_p^{(i)}(t)
  \,. 
\end{equation}
Furthermore, in (\ref{equ-def-relative-velocity}) 
\revision{}{$\mathbf{u}_f(\tilde{\mathbf{x}},t)$} 
refers to the fluid velocity vector field and $\mathbf{u}_p^{(i)}(t)$ is
the $i$th particle's velocity vector at a given time $t$. 
The average relative velocity
$\tilde{\mathbf{u}}(\tilde{\mathbf{x}},y^{(s)})$ in the slab centered at
a wall-distance $y^{(s)}$ is computed from 
\begin{equation}\label{equ-def-avg-relative-velocity}
  \tilde{\mathbf{u}}(\tilde{\mathbf{x}},y^{(s)})
  =
  \langle
  \mathbf{u}_{r}^{(i)}(\tilde{\mathbf{x}},t)
  \rangle_{p,t}
  \,,
\end{equation}
where the operator $\langle\cdot\rangle_{p,t}$ denotes averaging over
all particles 
\revision{(located in the respective slab)}{%
  (whose center position is located in the respective slab)
} 
and over time (i.e.\
over a number of snapshots). 
Please note that averaging is performed over the actual volume
instantaneously occupied by the fluid. This means that the volume
occupied by other particles located in the neighborhood of a
particular sphere is disregarded in the averaging
procedure shown in~(\ref{equ-def-avg-relative-velocity}).  
The reader is referred to Appendix~\ref{app-averaging} for a 
precise definition of the above averaging operator. 
%

We can now decompose the instantaneous value of the relative velocity
w.r.t.\ the $i$th particle 
(\ref{equ-def-relative-velocity}) into an average contribution
(\ref{equ-def-avg-relative-velocity}) and a fluctuation
$\mathbf{u}_{r}^{(i)\prime\prime}$, viz.   
\begin{equation}\label{equ-def-decompose-relative-velocity}
  \mathbf{u}_{r}^{(i)}(\tilde{\mathbf{x}},t)=
  \tilde{\mathbf{u}}(\tilde{\mathbf{x}},y^{(s)})
  +
  \mathbf{u}_{r}^{(i)\prime\prime}(\tilde{\mathbf{x}},t)
  \,,
\end{equation}
where the average of the fluctuation vanishes identically (i.e.\ 
$\langle\mathbf{u}_{r}^{(i)\prime\prime}\rangle_{p,t}=0$). 
%
From the fluctuations $\mathbf{u}_{r}^{(i)\prime\prime}(\mathbf{x},t)$
implicitly defined in (\ref{equ-def-decompose-relative-velocity}) we
can compute second moments; for a component in the $\alpha$-direction
the definition reads: 
\begin{equation}\label{equ-def-correlation-relative-velocity-fluct}
  \widetilde{
    {u}_{r,\alpha}^{\prime\prime}
    {u}_{r,\alpha}^{\prime\prime}
  }(\tilde{\mathbf{x}},y^{(s)})
  =
  \langle
  {u}_{r,\alpha}^{(i)\prime\prime}(\tilde{\mathbf{x}},t)
  \,
  {u}_{r,\alpha}^{(i)\prime\prime}(\tilde{\mathbf{x}},t)
  \rangle_{p,t}
  \,,
\end{equation}
where no summation is implied for greek indices. 
Furthermore, we can define velocity fluctuations in the
particle-centered frame of reference for each of the phases, viz. 
\begin{subequations}
  \label{equ-wakes-def-decompose-relative-velocity-fluid-particle} 
  \begin{eqnarray}
    \mathbf{u}_{f}^{(i)\prime\prime}(\tilde{\mathbf{x}},t)
    &=&
    \mathbf{u}_{f}(\tilde{\mathbf{x}},t)
    -\langle\mathbf{u}_f(\tilde{\mathbf{x}},t)\rangle_{p,t}
    \,,\\
    \mathbf{u}_{p}^{(i)\prime\prime}(t)
    &=&
    \mathbf{u}_{p}^{(i)}(t)
    -\langle\mathbf{u}_p(t)\rangle_{p,t}
    \,,
  \end{eqnarray}
\end{subequations}
such that $\mathbf{u}_{r}^{(i)\prime\prime}=
\mathbf{u}_{f}^{(i)\prime\prime}-\mathbf{u}_{p}^{(i)\prime\prime}$. 
%
\begin{figure}
  \centering
  \begin{minipage}{4ex}
    $\displaystyle\frac{\tilde{u}}{u_{lag}}$
  \end{minipage}
  \begin{minipage}{.45\linewidth}
    \includegraphics[width=\linewidth]
    {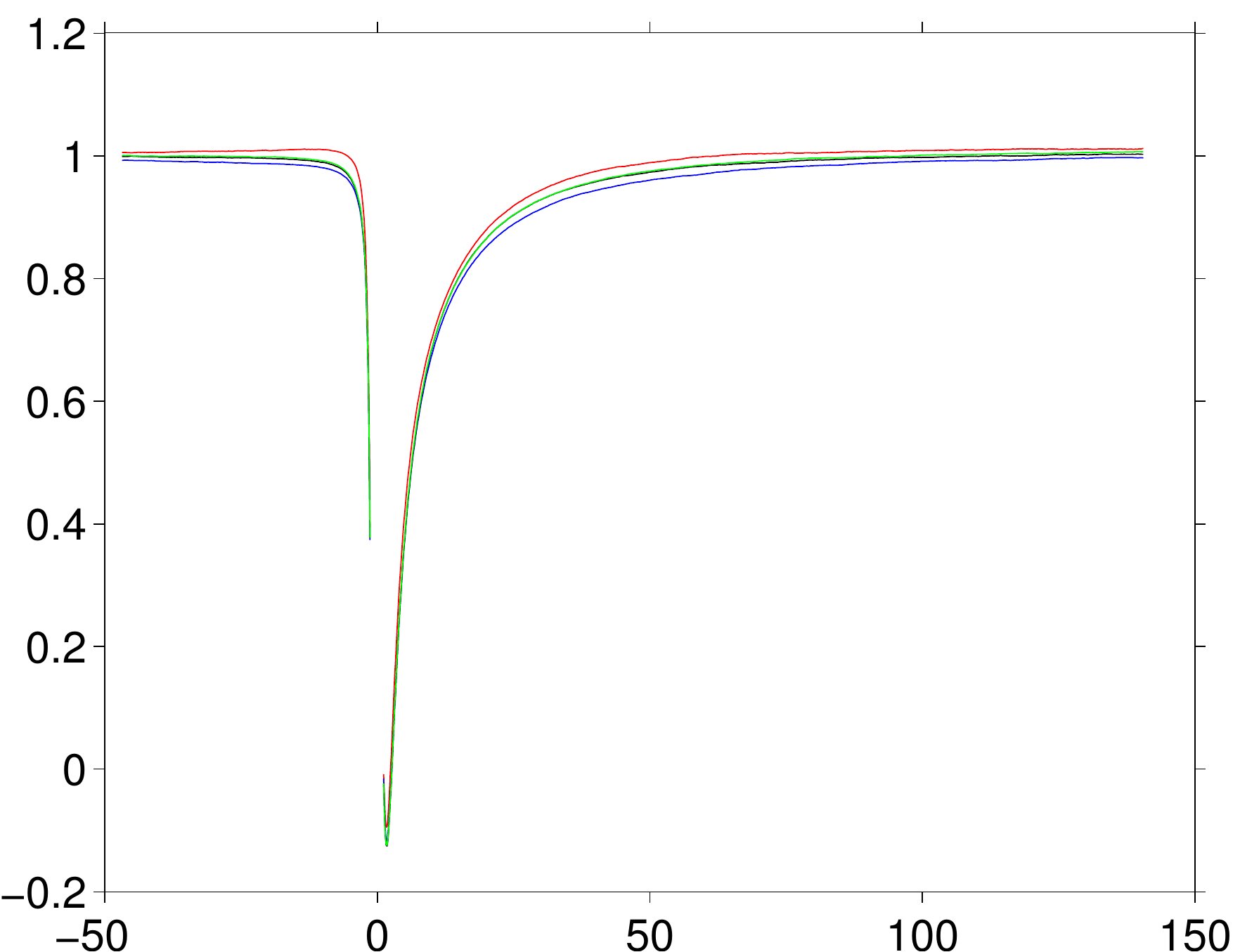}
    \\
    \centerline{$\tilde{x}/R$}
  \end{minipage}
  \caption{
    Average relative streamwise velocity $\tilde{u}$ along the streamwise axis
    through the center of the particles.
    The velocity $\tilde{u}$ is normalized by 
    the apparent velocity lag $u_{lag}=|\langle u_f\rangle-\langle
    u_p\rangle|$.
    %
    %
    The solid curves correspond to particles being located at
    different wall-distances, averaged over bins centered at:
    {\color{red}\solid}~$y^{(s)+}=11$, 
    {\color{blue}\solid}~$y^{(s)+}=33$, 
    {\color{black}\solid}~$y^{(s)+}=77$, 
    {\color{green}\solid}~$y^{(s)+}=208$. 
  }
  \label{fig-wake-avg-urel}
\end{figure}
\subsubsection{Mean relative velocity field}
%
Figure~\ref{fig-wake-avg-cont} shows the streamwise component of the
average relative velocity, $\tilde{u}$, in wall-parallel planes
through the particle center for averaging slabs at different
wall-distances. 
It can be seen that the overall wake pattern in those planes is
similar across the channel. 
However, the extent of the recirculation region as well as
the deceleration (acceleration) upstream (downstream) of the particle
are found to depend upon the wall distance.  
In particular, the region where the average streamwise velocity
exhibits negative values is somewhat smaller at $y^{(s)+}=11$ than in
the outer region. Likewise, the stagnation region upstream of the
particle is more compact at this near-wall location ($y^{(s)+}=11$)
than at larger wall-distances. 
It can also be observed in figure~\ref{fig-wake-avg-cont} that there
are no significant differences between the wake pattern of particles
in the logarithmic layer and those located in the outer region or at
the center of the channel. 

\begin{figure}
  \begin{minipage}[t]{.47\linewidth}
    \begin{minipage}{3ex}
      $\displaystyle u_{d0}$
    \end{minipage}
    \begin{minipage}{.85\linewidth}
      \includegraphics[width=\linewidth]
      {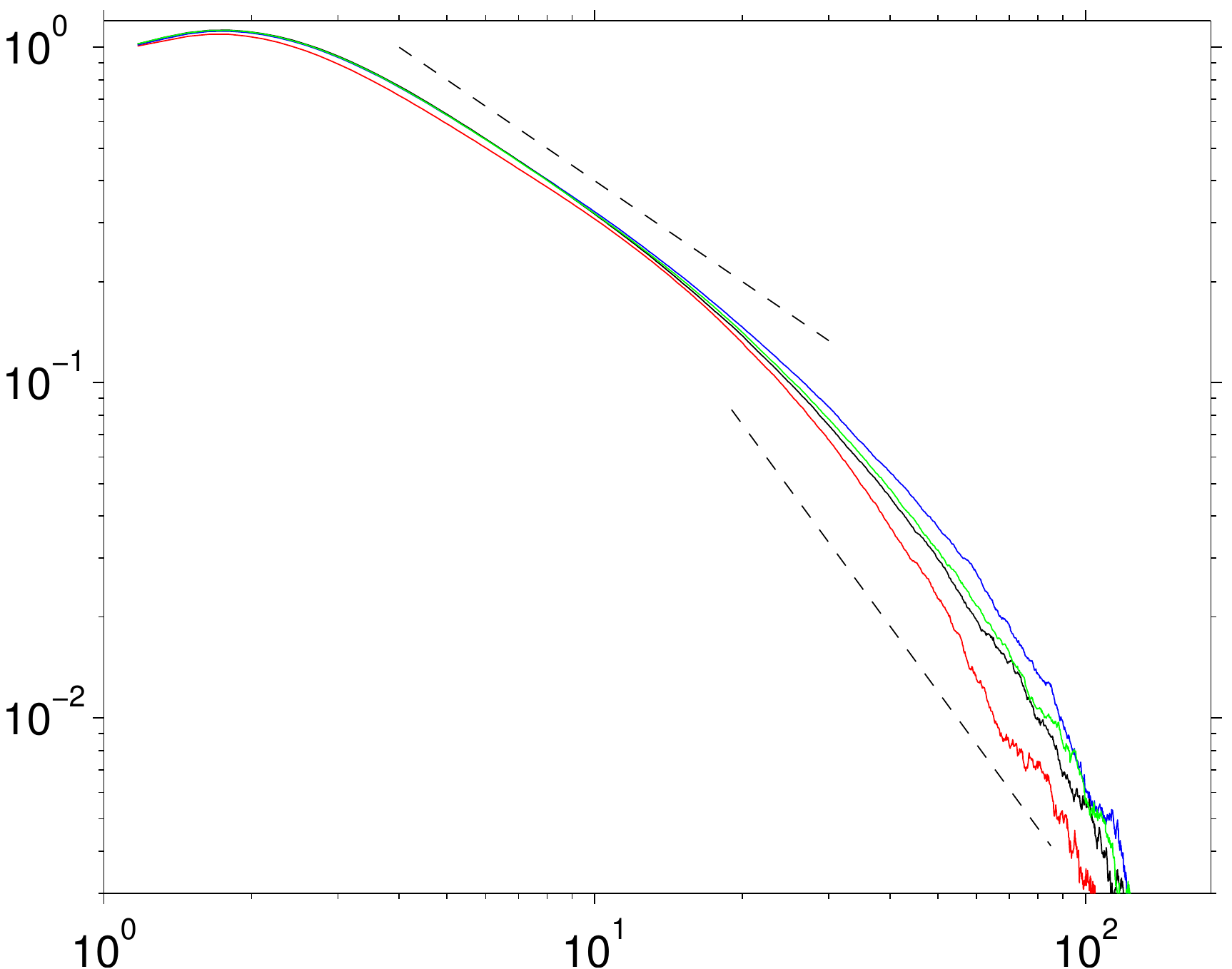}
      \\
      \centerline{$\tilde{x}/R$}
    \end{minipage}
  \caption{
    The streamwise velocity deficit $u_{d0}$ along the streamwise axis
    through the center of the particles (on their downstream side).
    Color coding as in figure~\ref{fig-wake-avg-urel}. 
    The dashed straight lines indicate decay rates
    proportional to $x^{-1}$ 
    and $x^{-2}$. 
  }
  \label{fig-wake-avg-1}
    \end{minipage}
  \hfill
  \begin{minipage}[t]{.47\linewidth}
    \begin{minipage}{6ex}
      $\displaystyle u_{d0}\,\tilde{x}^n$
    \end{minipage}
    \begin{minipage}{.85\linewidth}
      \includegraphics[width=\linewidth]
      {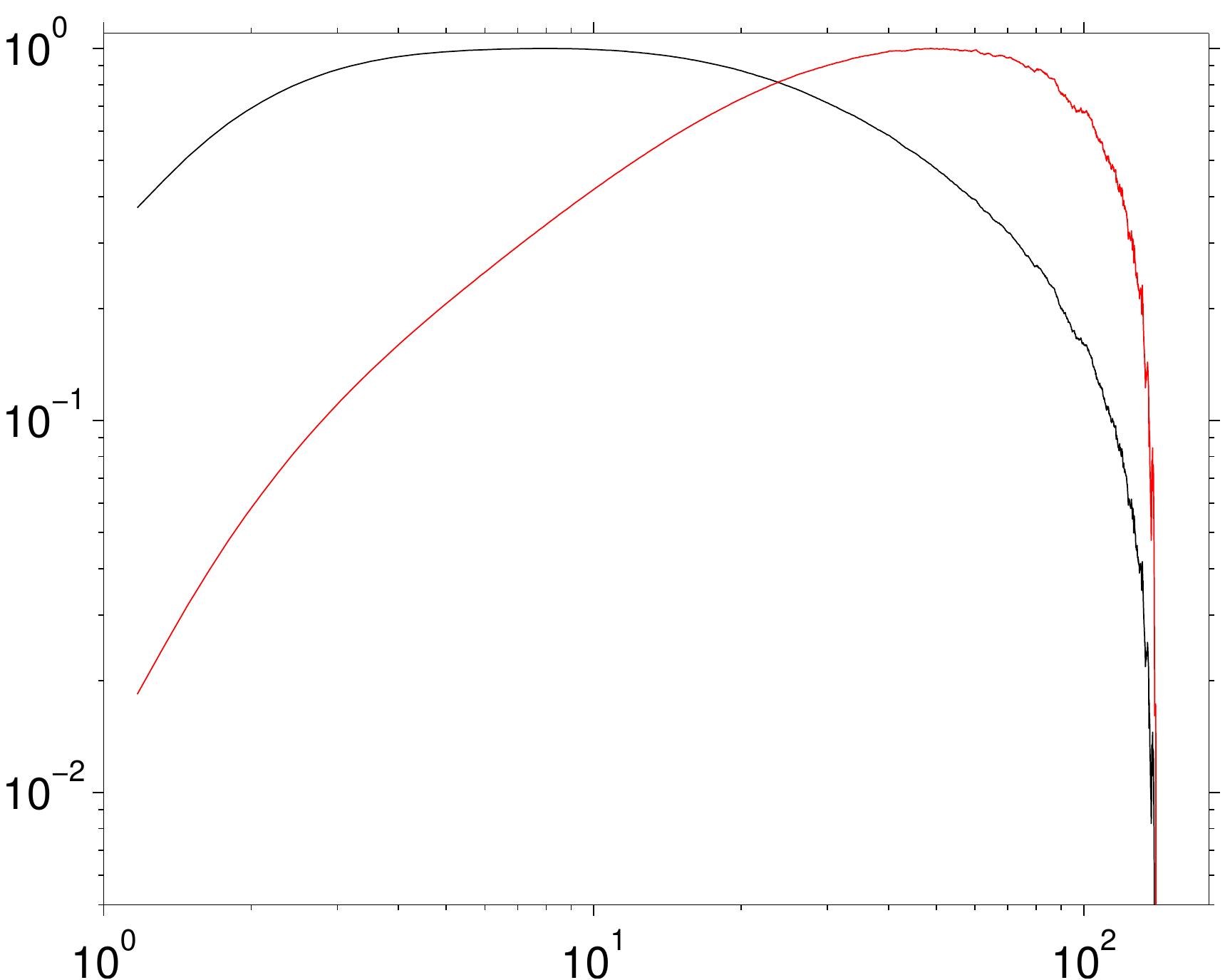}
      \\
      \centerline{$\tilde{x}/R$}
    \end{minipage}
  \caption{
    As in figure~\ref{fig-wake-avg-1}, but the velocity deficit is
    averaged over all particles located at $y/h\geq0.2$ ($y/D\geq4$),
    and the result is compensated in order to highlight 
    power-law regions, i.e.\ the quantity $u_{d0}\,\tilde{x}^n$ is
    plotted, where: 
    \solid, $n=1$;
    {\color{red}\solid}, $n=2$. 
  }
  \label{fig-wake-avg-1-comp}
    \end{minipage}
\end{figure}
The length of the recirculation region $L_e$, defined as the
streamwise distance
from the rear stagnation point to the downstream location where the
average relative velocity changes sign, is shown in
figure~\ref{fig-wake-avg-recirc} as a function of 
wall-distance. 
Please note that $L_e$ is measured
in the wall-parallel plane passing through the particle center;
therefore, the value of $L_e$ does not necessarily reflect the maximum
length of the three-dimensional recirculation region, which might be
located off-center. 
The figure shows an approximately constant value of $L_e/R\approx1.72$ 
for wall-distances $y/h\geq0.2$ ($y^+\geq45$), where 
$R=D/2$ is the particle radius. 
For smaller wall-distances, the length of the recirculation zone
decreases, down to a value of $L_e/R=1.48$
in the first slab at $y^{(s)}/h=0.05$ ($y^{(s)+}=11$).
%
%

%
Figure~\ref{fig-wake-avg-recirc} also includes data points for
single fixed spheres in laminar channel and uniform flow \citep[i.e.\
the laminar results reported by][]{bagchi:04,zeng:10} at
comparable average-flow particle Reynolds numbers. Since
homogeneous-isotropic inflow was considered by \cite{bagchi:04}, we
relate their data to our results at the channel center. 
It can be seen that the presently obtained recirculation lengths fall
below the reference values in laminar flow. 
This result is consistent with previous observations made with fixed
spheres swept by turbulent flow: both \cite{bagchi:04} as well as
\cite{zeng:10} observed a 
reduction of the recirculation length due to background turbulence. 
%
Compared to the values of those authors (cf.\ black symbols in
figure~\ref{fig-wake-avg-recirc}) our present results show somewhat
larger recirculation lengths in turbulent background flow at
comparable turbulence intensity. 
In particular, the present recirculation length in the buffer layer is
approximately 26\% higher than the value reported for case 2 of
\cite{zeng:10}; 
the present results in the core of the channel are 6\% higher than
case 4 of \cite{bagchi:04}.
Although the differences are not extremely large, they might reflect
the difference in the physics between our case and the single fixed
particle configurations.
Apart from possible effects of particle mobility, it should be
remembered that the present flow field is significantly altered by the
presence of particles, i.e.\ particle wakes are prominent
features. Consequently, particles 
experience 
a turbulent flow field which
is structurally significantly different from both the canonical channel
flow and homogeneous-isotropic turbulence swept over
fixed particles \citep{bagchi:04,zeng:10}. 
%
%

\begin{figure}
  \centering
    \begin{minipage}{5ex}
      $\displaystyle\frac{\tilde{z}}{h_{wz}}$
    \end{minipage}
    \begin{minipage}{.45\linewidth}
      \includegraphics[width=\linewidth]
      {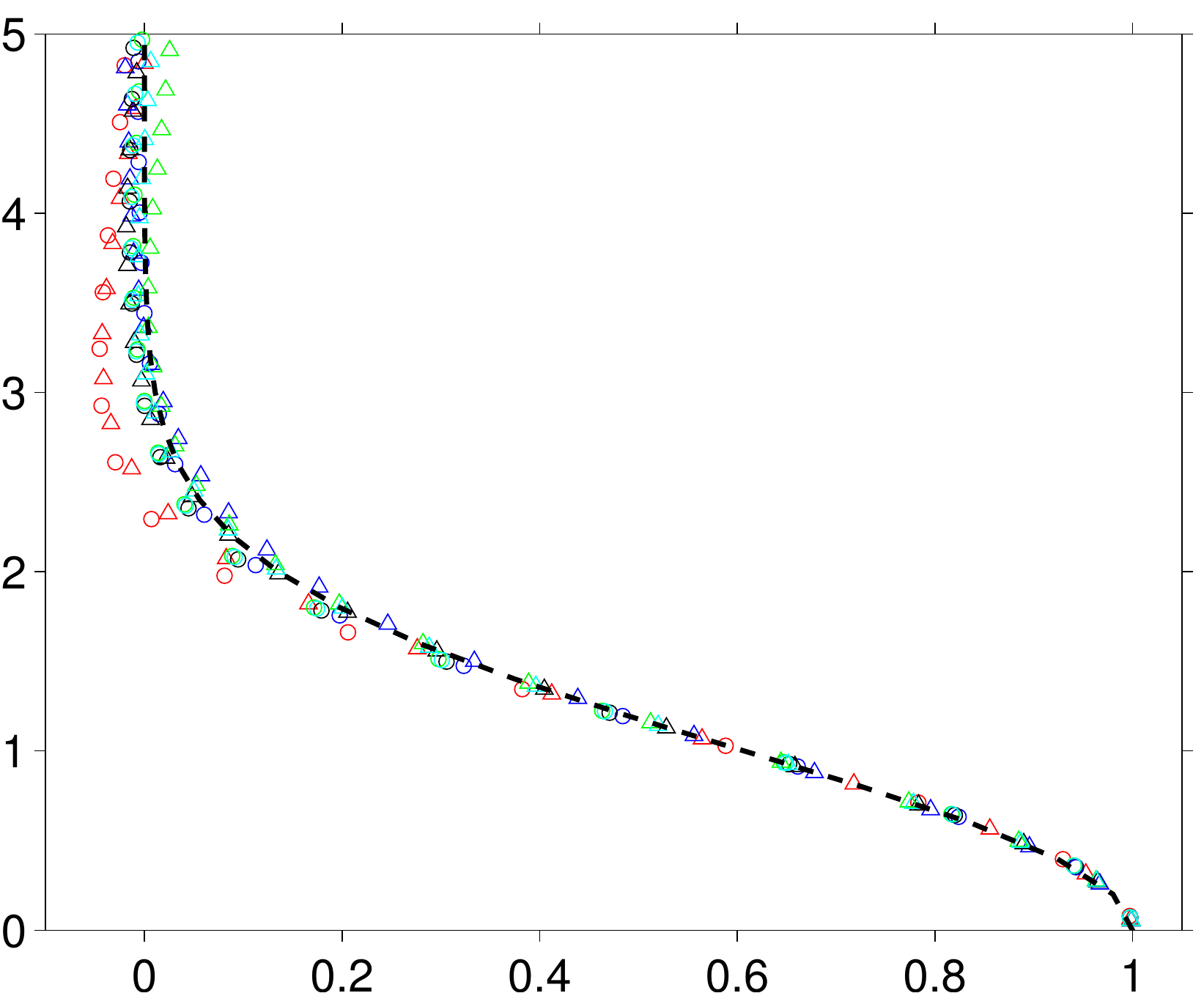}
      \\
      \centerline{$u_d/u_{d0}$}
    \end{minipage}
  \caption{
    Average velocity deficit in the particle wakes as a function of
    the normalized spanwise coordinate, given at two downstream
    locations: $\circ$, $x/R=10$; $\vartriangle$, $x/R=20$. The color
    code indicates the wall-distance of the averaging slab (as given
    in figure~\ref{fig-wake-avg-urel}).  
    The dashed line corresponds to a Gaussian function as defined in
    (\ref{equ-wakes-gaussian-fit-half-width}). 
    %
  }
  \label{fig-wake-avg-self-sim}
\end{figure}
%
%
Figure~\ref{fig-wake-avg-urel} 
shows the average relative
velocity on the streamwise axis through the particle center. It should
be noted that the curves do not quite reach the particle surface due
to the interpolation during particle-centered averaging, as explained
in Appendix~\ref{app-averaging-particle-centered}. 
In figure~\ref{fig-wake-avg-urel} 
the relative velocity
$\tilde{u}$ is normalized by the apparent lag $u_{lag}$.
It can be seen that with increasing distance from the particle, the
average relative velocity tends to unity for all wall-distances. 
It is also visible from figure~\ref{fig-wake-avg-urel} 
that 
for particles located in the slab centered at $y^{(s)+}=11$ the
approach to unity appears slightly faster (both on the upstream and
the downstream side) as compared to particles at larger wall
distances.    

In order to further investigate the recovery of the average velocity
in the wake, we define the normalized average velocity deficit in the
wall-parallel plane passing through the particle center, viz.
\begin{equation}\label{equ-def-avg-velocity-deficit}
  u_{d}(\tilde{x},y^{(s)},\tilde{z})=\frac{\tilde{u}_{\infty}(y^{(s)})
    -\tilde{u}((\tilde{x},0,\tilde{z})^T,y^{(s)})}
  {\tilde{u}_{\infty}(y^{(s)})}
  \,,
\end{equation}
where the average incoming relative velocity $\tilde{u}_{\infty}(y^{(s)})$
is defined as the maximum (over $\tilde{x}$) of
$\tilde{u}((\tilde{x},0,0)^T,y^{(s)})$.  
The velocity deficit along the streamwise axis through the particle
center, $u_{d0}$, is then simply defined as follows
\begin{equation}\label{equ-def-avg-velocity-deficit-center}
  u_{d0}(\tilde{x},y^{(s)})=u_{d}(\tilde{x},y^{(s)},0)
  \,.
\end{equation}
Figure~\ref{fig-wake-avg-1} shows the downstream evolution of the
normalized velocity deficit $u_{d0}$ in double logarithmic scale for
distances of up to $140R$. 
As observed in previous studies on turbulent flow around fixed single
spheres \citep{wu:94b,legendre:06,amoura:10}, roughly two regions can
be distinguished: a near-wake zone with a decay rate approximately
proportional to $x^{-1}$, and a far-wake with a decay of approximately
$x^{-2}$.  
The velocity deficit for particles in the averaging slab adjacent to
the wall ($y^{(s)+}=11$) is somewhat smaller for distances up to
$\tilde{x}/R\approx10$, as already observed above. Otherwise, the
curves corresponding to different wall-distances are equivalent to
within statistical uncertainty. 
It has been observed by \cite{legendre:06} that the
change in slope (from $x^{-1}$ to $x^{-2}$) takes place at a
downstream location where the velocity deficit and the turbulence
intensity are of the same order ($u_{d0}\approx\langle u^\prime
u^\prime\rangle^{1/2}$).   
The same is true in the present case: the change in slope
occurs at approximately $\tilde{x}/R=25$, where the velocity
deficit measures $u_{d0}\approx0.1$, which is indeed a value
comparable to the turbulence intensity at the corresponding
wall-normal locations (cf.\
figure~\ref{fig-stat-uu}$a$).   

\begin{figure}
  \begin{minipage}{3ex}
    $\displaystyle\frac{h_{wz}}{R}$
  \end{minipage}
  \begin{minipage}{.45\linewidth}
    \centerline{$(a)$}
    \includegraphics[width=\linewidth]
    {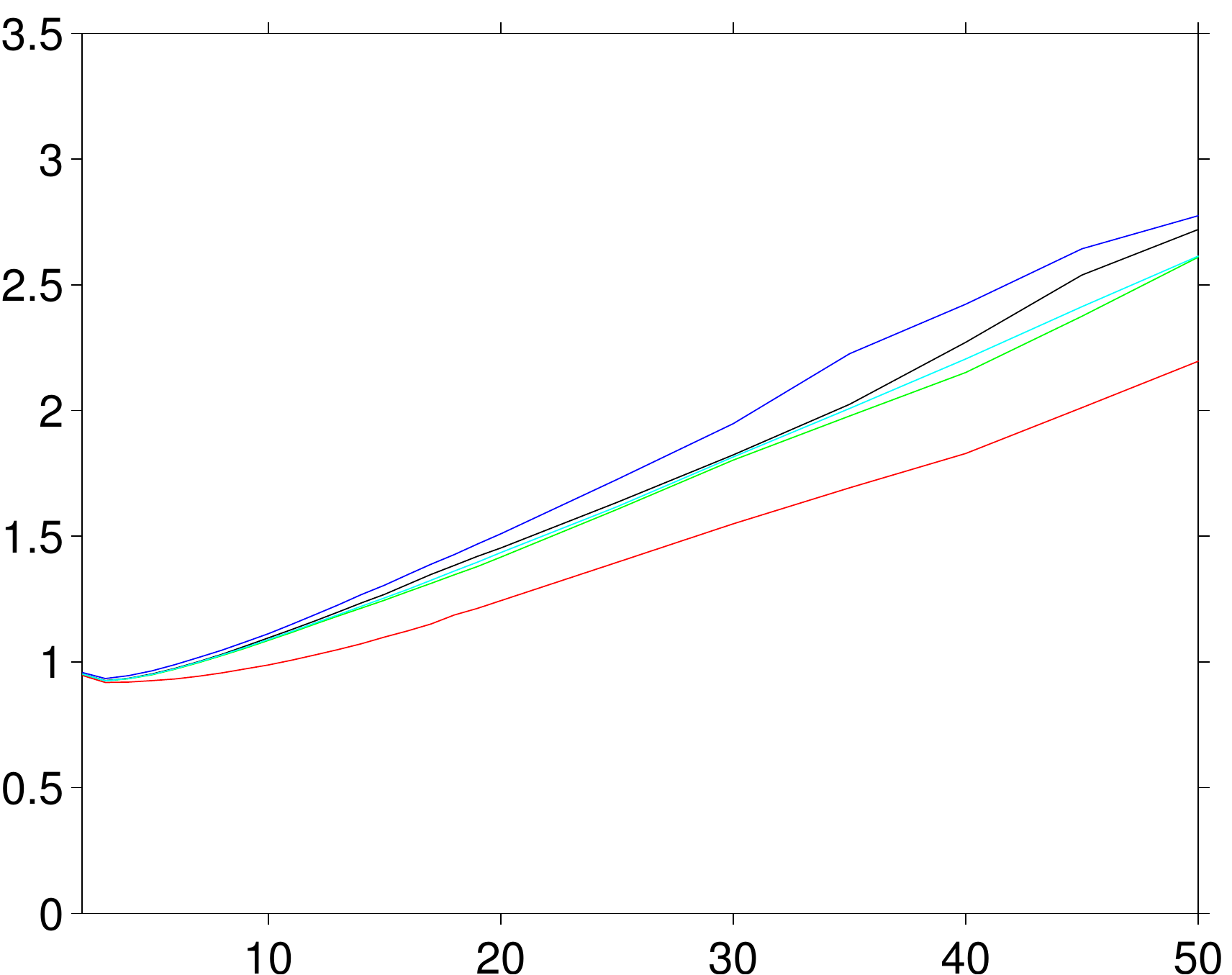}
    \\
    \centerline{$\tilde{x}/R$}
  \end{minipage}
  \begin{minipage}{3ex}
    $\displaystyle\frac{h_{wz}}{R}$
  \end{minipage}
  \begin{minipage}{.45\linewidth}
    \centerline{$(b)$}
    \includegraphics[width=\linewidth]
    {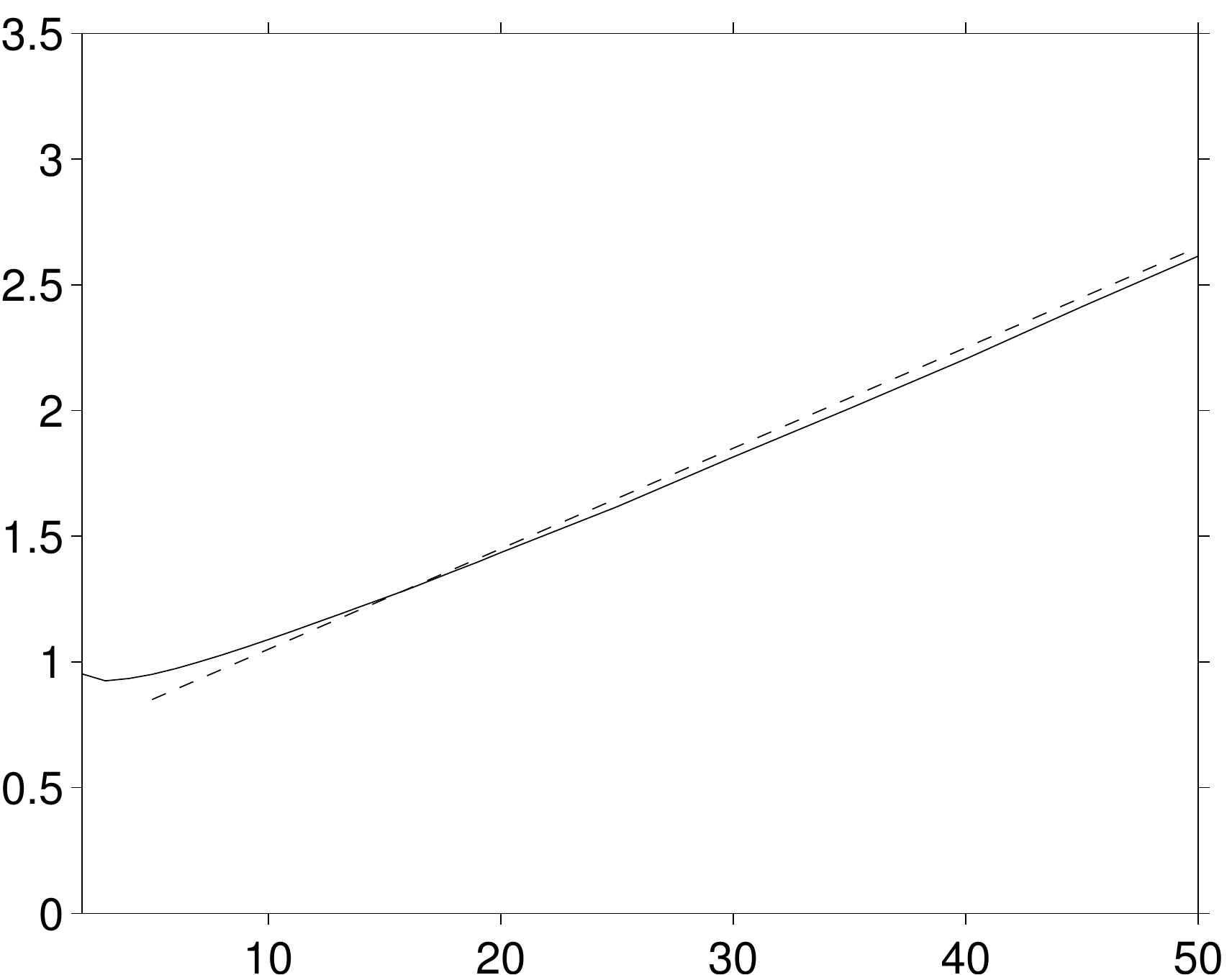}
    \\
    \centerline{$\tilde{x}/R$}
  \end{minipage}
  \caption{
    Spanwise half-width of the average particle wakes as a function of
    downstream distance. 
    $(a)$ shows data for individual wall-normal slabs, with
    color-coding as in figure~\ref{fig-wake-avg-urel}. 
    $(b)$ shows data averaged over the core of the channel (for
    $y/h\geq0.2$). 
    The dashed line in $(b)$ is given by the function
    $h_{wz}/R=0.04\,\tilde{x}/R+0.65$. 
  }
  \label{fig-wake-avg-half-width}
\end{figure}
The extent of the regions where power-law behavior is observed can be
deduced from figure~\ref{fig-wake-avg-1-comp} which shows the
compensated velocity deficit $u_{d0}\,\tilde{x}^n$ with exponents
$n=1$ and $n=2$. In order to further increase the number of available
samples, the data of all particles located at $y/h>0.2$ (i.e.\ slabs
centered at $y^{(s)}/h\geq0.25$) has been averaged. The figure shows
that a decay of the velocity deficit in the particle wakes according to
$x^{-2}$ takes place in a region approximately delimited by
$40\leq\tilde{x}/R\leq80$. 
By way of comparison, in relatively low-intensity turbulence ($I_r=0.037$)
\cite{legendre:06} found the $x^{-2}$ law to hold for
distances above $x_2=50R$, while \cite{amoura:10}
obtain the same evolution for $x_2\geq5R$ at higher turbulence intensity
($I_r\geq0.26$).   
The present results can therefore be qualified as consistent with the
trend exhibited in these two studies. 
%
%

Let us now turn to the question of self-similarity of the particle
wakes. 
\cite{wu:94b} and subsequent
authors \citep{bagchi:04,legendre:06} observed that cross-stream
profiles of the velocity deficit $u_{d}$ in the wake of isolated fixed 
spheres in turbulent surroundings follow a Gaussian function, viz.
\begin{equation}\label{equ-wakes-gaussian-deficit}
  \frac{u_d}{u_{d0}}=\exp\left(
    -\frac{\tilde{z}^2}{2h_{wz}^2}
    \right)
    \,.
\end{equation}
An appropriate length scale in (\ref{equ-wakes-gaussian-deficit}) is
the half-width $h_{wz}$ defined as the lateral position where the
following relation holds: 
\begin{equation}\label{equ-wakes-gaussian-fit-half-width}
  u_d(\tilde{z}=h_{wz})/u_{d0}=\exp(-1/2)
  \,.
\end{equation}
Figure~\ref{fig-wake-avg-self-sim} shows that the cross-wake profiles
of the velocity deficit in the present case do indeed follow the
Gaussian function (\ref{equ-wakes-gaussian-deficit}) reasonably well;
this is true for all wall-distances and over a substantial axial
distance downstream of the particles. 
The streamwise evolution of the spanwise half-width
$h_{wz}$ as defined in (\ref{equ-wakes-gaussian-fit-half-width}) is
shown in figure~\ref{fig-wake-avg-half-width}. 
The graph in figure~\ref{fig-wake-avg-half-width}$(a)$ demonstrates
that the particles' wall-distance has a minor effect upon the wake
half-width, as all curves have very similar evolutions;
again only the averaging slab adjacent to the wall ($y^{(s)+}=11$)
represents a small exception, with a computed half-width which is
systematically at slightly smaller values. 

In figure~\ref{fig-wake-avg-half-width}$(b)$ the core-averaged
half-width (i.e.\ found upon averaging over all particles located at
$y/h>0.2$) is shown. A region of linear growth, i.e.\ 
\begin{equation}\label{equ-wakes-expansion-law-fit}
  h_{wz}/R=\alpha\tilde{x}/R+\beta
  \,,
\end{equation}
which extends over $15\lesssim\tilde{x}/R\lesssim50$ is observed. 
Let us recall that a linear expansion $h_{zw}\sim\tilde{x}$ is
different from the case of spheres in uniform flow, in which laminar
wakes obey $h_{zw}\sim\tilde{x}^{1/2}$ and turbulent wakes exhibit
$h_{zw}\sim\tilde{x}^{1/3}$. 
The streamwise evolution in the present case (many mobile particles in
turbulent background flow) is quite accurately represented by the
coefficient values $\alpha=0.04$ and $\beta=0.65$ in the linear
expansion law (\ref{equ-wakes-expansion-law-fit}).   
At much lower relative turbulence intensity ($\tilde{I}_r=0.037$),
\cite{legendre:06} have found a value of $\alpha=0.024$ for the
wake expansion rate.  
\cite{bagchi:04}, on the other hand, have obtained a value of
$\alpha=0.135$ for their cases with relative turbulence intensity of
$\tilde{I}_r=0.1$, independently of the particle Reynolds number.
%
%
%
%
The analytic results of \cite{eames:11} for wake spreading in
turbulent surroundings do suggest a value of the expansion factor (in
the linear spreading regime) comparable to the value of the relative
turbulence intensity, viz.\ $\alpha\approx\tilde{I}_r$. 
This is apparently not the case in our present flow. 
However, it should be kept in mind that in the core of the channel,
the dominant contribution of turbulent kinetic energy stems from the
particle wakes themselves and not from `incoming' turbulence in the
classical sense. Therefore, some of the assumptions made in the
derivation of their model (in particular the homogeneous-isotropic
structure of the turbulent flow field) do not appear to apply. 
%
\begin{figure}
  \centering
  \begin{minipage}{10ex}
    $\displaystyle \frac{\widetilde{
        {u}_{r,i}^{\prime\prime}
        {u}_{r,i}^{\prime\prime}
      }/2}
    {k_{r}^{ref}}$
  \end{minipage}
  \begin{minipage}{.45\linewidth}
    \includegraphics[width=\linewidth]
    {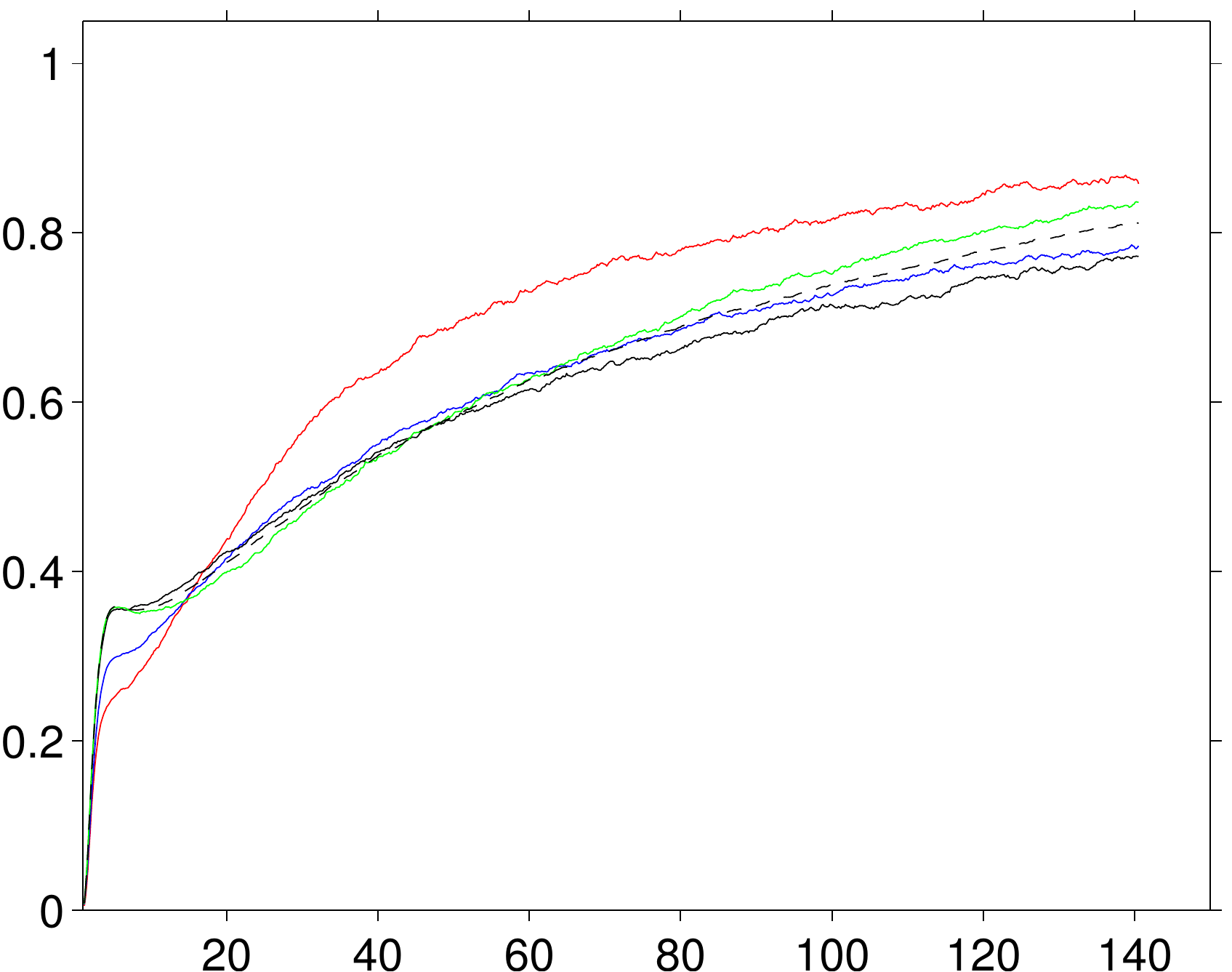}
    \\
    \centerline{$\tilde{x}/R$}
  \end{minipage}
  \caption{%
    Energy of 
    covariances 
    between relative velocity fluctuations 
    $\widetilde{u_{r,i}^{\prime\prime}
      u_{r,i}^{\prime\prime}}$ downstream of the particles, 
    %
    normalized by the reference kinetic energy given in
    (\ref{equ-wakes-def-ref-relative-vel-tke})  
    %
    The color code indicates the wall-distance of the averaging slab
    (as given in figure~\ref{fig-wake-avg-urel}).
    The dashed line corresponds to the data averaged over the core of
    the channel (for $y/h\geq0.2$).
    %
   }
  \label{fig-wake-avg-tke-ax}
\end{figure}
\subsubsection{Fluctuations}
By using the definition of the fluctuating velocities of each phase 
(\ref{equ-wakes-def-decompose-relative-velocity-fluid-particle}) the
definition of the covariances between fluctuating relative velocities
(\ref{equ-def-correlation-relative-velocity-fluct}) can be rewritten
as follows:   
\begin{equation}\label{equ-def-correlation-relative-velocity-fluct-2}
  \widetilde{
    {u}_{r,\alpha}^{\prime\prime}
    {u}_{r,\alpha}^{\prime\prime}
  }(\tilde{\mathbf{x}},y^{(s)})
  =
  \langle
  {u}_{f,\alpha}^{\prime\prime}(\tilde{\mathbf{x}},t)
  \,
  {u}_{f,\alpha}^{\prime\prime}(\tilde{\mathbf{x}},t)
  \rangle_{p,t}
  +
  \langle
  {u}_{p,\alpha}^{(i)\prime\prime}(t)
  \,
  {u}_{p,\alpha}^{(i)\prime\prime}(t)
  \rangle_{p,t}
  -2
  \langle
  {u}_{f,\alpha}^{\prime\prime}(\tilde{\mathbf{x}},t)
  \,
  {u}_{p,\alpha}^{(i)\prime\prime}(t)
  \rangle_{p,t}
  \,.
\end{equation}
This relation shows that the covariances of fluctuations w.r.t.\ the
average defined in (\ref{equ-def-avg-relative-velocity}) result from
three contributions: (i) covariances of the fluctuating fluid
velocity field (conditioned on particle presence), 
(ii) covariances of particle velocity fluctuations, and
(iii) covariances between particle velocity
fluctuations and particle-conditioned fluid velocity fluctuations. 
For fixed particles the last two contributions in
(\ref{equ-def-correlation-relative-velocity-fluct-2}) vanish
identically.   
Furthermore, for large distances from the particle (i.e.\ for large
values of $\tilde{x}$ and/or $\tilde{z}$) we expect the last
contribution (covariances between fluid and particle velocity
fluctuations) to vanish. 
In that long-distance limit, the contribution from the
particle-conditioned fluid velocity fluctuation covariances (first
term on the r.h.s.\ of
\ref{equ-def-correlation-relative-velocity-fluct-2}) is expected to
tend towards the simple (unconditioned) plane-and-time-averaged fluid
covariance $\langle{u}_{f,\alpha}^{\prime}{u}_{f,\alpha}^{\prime}\rangle$. 
Therefore, we define the following velocity scale:
\begin{equation}\label{equ-wakes-def-ref-relative-velocity}
  u_{r,\alpha}^{ref}
  =
  \left(
  \langle
  {u}_{f,\alpha}^{\prime}{u}_{f,\alpha}^{\prime}  
  \rangle
  +
  \langle
  {u}_{p,\alpha}^{(i)\prime\prime}(t)
  \,
  {u}_{p,\alpha}^{(i)\prime\prime}(t)
  \rangle_{p,t}
  \right)^{1/2}
  \,,
\end{equation}
which is expected to allow for a reasonable normalization of the
covariances of particle-conditioned relative velocities 
(\ref{equ-def-correlation-relative-velocity-fluct}).
From (\ref{equ-wakes-def-ref-relative-velocity}) we can derive a
reference fluctuation energy  $k_{r}^{ref}$, viz. 
\begin{equation}\label{equ-wakes-def-ref-relative-vel-tke}
  k_{r}^{ref}
  =
  u_{r,i}^{ref}\,u_{r,i}^{ref}/2
  \,.
\end{equation}
%
The downstream evolution of fluctuation energy 
$\widetilde{
  {u}_{r,i}^{\prime\prime}
  {u}_{r,i}^{\prime\prime}}/2$ 
on the axis through the particle center is shown in
figure~\ref{fig-wake-avg-tke-ax}.  
At all wall-distances, the fluctuation energy increases from zero at
the rear particle surface and tends towards unity at large downstream
distances. 
For intermediate distances, two regimes can be distinguished. 
In the near-wake ($\tilde{x}/R\lesssim10$, corresponding to the region
where $u_{d0}\sim\tilde{x}^{-1}$) a rapid increase of fluctuation
energy is observed, which evolves approximately linearly with
$\tilde{x}$.  
Further downstream, the approach towards unity is much slower,
approximately following a logarithmic dependency on the downstream
distance $\tilde{x}$. 
%
%
%
%

%% file: conclusion.tex
\section{Conclusion}
\label{sec-conclusion}
In the present study we have revisited the case of vertical plane
channel flow seeded with finite-size heavy particles already
investigated by \cite{uhlmann:08a}. An additional DNS has been carried
out with identical parameters as in the main simulation of that
reference, except for the streamwise period of the computational
domain which has been doubled. 

The new dataset has first been compared to 
the previous simulation data of \cite{uhlmann:08a}, 
in order to determine the influence of the box size on the
largest flow structures. It was observed that the columnar flow
structures, which are induced by the presence of particles, are still 
not fully decorrelated in the prolonged domain. 
As a consequence, the Lagrangian auto-correlation of particle
velocity is still biased by the fact that particles re-encounter
long-lived flow structures after one return time (based upon the
apparent slip velocity and the domain size). 
Results for average Eulerian quantities such as the mean fluid and
particle velocity profiles are not strongly affected by the streamwise 
extension of the domain.  

Furthermore, we have analyzed the new dataset with respect to
additional aspects not previously considered in the context of
vertical particulate channel flow. 
First, we have conducted an analysis of the spatial distribution of
the disperse phase based upon Voronoi tessellation. It was found that
the particles are less disorderly distributed than in a random case,
slightly tending towards a homogeneous distribution. In addition, 
examining the aspect ratio of Voronoi cells has revealed that the
particles exhibit a weak tendency to align in the streamwise
direction, a trend which may be related to wake sheltering. 

Second, we have carried out an analysis of the statistics related to 
particle acceleration. 
Near the wall, mean particle acceleration is found to significantly
deviate from the mean acceleration of fluid particles. 
The standard deviation of particle acceleration, when normalized in
wall units, differs among the three spatial components. 
When expressed as fluctuations of force coefficients (with approximate
velocity scales based on the mean and rms fluid velocity of the
corresponding components), this variability can be interpreted in
terms of the standard drag law. 
Concerning the temporal correlation of particle acceleration data, we
have observed a first zero-crossing after several viscous time units,
followed by a slower decorrelation over several bulk time units. 
The p.d.f.'s of particle acceleration in the wall-normal and spanwise
directions are found to be consistent with a lognormal distribution as
proposed by \cite{qureshi:08}. The streamwise component, however,
deviates from that lognormal fit, exhibiting significant positive
skewness. One possible explanation for this positively-skewed
acceleration p.d.f.\ is through a non-linear drag mechanism which
would solely affect the streamwise component since it is the only
spatial component with a finite apparent slip velocity. 
This point certainly merits further investigation. 

Finally, we have performed particle-conditioned averaging of the flow
field in the vicinity of the particles. It was found that the
characteristics of the particle wakes in the present case are nearly
independent of the wall distance, except for the near-wall region
($y/h\lesssim0.2$). 
The average length of the recirculation zone in the wake of the
present particles (with relative flow Reynolds number based upon
apparent slip velocity of $132$ in the core of the channel) is
consistent with data for single 
fixed spheres investigated by \cite{bagchi:04} and \cite{zeng:10} at
comparable Reynolds number and turbulence intensity.  
In particular, our results in the buffer layer
\citep[compared to case 2 of][]{zeng:10} 
and on the centerline \citep[compared to case 4 of][]{bagchi:04}
feature a somewhat larger mean wake length. 
The streamwise velocity deficit on the axis through the particles is
found to decay as $x^{-1}$ in the near-wake and as $x^{-2}$ for
distances beyond approximately $40$ particle radii. 
As in previous studies involving fixed spheres in turbulent flow
\citep{wu:94b,bagchi:04,legendre:06}, 
the present average streamwise velocity profiles in the particles'
wakes can be fit with reasonable agreement to an exponential
function. The wake half-width resulting from the fit evolves almost
linearly with the downstream distance over a considerable interval. 
The energy of the velocity fluctuations with respect to their 
particle-conditioned average, which is by definition zero on the
particle surface, is found to evolve approximately linearly with
downstream distance in the near-wake, and according to a logarithmic
law in the far-wake. 
%
%

%% file: ack.tex
\section*{Acknowledgments}
The simulations were partially performed at the Barcelona
Supercomputing Center, LRZ M\"unchen and SCC Karlsruhe. 
The computer resources, technical expertise and assistance provided by  
these centers are thankfully acknowledged. 
AGK has received financial support through a FYS grant from KIT within
the framework of the German Excellence Initiative.
Thanks is also due to V\'eronique Roig, Fr\'ed\'eric Risso and
Dominique Legendre for suggesting the analysis of particle wakes. 

%% file: app_averaging.tex
\section{Averaging procedures}
\label{app-averaging}
\subsection{Particle-centered averaging}
\label{app-averaging-particle-centered}
Let us define an indicator function $\phi_{bin}^{(j)}(y)$ which
signals whether a given wall-normal position $y$ is located inside or
outside a particular wall-normal slab with index $j$, viz.
\begin{equation}\label{equ-def-ybin-indicator-fct}
  \phi_{bin}^{(j)}(y)=
  \left\{
    \begin{array}{lll}
      1&\mbox{if}&
      (j-1)\frac{2h}{N_{bin}}\leq y<j\frac{2h}{N_{bin}}
      \\
      0&\mbox{else}&
    \end{array}
  \right.
  \,,
\end{equation}
where $N_{bin}$ is the number of slabs used to span the channel
width. 
%
Similarly, we define an indicator function $\phi_{f}(\mathbf{x},t)$ for the
fluid phase: 
\begin{equation}\label{equ-def-fluid-indicator-fct}
  \phi_{f}(\mathbf{x},t)
  =
  \left\{
    \begin{array}{lll}
      1&\mbox{if}&
      \mathbf{x}\in\Omega_f(t)
      \\
      0&\mbox{else}&
    \end{array}
  \right.
  \,,
\end{equation}
where $\Omega_f(t)$ is the part of the computational domain $\Omega$
which is occupied by fluid at time $t$. 
%
We can now define a discrete counter field $\tilde{n}_{ijk}^{(s)}$ which holds
the number of samples obtained through averaging at a given grid node
with indices $i,j,k$ for a given $y$-slab with index $s$, viz.
\begin{equation}\label{equ-def-sample-counter}
  \tilde{n}_{ijk}^{(s)}
  =
  \sum_{m=1}^{N_{snap}}
  \sum_{l=1}^{N_p}
  \phi_{bin}^{(s)}(y_p^{(l)}(t^m))\,
  \phi_{f}(\tilde{\mathbf{x}}_{ijk}^{(l)}(t^m),t^m)
  \,,
\end{equation}
In equation (\ref{equ-def-sample-counter}) the symbol $t^m$ indicates
the time corresponding to the $m$th snapshot in the database (comprising a
total of $N_{snap}$ snapshots) and 
$\tilde{\mathbf{x}}_{ijk}^{(l)}(t^m)$ is a coordinate relative to the
$l$th particle's center position at time $t^m$, as defined in 
(\ref{equ-def-relative-position}). 

The actual averaging can now be performed analogously to
(\ref{equ-def-sample-counter}), including a division by the local
number of samples. For a vector field
$\boldsymbol{\xi}(\mathbf{x},t)$ we define the averaging operator as follows:
\begin{equation}\label{equ-def-avg-operator-wake-time-particles-ybinned}
  \langle
  \boldsymbol{\xi}
  \rangle_{p,t}(\tilde{\mathbf{x}}_{ijk},y^{(s)})
  =
  \frac{1}{\tilde{n}_{ijk}^{(s)}}
  \sum_{m=1}^{N_{snap}}
  \sum_{l=1}^{N_p}
  \phi_{bin}^{(s)}(y_p^{(l)}(t^m))\,
  \phi_{f}(\tilde{\mathbf{x}}_{ijk}^{(l)}(t^m),t^m)\,
  \boldsymbol{\xi}(\tilde{\mathbf{x}}_{ijk}^{(l)}(t^m),t^m)
  \,.
\end{equation}
In practice the coordinates $\tilde{\mathbf{x}}_{ijk}^{(l)}(t^m)$ of the
grid nodes covering the predefined averaging volume do not coincide
with the grid used in the direct numerical simulation. This means that
the vector field to be averaged ($\boldsymbol{\xi}$ in equation 
\ref{equ-def-avg-operator-wake-time-particles-ybinned}) is not
directly available at the coordinates $\tilde{\mathbf{x}}_{ijk}^{(l)}(t^m)$. 
Therefore, the averaging process defined
in (\ref{equ-def-avg-operator-wake-time-particles-ybinned}) involves
spatial interpolation, which has been realized with a tri-linear
formula. 
As a consequence of the use of the fluid indicator function, the
particle-centered average fields do not quite reach the particle
surface.  

The number of (uniformly spaced) bins for evaluating averages defined by
(\ref{equ-def-avg-operator-wake-time-particles-ybinned}) was chosen as
$N_{bin}=20$. The number of snapshots amounts to
%
$N_{snap}=97$,  
yielding approximately 
80000
samples per slab. 
Further averaging over all bins with $y^{(s)}/h\geq0.25$ (also
referred to as ``core-averaging'' in the main text) yields
approximately 
%
$650000$ 
samples for the quantities shown in 
figures~\ref{fig-wake-avg-1-comp} and
\ref{fig-wake-avg-half-width}$(b)$.
\begin{figure}
  \begin{minipage}{2ex}
    \rotatebox{90}{
      $\langle u_f\rangle/u_b$, $\langle u_p\rangle/u_b$}
  \end{minipage}
  \begin{minipage}{.45\linewidth}
    \centerline{$(a)$}
    \includegraphics[width=\linewidth]
    {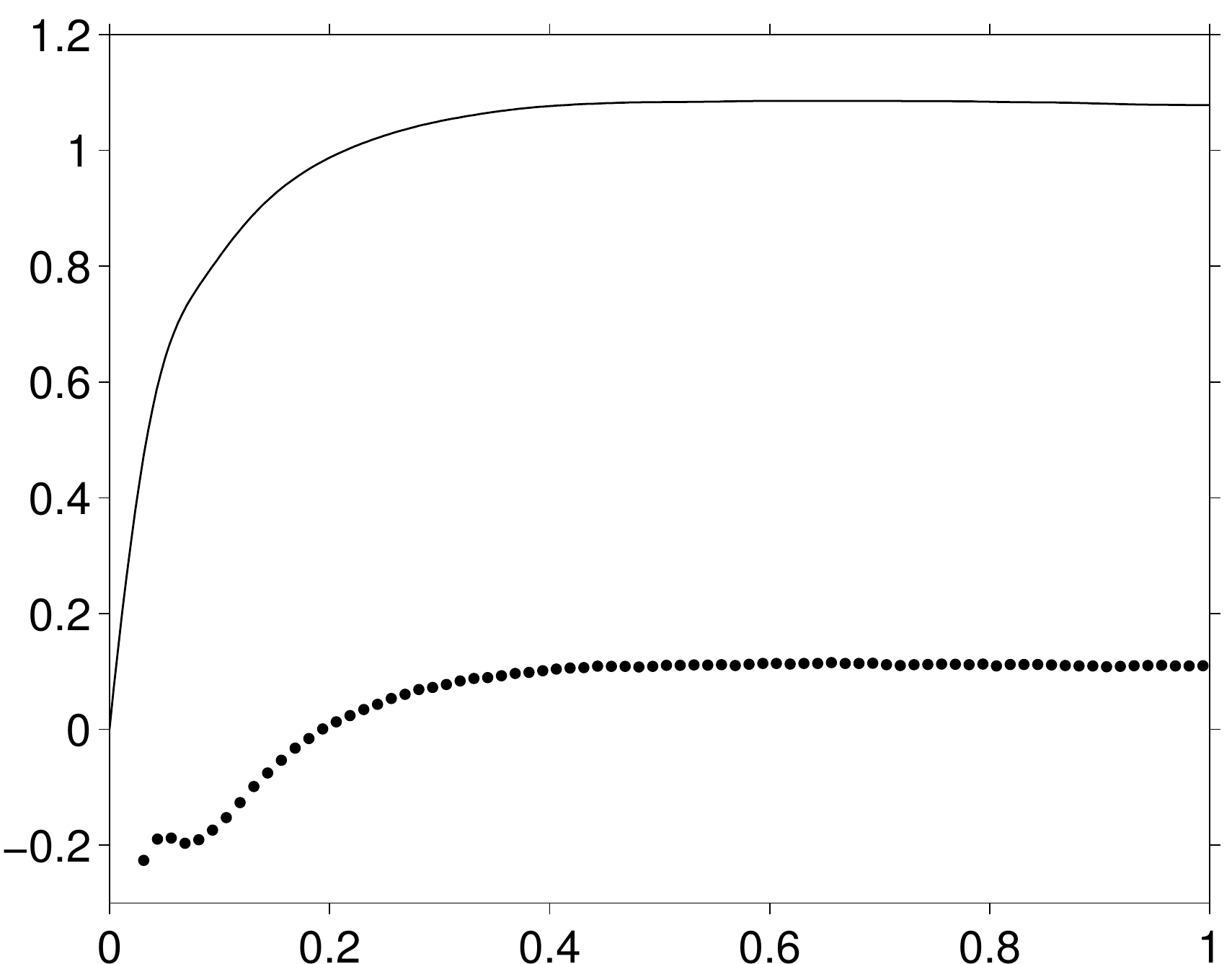}
    \\
    \centerline{$y/h$}
  \end{minipage}
  \begin{minipage}{2ex}
    \rotatebox{90}{
      $(\langle u_p\rangle-\langle u_f\rangle)/u_b$}
  \end{minipage}
  \begin{minipage}{.45\linewidth}
    \centerline{$(b)$}
    \includegraphics[width=\linewidth]
    {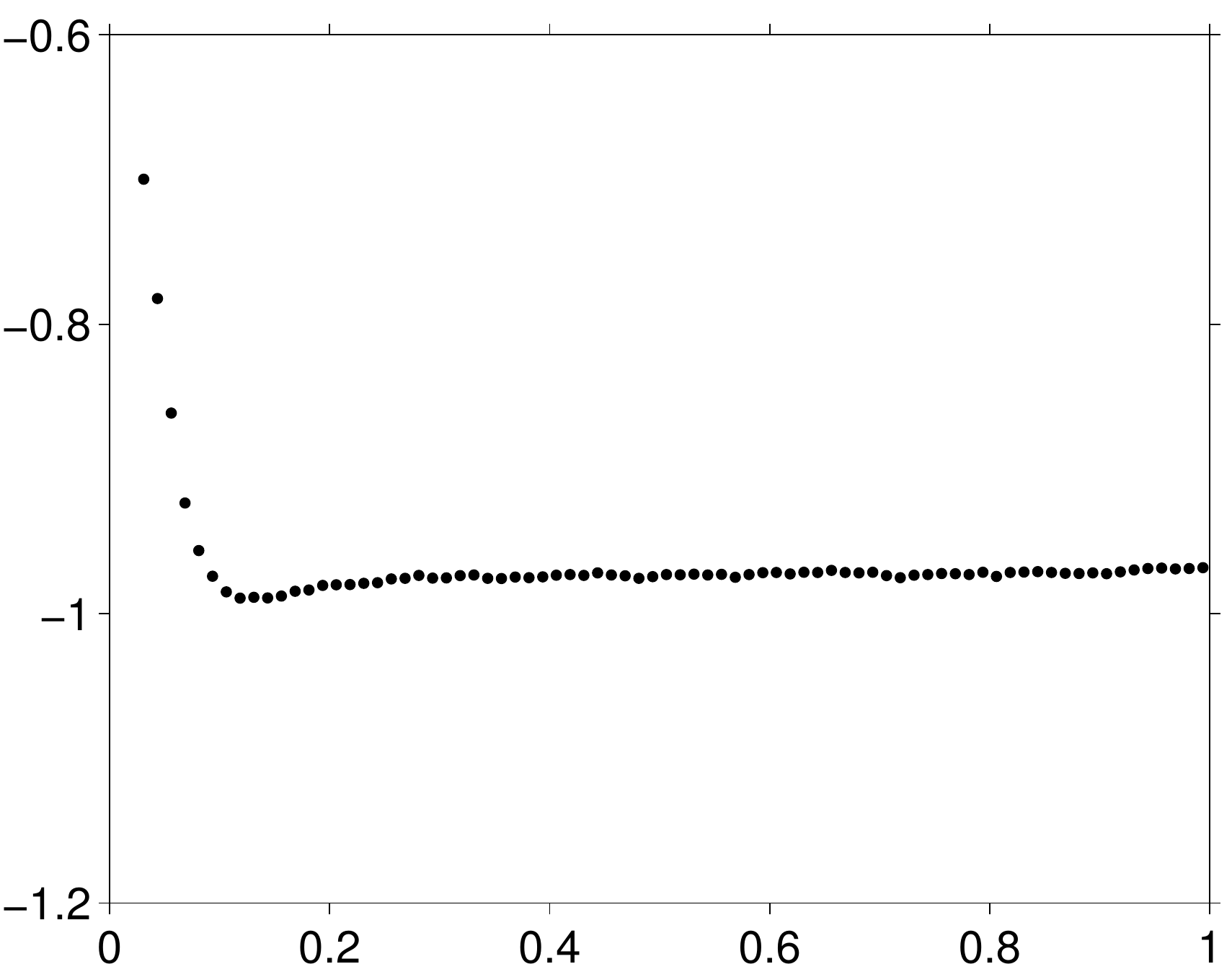}
    \\
    \centerline{$y/h$}
  \end{minipage}
  \caption{%
    $(a)$ Mean velocity profiles of both phases in the present
    simulation: \solid, fluid phase; 
    $\bullet$, solid phase. 
    $(b)$ Apparent slip velocity between the phases. 
    The fluid phase data is averaged according to the operator defined
    in (\ref{equ-def-avg-operator-plane-and-time-fluid-only}). 
}
  \label{fig-stat-pure-um}
\end{figure}
\subsection{Wall-parallel plane and time averaging}
\label{app-averaging-plane-and-time}
The standard averaging over wall-parallel planes and time (not
conditioned upon particle presence) can now conveniently be defined
using the notation and the indicator functions already introduced in 
\ref{app-averaging-particle-centered}.  
Let us first define a counter of fluid sample points in a
wall-parallel plane at a given wall-distance with index $j$, viz.
\begin{equation}\label{equ-def-sample-counter-plane-and-time-fluid-only}
  n_{j}
  =
  \sum_{m=1}^{N_{time}}
  \sum_{i=1}^{N_x}
  \sum_{k=1}^{N_z}
  \phi_{f}(\mathbf{x}_{ijk},t^m)
  \,.
\end{equation}
Note that the number of data-sets $N_{time}$ used for this quantity is
much larger than $N_{snap}$ in the case of particle-centered averaging
(equations \ref{equ-def-sample-counter} and
\ref{equ-def-avg-operator-wake-time-particles-ybinned}), since
standard averaging is performed at runtime, while the latter is
carried out at a post-processing stage. 
Averaging over wall-parallel planes and in time, while considering
only grid points being located in the fluid domain, is defined
as follows:
\begin{equation}\label{equ-def-avg-operator-plane-and-time-fluid-only}
  \langle
  \boldsymbol{\xi}
  \rangle(y_j)
  =
  \frac{1}{{n}_{j}}
  \sum_{m=1}^{N_{time}}
  \sum_{i=1}^{N_x}
  \sum_{k=1}^{N_z}
  \phi_{f}(\mathbf{x}_{ijk},t^m)\,
  \boldsymbol{\xi}(\mathbf{x}_{ijk},t^m)
  \,.
\end{equation}
Consequently, $\langle\boldsymbol{\xi}\rangle$ is a function of
wall-distance alone, whereas $\langle\boldsymbol{\xi}\rangle_{p,t}$ is
a three-dimensional field for each wall-normal slab $y^{(s)}$. 

Note that purely for the purpose of strict comparison between the
present simulation results and data from \cite{uhlmann:08a}, the
following averaging operator which does not distinguishes between the
solid and fluid phases is used ('composite' averaging, cf.\
figures~\ref{fig-stat-um} and \ref{fig-stat-uu}):
\begin{equation}\label{equ-def-avg-operator-plane-and-time-COMPOSITE}
  \langle
  \boldsymbol{\xi}
  \rangle_c(y_j)
  =
  \frac{1}{N_{time}N_xN_z}
  \sum_{m=1}^{N_{time}}
  \sum_{i=1}^{N_x}
  \sum_{k=1}^{N_z}
  \boldsymbol{\xi}(\mathbf{x}_{ijk},t^m)
  \,.
\end{equation}
\subsection{Binned averages over particle-related quantities}
\label{app-averaging-binned-particles}
Concerning Lagrangian quantities, we employ averages over wall-normal
bins, using the indicator function given in
(\ref{equ-def-ybin-indicator-fct}). 
The sample counter for each bin with index $s$ is computed over the
number $N_{time}^{(p)}$ of available particle fields and summing over
all particles, viz.
\begin{equation}\label{equ-def-sample-counter-binned-particles}
  \hat{n}^{(s)}
  =
  \sum_{m=1}^{N_{time}^{(p)}}
  \sum_{l=1}^{N_p}
  \phi_{bin}^{(s)}(y_p^{(l)}(t^m))\,
  \,.
\end{equation}
The binned average (over time and the number of particles) of a
Lagrangian quantity $\zeta_p$ is defined as follows: 
\begin{equation}\label{equ-def-avg-operator-binned-particles}
  \langle
  \boldsymbol{\zeta_p}
  \rangle
  (y^{(s)})
  =
  \frac{1}{\hat{n}^{(s)}}
  \sum_{m=1}^{N_{time}^{(p)}}
  \sum_{l=1}^{N_p}
  \phi_{bin}^{(s)}(y_p^{(l)}(t^m))\,
  \boldsymbol{\zeta_p}^{(l)}(t^m)
  \,.
\end{equation}
%
%
%
In this manuscript we have chosen the number of (uniformly spaced)
bins for evaluating averages defined by
(\ref{equ-def-avg-operator-binned-particles}) as 
$N_{bin}=160$. 
The present data corresponds to a total number of approximately
%
$1.8\cdot10^7$ 
samples. 

%% file: app_pure_stats.tex
\begin{figure}
  \begin{minipage}{3ex}
    \rotatebox{90}{
      $\langle u_{f,\alpha}^\prime u_{f,\alpha}^\prime\rangle^{1/2}/u_b$, 
      $\langle u_{p,\alpha}^\prime u_{p,\alpha}^\prime\rangle^{1/2}/u_b$
      }
  \end{minipage}
  \begin{minipage}{.45\linewidth}
    \centerline{$(a)$}
    \includegraphics[width=\linewidth]
    {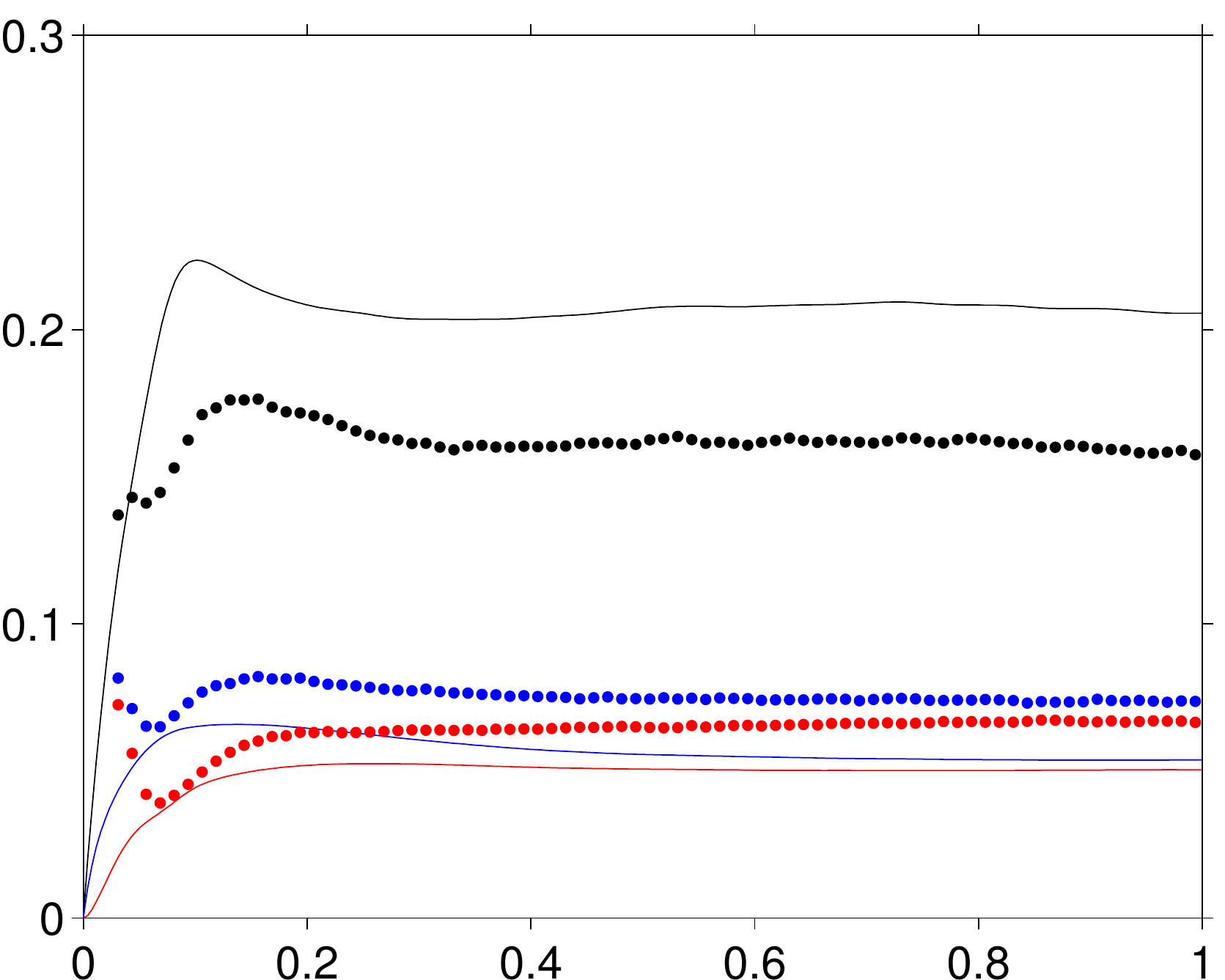}
    \\
    \centerline{$y/h$}
  \end{minipage}
  \begin{minipage}{3ex}
    \rotatebox{90}{
      $\langle u_{f}^\prime v_{f}^\prime\rangle/u_b^2$, 
      $\langle u_{p}^\prime v_{p}^\prime\rangle/u_b^2$
    }
  \end{minipage}
  \begin{minipage}{.45\linewidth}
    \centerline{$(b)$}
    \includegraphics[width=\linewidth]
    {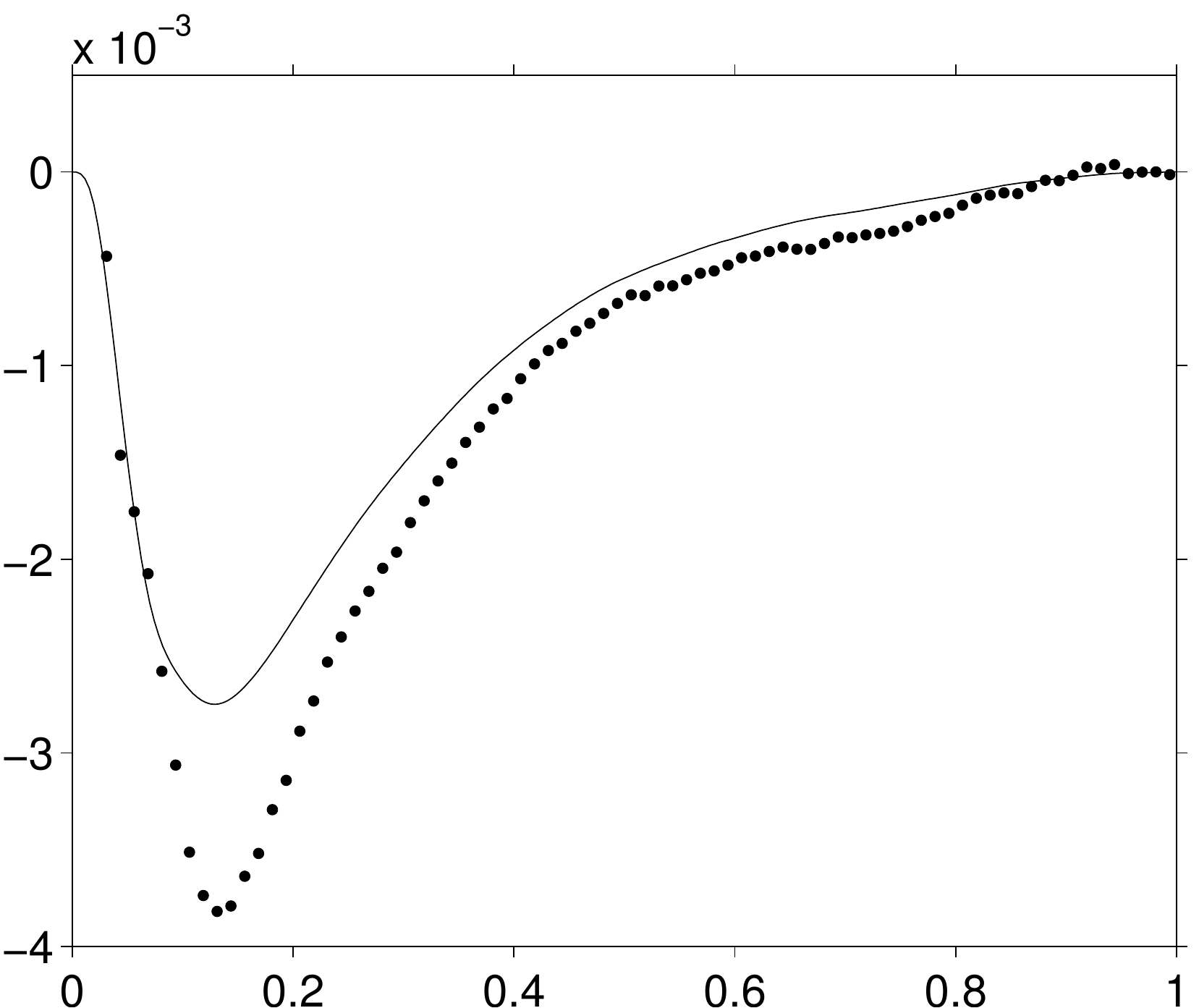}
    \\
    \centerline{$y/h$}
  \end{minipage}
  \caption{%
    $(a)$ R.m.s.\ of velocity fluctuations of both phases in the
    present simulation, with lines
    corresponding to the fluid phase, 
    and symbols to the particulate phase. 
    The color coding indicates streamwise (black),
    wall-normal (red), spanwise (blue) components.
    $(b)$ Reynolds shear stress of fluid velocity fluctuations as well
    as corresponding velocity correlation of the particle motion.
    All quantities are normalized in bulk units. 
    The fluid phase data is averaged according to the operator defined
    in (\ref{equ-def-avg-operator-plane-and-time-fluid-only}). 
}
  \label{fig-stat-pure-uu}
\end{figure}
\section{Eulerian statistics for the present simulation}
\label{app-pure-stats}
Figures~\ref{fig-stat-pure-um} and \ref{fig-stat-pure-uu} show the 
same quantities as presented in figures~\ref{fig-stat-um} and
\ref{fig-stat-uu} (for the present simulation), but defined with the 
correct averaging operator which only considers grid nodes being
instantaneously located in the fluid domain (cf.\ definition in
equation~\ref{equ-def-avg-operator-plane-and-time-fluid-only}). 
These quantities are presented for future reference. 